\documentclass[useAMS,usenatbib]{mn2e}
\usepackage{graphicx}
\usepackage{graphics,float}                         % graphics
\usepackage{epsf}
\usepackage{amsmath}                                % Eqn environments
\usepackage{amssymb}
\usepackage{picinpar}                               % allows windows
\usepackage{relsize,setspace,subfigure}
\usepackage{tabularx}
\usepackage[innercaption]{sidecap}                  % side captions
\usepackage[figuresright]{rotating}
\usepackage{url}
\usepackage{aas_macros}
\usepackage{natbib}

\onecolumn
\title[Merging Tree Algorithm of Growing Voids]{Merging Tree Algorithm of Growing Voids in Self Similar and CDM Models}
\author[Russell]{Esra Russell$^{1,2}$\thanks{E-mail: esrarussell@iyte.edu.tr}
\\
$^{1}$Kapteyn Astronomical Institute, University of Groningen, P.O.
Box 800, 9700 AM, Groningen, The Netherlands.
\\
$^{2}$Izmir Institute of Technology, Department of Mathematics, 35430, Gulbahcekoyu/Urla, Izmir, Turkey.}

\begin{document}
\date{Accepted .... Received ...; in original form ...}

\pagerange{\pageref{firstpage}--\pageref{lastpage}} \pubyear{2013}

\maketitle

\label{firstpage}

\begin{abstract}
Observational studies show that voids are prominent features of the large scale structure of the present day Universe. Even though their emerging from the primordial density perturbations and evolutionary patterns differ from dark matter halos, $N$-body simulations and theoretical models have shown that voids also merge together to form large void structures. In this study, following \cite{sw}, we formulate an analytical approximate description of the hierarchical void evolution of growing voids by adopting the halo merging algorithm given by \cite{lace} in the Einstein de Sitter (EdS) Universe. To do this, we take into account the general volume distribution of voids which consists of two main void processes: merging and collapsing. We show that the volume distribution function can be reduced to a simple form, by neglecting the collapsing void contribution since the collapse process is negligible for large size voids. Therefore, the void volume fraction has a contribution only from growing voids. This algorithm becomes the analogue of the halo merging algorithm. Based on this growing void distribution, we obtain the void merging algorithm in which we define and formulate void merging and absorption rates, as well as void size and redshift survival probabilities and also failure rates in terms of the self similar and currently favored dark energy dominated cold dark matter models in the EdS Universe.
\end{abstract}


\section{Introduction}

Early galaxy surveys have shown that voids are integral features of the observed Universe \citep{Rood,thompson,einasto1980,wp}. After the discovery of the Bo\"{o}tes void \citep{kir}, it has been shown that they are prominent features of the Cosmic Web \citep{bondmyers} filling $95\%$ of the total volume of the galaxy distribution \citep{kir,gehu,dac,shect,ei,plib}. From the perspective of the void-based description of the Cosmic Web, voids form the structure formation of the Universe \citep{icke,regge,PhDWeygaert,sw}. As voids expand, matter is squeezed in between them, and sheets and filaments form the void boundaries. This view is supported by numerical studies and computer simulations of the gravitational evolution of voids in more complex and realistic configurations \citep{wp}.

Voids can have a broad range of shapes and sizes. Observations \citep{plib,Hoyle02,hoylevogley2004,tully2008,tk2008,kraan,Panvogeley2012,sutter} and $N$-body simulations \citep{regge,wk,dubinski1993,benson,gott,colb,tinker2009,2013MNRAS.428.3409A,2012MNRAS.426..440B,2013MNRAS.434.1192R} show that voids can have sizes in the range $5-135 h^{-1}Mpc$. There are voids that possess sizes smaller than this range; on the basis of the Catalog of Neighbouring Galaxies \citep{Karachentsev2004}, \cite{tk2008} found $30$ minivoids with a range of sizes, $0.7 - 3.5 h^{-1}Mpc$.
%One of the most detailed works on void distribution and statistics is done by \cite{Panvogeley2012}. They define $1054$ voids in the northern galactic hemisphere with $R > 10 h^{-1} Mpc$ at redshifts $z < 0.107$ by using Sloan Digital Sky Survey Data Release $7$ (SDSS DR7). In their survey, they find approximately $30 h^{-1} Mpc$ as the largest void size. Recently, again using SDSS DR7, Sutter et al. released a void catalogue up to redshift $z = 0.2$ and the Luminous Red Galaxy sample out to $z = 0.44$ \citep{sutter}. They show that voids have radii in the range $5-135 h^{-1}Mpc$.
However, firm upper limits on the size of voids have not yet been set \citep{wp} due to two reasons; first, in a model, the void volume distribution is not expected to have a firm upper limit, but rather an exponential tail, and secondly this statement depends on the definition of a void. \cite{2013MNRAS.428.3409A} states that at the top of the hierarchy the peak of the void size distribution is approximately $11 h^{-1}Mpc$ in their simulation and this agrees with the radius of the Local Void estimated by \cite{olga2011}. Studies on the size of the Local Void based on its dimensions and the extent of its galaxy population are still debated. \cite{tully2008} note that the region of low density extends up to distances of $\approx 20-30$ $h^{-1}Mpc$. \cite{kraan} suggest that the local region of depression may be even larger, neighboring a more distant void in Microscopium/Sagittarius. Furthermore, these authors denote the existence of a few filaments inside this volume, dividing the supervoid into $2$ or $3$ voids with a size of $10$-$30$ $h^{-1}Mpc$. Recently \cite{Courtois2012} reconstructed the full linear density and three dimensional velocity
fields in terms of the dark energy dominated cold dark matter ($\Lambda$CDM) model with the cosmological parameters derived using data from Wilkinson Microwave Anisotropy Probe $5$ year data release (WMAP$5$). In this study, they show that the prominent structure of the Local Supercluster is wrapped in a horseshoe shape underdensity with the Local Void. \cite{piran2006} argue that voids selected from catalogues of luminous galaxies should be larger than those selected from faint and dark matter galaxies: the characteristic radii range is from $\approx 5$ to $10$ $h^{-1} Mpc$. Within large voids, the mass function is nearly independent of the size of the underdensity, but finite-size effects play a significant role in small voids, $R \approx 7 h^{-1} Mpc$ \citep{piran2006}. The work of \cite{hoylevogley2004} represents the most elaborate study of voids. They find that voids with characteristic radii $15$ $h^{-1} Mpc$ fill $\approx 35$ per cent of the Universe. In this study they restrict their search to voids with radii greater than $10$ $h^{-1} Mpc$ since smaller voids are difficult to identify due to confusion with random fluctuations in the galaxy distribution. They used a void finding algorithm given by \cite{Hoyle02}. They describe the steps of this algorithm which classified galaxies as wall or void galaxies, detecting empty cells in the distribution of wall galaxies, growth of the largest empty spheres and the enhancement of the void volume. Their work was extended into the definition of a void galaxy catalogue from the Sloan Digital Sky Survey $7$th data release (SDSS DR$7$) \citep{Panvogeley2012,Hoylevogeleypan2012}. \cite{Hoylevogeleypan2012} define $1054$ voids in the northern galactic hemisphere with $R > 10 h^{-1} Mpc$ at redshifts $z < 0.107$ using SDSS DR$7$. In their survey, they find approximately $30 h^{-1} Mpc$ as the largest void size. Recently, again using SDSS DR$7$, \cite{sutter} release a void catalogue up to a redshift of $z = 0.2$ and the Luminous Red Galaxy sample out to $z = 0.44$. They show that voids have radii in the range $5-135 h^{-1}Mpc$. In this study, \cite{sutter} identify voids by using a modified version of the parameter free void finder ZOBOV \citep{marc2008,Lavaux2010}, which is based on a
Voronoi tessellation that reconstructs the density field \citep{rien2007,Platen2011} followed by a watershed algorithm to group Voronoi cells into zones and voids \citep{Platen2007}.

The first dynamical models of voids have been based on isolated spherical underdense regions in a uniform background. A very detailed study of void dynamics in the EdS Universe is achieved by \cite{be83,be85}. He formulated the scale free solutions in terms of the nonlinear evolution of isolated spherical voids in baryon, dark matter and mixed gases. In addition to this, \cite{filgold} reached the same results for dark matter and baryon matter. Later on it was shown that, similarly to their overdense counterparts, voids also merge together to construct large void structures hierarchically \citep{regge,wk,sahni1994,gott,colb}. \cite{sahni1994,Sahni95,Sahnishandarin1995} provides a significant contribution towards a proper theoretical insight into the unfolding void hierarchy describing void evolution in the context of a Lagrangian Adhesion model. Following this, \cite{sw} argued that void evolution is dictated by two processes: their {\it merging} into ever larger voids and the {\it collapse} of voids that are embedded in overdense regions. In the same study, by identifying these two evolutionary paths, \cite{sw} derived a mass fraction function to model the hierarchical evolution of merging and collapsing void populations. They show how void evolution is driven by the gradual hierarchical evolution of voids \citep{sw}. They were the first to point out that this void hierarchy can be modeled by adopting the Extended Press Schechter formalism with two critical barriers.

In this study, following up on \cite{sw}, we construct a merger tree algorithm of spherical growing voids. First we obtain the general mass fraction function of growing and collapsing voids given by \cite{sw} in terms of volume elements. This leads to obtain a realistic void merging algorithm, since void volume increases in time rather than mass, after reaching shellcrossing. In addition, it is shown that the general volume fraction function can be reduced to a simple form by showing that the collapse void contribution is very small compared to the merging, relatively large size voids. As a consequence of this, the void volume fraction has a contribution only from growing voids and this becomes the analogue of the halo mass fraction function but in terms of volume. This is an important result, since the void merging algorithm of growing voids can be constructed in the same way as dark matter halos. Therefore, in this study, we adopt the dark matter merging halo algorithm of \cite[][hereafter LC93]{lace} in order to obtain the void merging tree algorithm. We obtain the void merging algorithm in which we define and formulate void merging and absorption rates, void size and redshift survival probabilities and also failure rates in terms of the self similar and $\Lambda$CDM models in the EdS Universe. Note that here we limit ourselves to self similar models, that are hierarchical scenarios, with spectral indices $-3 < n < 1$. The case with spectral index $n= -1.5$ provides  an approximation to $\Lambda$CDM on megaparsec scales.

\subsection{Outline of Results on Growing Void Merging Tree}
In this paper, we study the void merging process called the `void in void problem' in excursion set theory. As a result we construct a merging tree algorithm of spherical growing voids by using the reduced void distribution \citep{sw} and LC93 halo merging tree algorithm in terms of self similar models ($n=0, -1, -2$) and the $\Lambda$CDM model in the EdS Universe. Here we state the outline of this paper and the general results that we obtain from this study;

\begin{itemize}
\item In Sec. \ref{sec:Origin}, we provide a basic framework of how voids evolve and form from the primordial density field. Also in the same section we give a description of the two main void process known as merging and collapsing. In Sec. \ref{sec:EPSV}, we introduce collapse and merging barriers in the EdS Universe as well as the normalization of power spectra that are used in the figures. Also, in Sec. \ref{sec:EPSV}, the derivation of the two-barrier mass fraction function of \cite{sw} is given.

\item In Sec. \ref{sec:LC93V}, the two-barrier mass fraction of the void population of \cite{sw} is reduced to the one-barrier one. The reason is  by taking into account that large voids are not affected by overdense regions. Therefore large voids that satisfy a certain criterion given by \cite{sw}, do not collapse in overdense region(s). Note that voids that do not feel the effect of their environment, only merge. As a result, collapse barrier disappears and one can obtain the one-barrier void merging tree algorithm, called the \emph{void in void} process in the EPS formalism. This is analogous to the one-barrier \emph{cloud in cloud} process.

\item Following this, in Sec. \ref{sec:LC93V}, based on the LC93 dark matter halo merging algorithm, the conditional volume and size distribution probabilities have been derived in terms of self similar models ($n=0, -1, -2$) and the $\Lambda$CDM model in the EdS Universe. Here we show that the void size distribution of relatively small size voids increases with decreasing spectral index. In addition, the exponential cutoff in the size of the void distribution moves to very large sizes with decreasing index and decreasing redshift values. Similarly, in the $\Lambda$CDM model, at all redshifts small size voids become dominant in the distribution. However this dominance becomes stronger with increasing redshift. Moreover, for the same spectral index, the conditional probability of small and large voids has a higher value for higher redshifts, compared to lower redshifts. This shows that high redshifts have small size voids, with fewer large size voids. This trend increases with decreasing spectral index.

\item In Sec. \ref{sec:VoidMerandAbs}, assuming void and halo merging events are analogous, void merging and absorption rates are defined and derived following the halo merging algorithm of LC93. While the merging of a void indicates incorporation of voids into another one, absorption of a void can be interpreted as a small void merging event that has a small contribution to the main merging event. Note that the absorption of a void is analogous to an accreting halo event.
    As a result, we show that void mergers with relatively small volume have very high absorption tendency compared to large volume mergers. The reason that small size voids have higher absorption rates is that they are absorbed by large voids. However, large voids have more merger events than small size voids. Due to the adoption of the LC93 algorithm, these results are similar to the LC93 results on the merging and accreting rates of halos.

\item Following LC93 we define and derive exact solutions of the survival probabilities and failure rates of the growing void population in terms of size for given a redshift values based on survival analysis in Sec.~\ref{sec:SurvFail}. Agreeing with the results on the void size distribution, at high redshifts small size voids have a higher probability of surviving compared to relatively large size voids. Also, when hierarchical clustering becomes stronger, the size of surviving voids at high redshifts becomes smaller and their survival probabilities decrease. Apart from survival probability, we define failure rate as the instantaneous probability of a void failing to double its size because of merging or growing events. Based on this definition, we show that in all models, the failure rate of a growing void increases with increasing size/volume at a given redshift up to a limit size value. A growing void above this limit value has zero failure rate, in other words, it will survive with $100\% $ confidence level, theoretically.

\item In addition, the approximate analytical void formation probabilities are obtained (see \ref{probanalyticsol} in Appendix \ref{appendix:exactsolutiononmergertree}). These formation probabilities give an insight of what to expect in the Monte Carlo simulations of voids. However, due to the simplification of the LC93 Monte Carlo merging tree method, void progenitors are overpredicted.

\item Finally, we show that there are analytical solutions for the expected void distribution. This distribution defines the void merging history for self similar models, which may be interpreted as an approximate analytical merger tree solution (see Appendix \ref{probanalyticsol2}).
\end{itemize}
A detailed discussion of these results and comparisons between them and previous studies given in the following sections. However, before giving the details of the void merging algorithm, we provide a general insight on how voids form and what their origin is in the following section.

\section{Origin and Dynamics}\label{sec:Origin}
Voids are prominent features of the Megaparsecscale structure of the Universe. It is impossible to formulate the dynamical characteristics of the Cosmic Web without understanding the origin and dynamics of voids. In the primordial density fluctuations, voids emerge out of density minima and they have negative density contrast \citep{rienerwin}. As a result of this negative density profile, voids represent a region of weaker gravity, resulting in an effective repulsive peculiar gravitational influence. Note that there are two correlated effects; evacuation and expansion of voids. Evacuation occurs due to the negative gravity which forces matter to move from the center to the boundary of the void. As a result of this, the density within voids gradually increases outward and void matter in the center moves outward faster than void matter towards the boundary. This results in a typical void density profile.

Similarly to evacuation, due to the negative gravity, initially underdense regions expand faster than the Hubble flow, and thus expand with respect to the background Universe. This continuous matter evacuation causes voids to become emptier and emptier \citep{rienerwin}. Numerical calculations and $N$-body simulations show that voids tend to become spherical with respect to their time evolution \citep{centme,fuji,be85}. This tendency on the basis of the expansion of voids was explained by the bubble theorem by \cite{icke}. Computer simulations of gravitational evolution of voids in realistic cosmological environments do show a considerably more complex situation than that described by idealized spherical or ellipsoidal models \citep{wk,colb,2013MNRAS.428.3409A}. In recent years the huge increase in computational resources has enabled $N$-body simulations to resolve in detail the intricate substructure of voids within the context of hierarchical cosmological structure formation scenarios. They confirm the theoretical expectation of voids having a rich substructure as a result of their hierarchical buildup \citep{regge,wk,gott,colb,sw,2013MNRAS.428.3409A}.

This leads to a considerably modified view of the evolution of voids. One aspect concerns the dominant environmental influence on the evolution of voids. To a large extent the shape and mutual alignment of voids is dictated by the surrounding large scale structure and by large scale gravitational tidal
influences \citep{erwin2008,2012MNRAS.426..440B,2013MNRAS.434.1192R}. Equally important is the role of substructure within the interior of voids. This, and the interaction with the surroundings, turn out to be essential aspects of the hierarchical evolution of the void population in the Universe.

\subsection{The Complex Evolutionary Path of Voids}\label{subsec:evolution}
Voids have a more complex evolutionary path than their overdense counterparts due to their environments. Two main processes influence the evolution of voids depending on their surroundings: relatively large voids can {\it merge} into ever larger voids, and voids that are embedded in overdense regions can {\it collapse}.

\emph{Merging:} The merging of subvoids within a large void's interior usually follows the emergence of these small scale
depressions as true voids. Once they merge, their expansion tends to slow down. When the adjacent subvoids meet up,
the matter in between is squeezed into thin walls and filaments. The peculiar velocities perpendicular to the void walls are mostly suppressed, resulting in a primarily tangential flow of matter within their mutual boundaries. Gradual fading of these structures occurs while matter evacuates along the walls and filaments towards the enclosing boundary of the emerging void \citep{dubinski1993}. The final result is the merging and
absorption of the subvoids in the larger void. As far as the void population is concerned,
only the large void counts, while the faint and gradually fading imprint of
the original outline of the subvoids remains as a reminder of the initial internal substructure. The timescale on
which the internal substructure of the encompassing void is erased is approximately the same as that on which it reaches its maximum
size defined by the survival probability.

\emph{Collapse:} The second void process, that of the collapse of mostly small and medium sized voids, is responsible for
the radical dissimilarity between void and halo populations. If a small scale minimum is embedded in a sufficiently
high large scale density maximum, then the collapse of the larger surrounding region will eventually squeeze the
underdense region it surrounds. Small scale voids will vanish when the region around them fully collapses.
The most frequent manifestation of this process is anisotropic shearing of collapsing voids near the boundaries of prominent voids, and is an indication of the important role of tidal forces in bringing about their demise \citep{weygaert2002,sw}.

\begin{figure}
\centering
\hspace{-4mm}\includegraphics[scale=0.6]{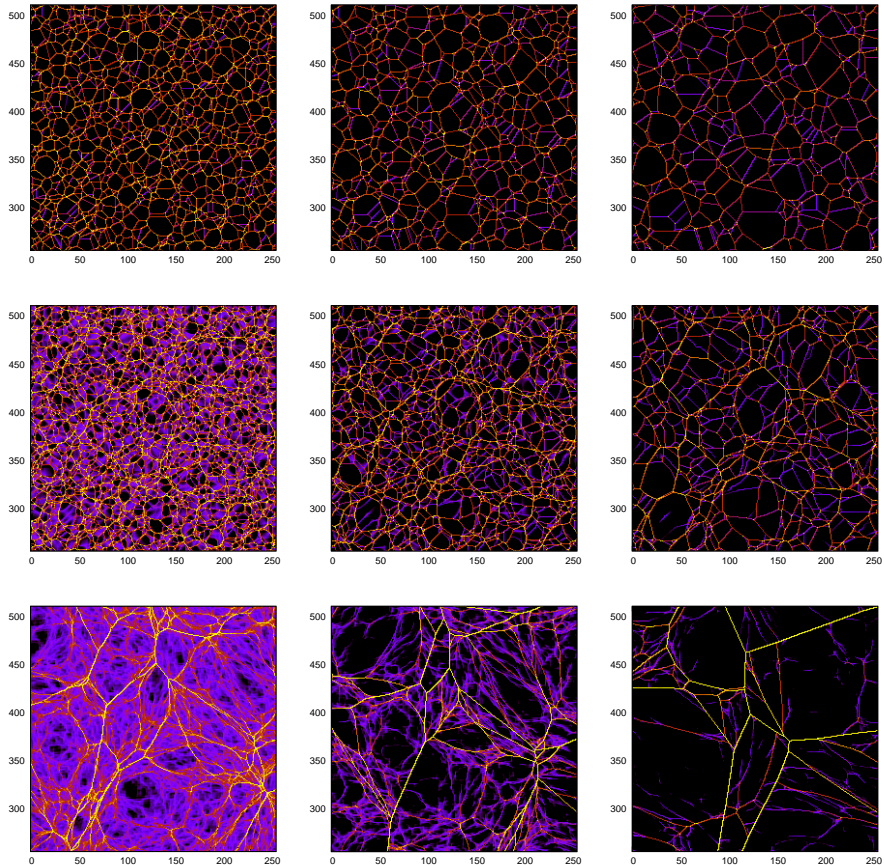}
\caption{Evolution of the Cosmic Web for three different powerlaw power spectra based on the Adhesion model. From top to bottom, simulations based on a $\Lambda$CDM Universe, in Lagrangian space, show the appearance of self similar models with a power law power spectrum $P(k)=k^{n}$ having the spectral index $n =$ $-2$, $-1$, $0$ at the time series from left to right. The color scale refers to density in which the yellow color presents the highest density.}
\label{adhs}
\end{figure}

Fig.\ref{adhs} illustrates the two main void processes in terms of power law power spectra with spectral index $n= -2, -1, 0$. The emerging weblike structure is in a $\Lambda$CDM Universe in three time steps, as predicted by Adhesion theory \citep{1988Natur.334..129K}.
\cite{1988Natur.334..129K} provides an explanation of the origin of the intricate structure formation due to inhomogeneities in the initial gravitational potential by applying the Burgers' equation. The Burger's equation that behaves as the gravitational sticking of matter at the nonlinear stage of gravitational instability. Therefore, in Fig.\ref{adhs}, the images were generated by solving Burgers' equation $v_t + (v . \nabla) v = n \nabla^2 v$ (in which $v$ is the velocity field), where the limit of $n \rightarrow 0$ is accepted. The computation was done by the method of discrete Legendre transformations by \cite{1994A&A...289..325V} and Hidding et al. (in preparation). In Fig.\ref{adhs},
the colors show the logarithm of the density. Here the color scale is not very important. There are no physical units implied for two reasons:
the power spectra are pure powerlaw, there is no characteristic scale to attach physical units. Also the concept of density is somewhat different in the Adhesion model than in N-body simulations. The yellow lines demonstrate high densities, part of the color scale. In fact the density is locally infinite. The sequence from top to bottom shows a progression of advancing time and the coordinates of the plots are Eulerian, although the figure is produced by using Adhesion model which is in Lagrangian coordinates.

Apart from this, in Fig.\ref{adhs}, one may see how the initially intricate weblike network in the interior of the large central underdense region gradually disappears as voids merge, while the internal boundaries (in blue) gradually fade away. In particular, near the boundaries of large voids we may see the second void process, that of the collapse of voids. It manifests itself in the form of a shearing and squeezing of less prominent voids, as a result of the expansion of prominent neighboring voids or of the tidally induced filamentary or planar collapse of the weblike mass concentrations at the edges of the voids.

\section{The Extended Press Schechter Formalism of Voids}\label{sec:EPSV}
$N$-body simulations of void evolution show a more complex situation than that described by idealized ellipsoidal or spherical models, and indicate the intricate substructure of voids within the context of hierarchical cosmological structure formation scenarios \citep{MartelW90,regge,dubinski1993,wk,mathis,ArbabiMuller02,benson,gott,goldbergvogeley,padilla05,colb,Hoeft06,2013MNRAS.428.3409A}. These studies confirm the theoretical expectation of voids having a rich substructure as a result of their hierarchical buildup. Apart from these studies, \cite{sahni1994,Sahni95,Sahnishandarin1995} describe void hierarchy in Lagrangian perturbation theory and its subsequent elaboration, the Adhesion approximation. \cite{sw} provide a considerably modified view of the evolution of voids in the context of hierarchical scenarios. They also show that the hierarchical evolution of voids, akin to the evolution of overdense halos, may be described by the EPS formalism \citep{sw}. \cite{piran2006} built a model to describe the distribution of galaxy underdensities. Our model is based on the `excursion set formalism', the same technique used to predict the dark matter halo mass function.

\subsection{Collapse and Merging Barriers}
The hierarchical evolution of complex voids can be modeled by a two-barrier excursion set formalism. Here the two barriers refer to the two main processes
that dictate void evolution: merging and collapse \citep{sw,vdWbond}. In the two-barrier excursion set, the merging threshold is the shell crossing value
of the spherical voids $\delta_{\mathbf{v}}=-2.81$ in the EdS Universe. The collapse threshold of voids that are embedded within a contracting overdensity is set
by the collapse barrier of the spherical model $\delta_{c}=1.686$ in the EdS Universe. In linear theory, the growing under- and overdensities of the
spherical model are written in terms of redshift, as follows,

\begin{eqnarray}
\label{twobarrierdensities}
\delta_{lin,c}(z) &=& \frac{\delta_{c}}{D(z)}={\delta_{c}}\left(1+z\right),
\nonumber\\
\\
\delta_{lin,\mathbf{v}}(z) &=& \frac{|\delta_{\mathbf{v}}|}{D(z)}={|\delta_{\mathbf{v}}|}\left(1+z\right).
\end{eqnarray}
\noindent
In this study, we use linear over- and underdensities of the spherical objects as time variables by following \cite{lace}. This is a natural choice by considering that the critical over- and critical underdensities are constant values in the EdS Universe ($\Omega=1$).

\subsection{Normalization of Power Spectra}
Here we give the core elements of the normalization of the power law power spectra and physical spectrum in the context of void hierarchy. To define the normalization of the power spectrum, it is important to convert the mass fraction into a void size distribution. It is possible to do this in terms of the spherical model. All the time dependence comes from the linearly extrapolated density, and mass is not time dependent. Therefore the comoving volume $V$ of the void is equal to,

\begin{eqnarray}
V=(M/\rho_{u})
\end{eqnarray}
\noindent
We can set the relation between void mass, volume and size by using the definition of the mass variance for the self similar models, as follows,

\begin{eqnarray}
\label{relation1}
\sigma^2(M)=S(M)=\delta^{2}_{lin,\mathbf{v}}\left(\frac{M}{M_{*}}\right)^{-\alpha}= \delta^{2}_{lin,\mathbf{v}} \left(\frac{V}{V_{*}}\right)^{-\alpha}=\delta^2_{lin,\mathbf{v}}\left(\frac{R}{R_{*}}\right)^{-3\alpha},
\end{eqnarray}
\noindent
in which $\alpha=n+3/3$. $M_{*}$, $V_{*}$ and $R_{*}$ are the characteristic mass, volume and radius respectively, while $\alpha$ is the constant that is defined by $(n+3)/3$ where $n$ is the spectral index. The self similar evolution of the mass scale is specified via the time development of the characteristic mass by following \cite{vdWbond},

\begin{eqnarray}M_{*}(z)=D^{\frac{6}{n+3}}(z) M_{*,0},\phantom{a}M_{*,0}=\left(\frac{\sigma_{8}}{|\delta_{\mathbf{v}}|}\right)^{6/n+3},
\label{characteristicmass}
\end{eqnarray}
\noindent
where $M_{*,0}$ is the present day value of the characteristic mass,
\begin{eqnarray}M_{*}(z)=\left(\frac{\sigma_{8}}{|\delta_{lin,\mathbf{v}}|}\right)^{6/n+3}.
\label{characteristicmass2}
\end{eqnarray}
\noindent
Here, for the normalization of the power spectra, the characteristic mass $M_{*}$ can be chosen as $M=M_{*}$. This indicates that in equation (\ref{relation1}), the characteristic normalization mass variance is equal to $\sigma^{2}(R=8 Mpc h^{-1})\sim |\delta_{\mathbf{v}}|$. Therefore, we choose the characteristic radius as $R_{*}=8 h^{-1} Mpc$. Following this, the characteristic radius $R_{*}=8 h^{-1} Mpc$, and the radius $R$ of a void region for a power spectrum approximated by a power law of slope $n$ is given by,

\begin{eqnarray}R= 8\left(\frac{\sigma_{8}}{|\delta_{\mathbf{v}}|}\right)^{2/n+3},
\label{voidcharacteristicsize}
\end{eqnarray}
\noindent
in which $\sigma_{8}$ is the variance of the density perturbation smoothed on $8 h^{-1} Mpc$. The correlation length is of the order of $8 h^{-1} Mpc$. As result, this makes the typical void diameter similar to the correlation length. This normalization of the power law power spectrum, as is seen in equation (\ref{voidcharacteristicsize}), allows us to compute the void volume distribution for a range of $\Lambda$CDM spectra with different cosmological parameters (e.g. those for WMAP$9$, WMAP$7$, etc.). Apart from this, equation (\ref{voidcharacteristicsize}) gives the identification of the initial comoving void size $R$ of a region.

\begin{figure}
\centering
\includegraphics[width=0.5\textwidth]{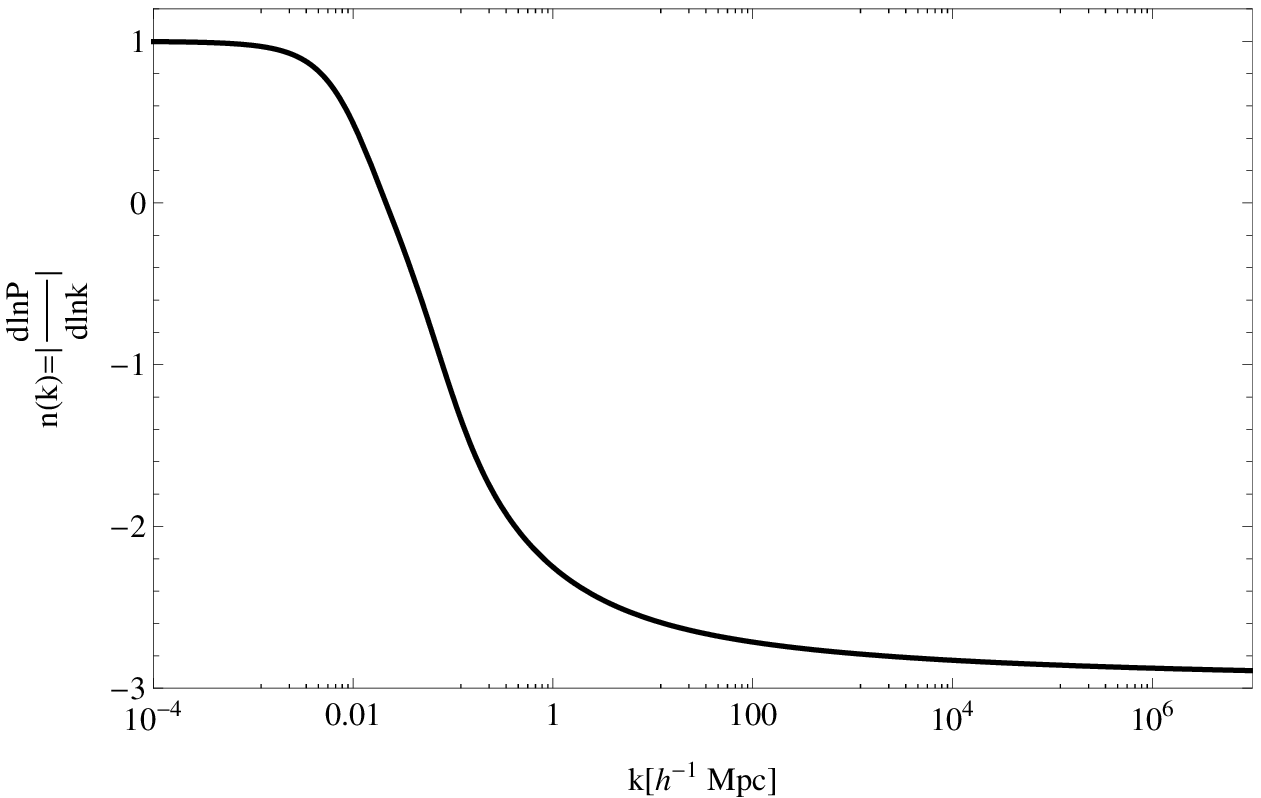}\\
\includegraphics[width=0.5\textwidth]{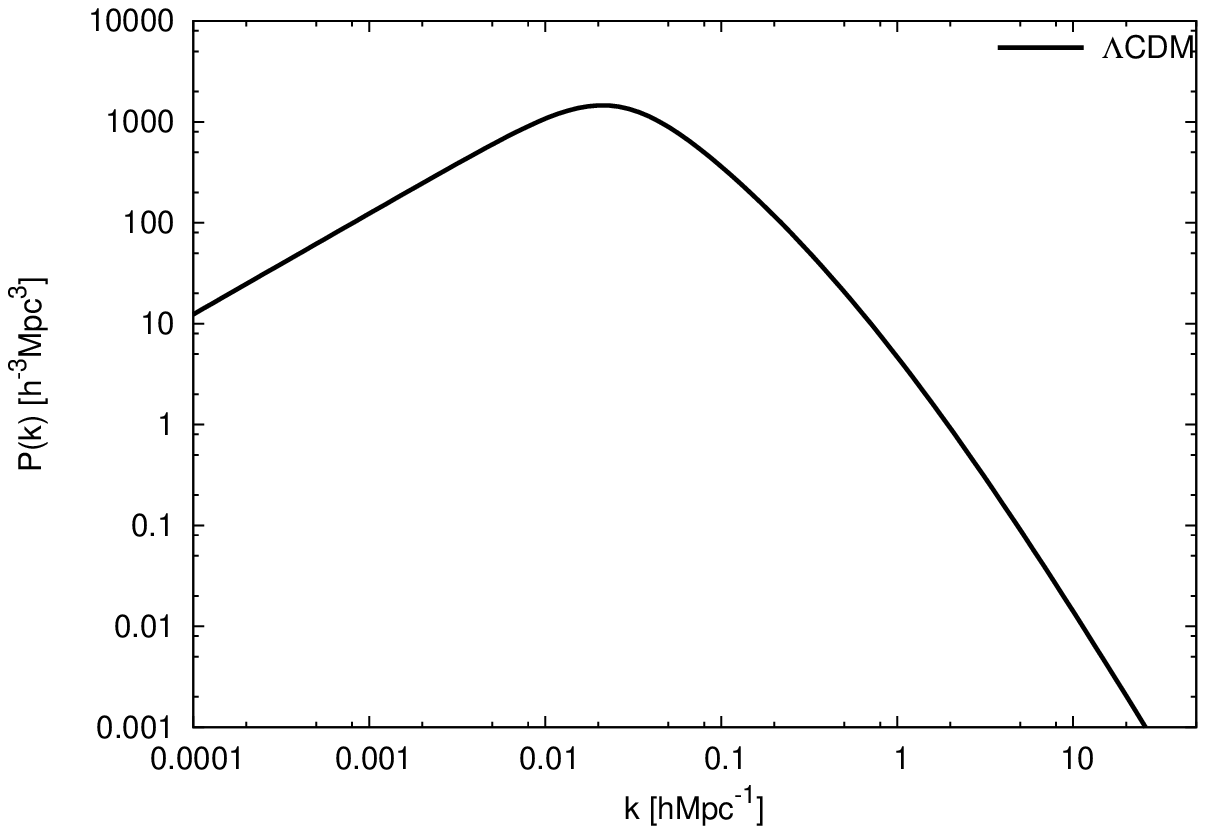}
\caption{{Upper:} The evolution of the power spectrum slope $n(k)$ in terms of wave number $k$ for the $\Lambda$CDM model. The slope of the power spectrum changes between $1$ and $-3$. Lower: The normalized $\Lambda$CDM power spectrum with parameters $\Gamma=0.21$, $n=1$ and $\sigma_{8}=0.9$.}
\label{fig:neff}
\end{figure}

To normalize the CDM spectrum, we limit ourselves to the currently favored $\Lambda$CDM model with $\sigma_{8}=0.9$. We use the transfer function by \cite{BondEfstathiou1984}. Hence the power spectrum is given by,

\begin{eqnarray}
P(k)=A \frac{k^{n}}{\left[1+\left[6.4q+{(3q)^{3/{2}}}+(1.7q)^{2}\right]^{1.13}\right]^{1/1.13}},
\label{powerspectrumCH}
\end{eqnarray}
\noindent
where $q$ is given by,
\begin{eqnarray}
q=\frac{k}{\Gamma}\phantom{a}\textstyle{h Mpc^{-1}},
\nonumber
\end{eqnarray}
\noindent
in which $k$ is the wavenumber and $\Gamma$ is called the shape parameter. The shape parameter $\Gamma=0.21$ and normalization $\sigma^2_{8}=0.9$ for the $\Lambda$CDM model of \cite{Jenkins98}. To normalize the $\Lambda$CDM spectra we take the ratio between the observed $\sigma_{8}=0.9$ and numerically calculated $\sigma_{8}$ by using the power spectrum equation (\ref{powerspectrumCH}) in terms of the numerical Romberg integration. The ratio between observed and numerically calculated mass variances gives us the amplitude of the power spectra, in other words the normalization constant. In $\Lambda$CDM- related plots we use the Romberg numerical integration to obtain the normalized mass and volume variance parameters $\sigma_{M}$ and $\sigma(V)$. Care should also be taken when deciding the choice of the spectral index $n$ in the power spectrum; equation (\ref{powerspectrumCH}). The spectral index gives the slope of the power spectrum and it varies between $1$ and $-3$ depending on the scale/wavenumber of the power spectrum. The amplitude of the power spectrum changes in terms of scale/wave numbers (Fig.\ref{fig:neff}). Small wavenumbers indicate large scales while large wavenumbers show small scales. As a result, the slope of the spectral index varies between very large and very small scales (Fig.\ref{fig:neff}). Since our goal is to construct merging dark matter voids, our interest is large scales, that is why we choose $n=-1.5$. That is why we approximated self similar model $n=-1.5$ by using $\sigma_{8}=0.9$ to the $\Lambda$CDM by following \cite{sw}.

\subsection{Derivation of Void Excursion Set}
The analytical description of the two-barrier excursion set is based on a void distribution function on a mass scale $M$ derived by \cite{sw}. By following their formalism, here we first introduce the related functions and parameters of the two-barrier void excursion set in terms of mass element $M$. However to obtain a physically realistic description of the void population in the context of hierarchical build up scenarios, it is crucial to define the void distribution function in terms of volume/size since the volume of voids tends to grow in time. This will be taken care of at the end of this section by using the relations between volume, mass and size parameters which we have shown above.

\begin{figure}
\begin{center}
\includegraphics[width=0.65\textwidth]{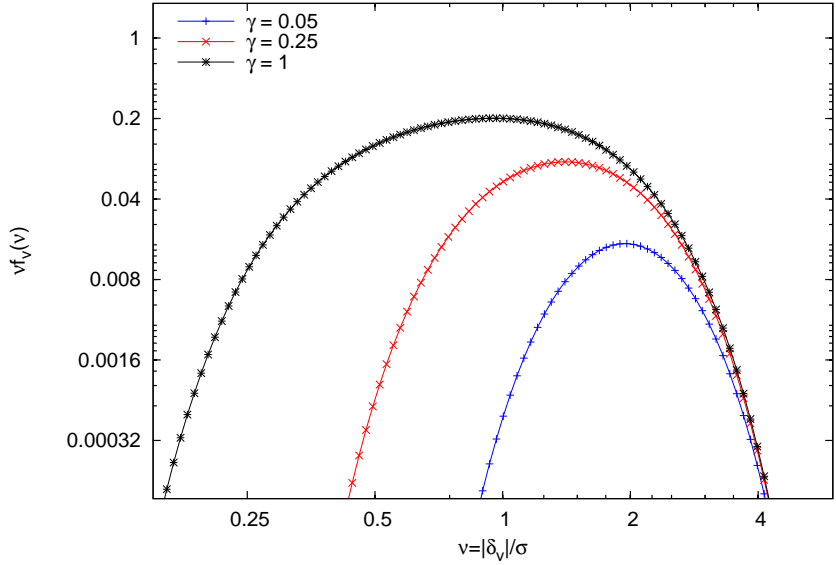}
\caption{The scaled fraction of the void population $f_{\mathbf{v}}$ in terms of different barrier ratios $\gamma =$ $0.05$, $0.25$, $1$.}
\label{fig:twobarrier}
\end{center}
\end{figure}

The void distribution function based on the two-barrier random walk problem on a mass scale $M$ was derived by \cite{sw} and is approximated to (see also \cite{vdWbond}),

\begin{eqnarray}
{n_{v}(M)}dM
\approx \sqrt{\frac{2}{\pi}}\frac{\rho_{u}}{M^2}{\nu_{\mathbf{v}}}
\exp\left[-\frac{\nu^2_{\mathbf{v}}}{2}\right]
\left|\frac{d\ln\sigma( M)}{d\ln
M}\right|\exp\left[-\frac{|\delta_{\mathbf{v}}|}{\delta_{c}}\frac{\mathcal{D}^2}{4\nu^{2}_{\mathbf{v}}}-
\frac{2\mathcal{D}^4}{\nu^{4}_{\mathbf{v}}}\right],
\label{massforvoids}
\end{eqnarray}
\noindent
where $\nu_{\mathbf{v}}$ corresponds to a fractional underdensity function \citep{sw},

\begin{eqnarray}
\nu_{\mathbf{v}}(M)=\frac{\delta_{\mathbf{v}}}{\sigma(M)}= \frac{|\delta_{\mathbf{v}}|}{\sqrt{S(M)}},
\label{voiddensityparameter}
\end{eqnarray}
\noindent
in which $\delta_{\mathbf{v}}$ is the void threshold while the mass dependence comes in via the mass variance function $\sigma(M)$ or the mass scale function $S$. In equation (\ref{massforvoids}), $\mathcal{D}$ is the void and cloud parameter \citep{sw} and it is defined as,

\begin{eqnarray}\label{Dparameter}
\mathcal{D}\equiv\frac{|\delta_{\mathbf{v}}|}{\left(\delta_{c}+|\delta_{\mathbf{v}}|\right)}.
\end{eqnarray}
\noindent
Here, void and cloud parameter $\mathcal{D}$ parameterizes the impact of the halo evolution on the evolving population of voids for overdense $\delta_{c}$ and underdense $\delta_{\mathbf{v}}$ regions \citep{sw}. The void mass distribution function (or void distribution) can be obtained using the following expression,

\begin{eqnarray}
{n_{\mathbf{v}}(M)}{dM}
=2{\sigma^2}\frac{\rho_{u}}{M^2}f_{\mathbf{v}}(\nu_{\mathbf{v}})\left|\frac{d\ln\sigma}{d\ln
M}\right|dM.
\label{massfraction1}
\end{eqnarray}
\noindent
The mass distribution of voids on mass scale $M$ with respect to the two barriers can be derived from equation (\ref{massfraction1}) as follows,

\begin{eqnarray}
f_{\mathbf{v}}(M) dM =
\frac{1}{\sqrt{2\pi}}\frac{{\nu_{\mathbf{v}}}}{\sigma^2}\exp\left[-\frac{\nu^2_{\mathbf{v}}}{2}\right]
\exp\left[-\frac{|\delta_{\mathbf{v}}|}{\delta_{c}}\frac{\mathcal{D}^2}{4{\nu^2_{\mathbf{v}}}}-\frac{2\mathcal{D}^4}{\nu^{4}_{\mathbf{v}}}\right]dM.
\label{massfractionm}
\end{eqnarray}
\noindent
As is seen, there are two cutoffs in the void mass fraction equation (\ref{massfractionm}) at large and small values of the fractional density $\nu_{\mathbf{v}}$ (see Fig. \ref{fig:twobarrier}). \cite{sw} mention that this expression is accurate for values satisfying $\gamma\equiv \delta_{c}/\delta_{\mathbf{v}}\geq 0.25$. Choosing the ratios of linearly extrapolated densities larger than $0.25$ guarantees that the void distribution is well peaked at the characteristic value $v\sim1$ ($\sigma(M_{*})\sim |\delta_{\mathbf{v}}|$). Fig. \ref{fig:twobarrier} shows the different choice of $\gamma$ parameters. As is seen, curves with $\gamma \geq 0.25$ seem well peaked at the characteristic void mass which is $v\sim1$. In addition, Fig. \ref{fig:twobarrier} indicates two cutoffs of the distribution function at small and large scaled mass.

Up until here, we give the important parameters and functions of void evolution that allow us to construct a hierarchical evolution of voids by taking into account their complex evolutionary paths. In this concept, the void distribution is especially important since it becomes the backbone of the void merging tree algorithm, based on the two-barrier excursion set.

\begin{figure}
\begin{center}
\includegraphics[width=0.55\textwidth]{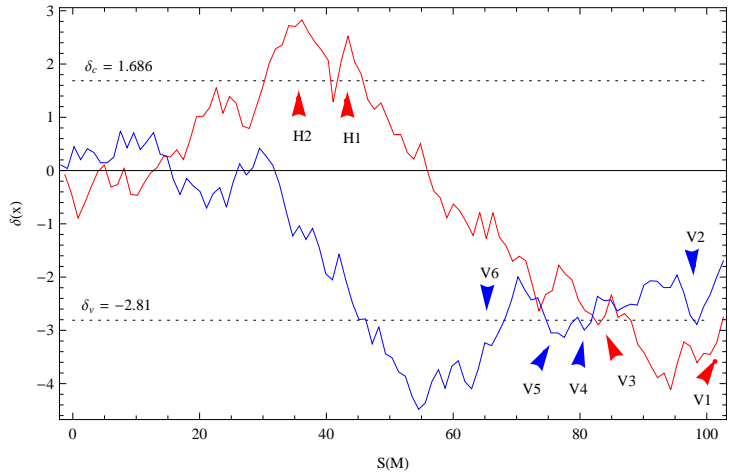}
\caption{Two-barrier EPS formalism of the two void processes: void merging (blue)
and void collapse (red). The random walk is represented by the average spherical smoothed density $\delta(x)$ by the mass scale $S$ (large volumes are on the left, small volumes on the right). The smoothed density is centered on a randomly chosen position in a Gaussian random field. Dashed horizontal
lines indicate the collapse barrier $\delta_{c}$ and the void shell crossing barrier $\delta_{\mathbf{v}}$ of the spherical collapse model. Arrows represent the mass scales at barriers where voids and halos are formed.}
\label{fig:halovoid}
\end{center}
\end{figure}

In this study we construct a void merging tree by taking into account void volume and size distribution instead of mass distribution. As was pointed out before, voids tend to grow in size and due to their peculiar gravitational field, their mass content is accumulated in a thin mass shell surrounding them. Therefore, constructing a model in order to obtain their evolution from a volume or size perspective seems to be a natural and realistic approach. To do this, as a first step, the mass-dependent void distribution equation (\ref{massfractionm}) is obtained in terms of scale $S$ by using the relation between mass variance and mass scale $\sigma^2(M)=S(M)$ in equation (\ref{massfraction1}), which leads to,

\begin{eqnarray}
{n_{\mathbf{v}}(M)}{dM}=S\frac{\rho_{u}}{M^2}f_{\mathbf{v}}(S,\delta_{c},|\delta_{\mathbf{v}}|)\left|\frac{d\ln S}{d\ln M}\right|dM.
\label{massfraction2}
\end{eqnarray}
\noindent
Then by using equation (\ref{massfraction2}), we obtain the mass scale $S$ dependent void distribution function as follows,

\begin{eqnarray}
f_{\mathbf{v}}(S) dS \approx \frac{1}{\sqrt{2\pi}}\frac{{\delta_{\mathbf{v}}}}{S^{3/2}}\exp\left[-\frac{\delta^2_{\mathbf{v}}}{2S}\right]
\exp\left[-\frac{1}{4}\frac{|\delta_{\mathbf{v}}|}{\delta_{c}}{\mathcal{D}^2}\frac{S}{{\delta^2_{\mathbf{v}}}}-2 {\mathcal{D}^4}\frac{S^2}{{\delta^4_{\mathbf{v}}}}\right]dS.
\label{massfraction2fs}
\end{eqnarray}
\noindent
This void distribution is particularly important since it defines the transition between barriers, which is the base of the void merging algorithm. \cite{sw} indicated the distribution function can be transformed into void mass, void volume and void size distributions with respect to the definition of the variance in terms of simple self similar spectra (equation (\ref{relation1})),

\begin{eqnarray}f_{\mathbf{v}}(S) d S\propto f_{\mathbf{v}}(M)d M \propto f_{\mathbf{v}}(V)d V\propto f_{\mathbf{v}}(R)d R.
\label{relation}
\end{eqnarray}
\noindent
This simple approximation allows us to use the volume scale function $S(V)$ instead of the mass scale function $S(M)$. Due to this approximation, void distribution equation (\ref{massfraction2fs}) represents the void volume distribution. Under these circumstances, the void volume distribution function has contributions from both growing and collapsing void populations. These two dynamical characteristics are encapsulated by collapse and shell crossing barriers in the void volume distribution function. The description of the two barriers in the void excursion set can be given as follows: a trajectory crosses the collapse threshold $\delta_{c}$ on a volume scale $S$ and then crosses $\delta_{\mathbf{v}}$ on a larger volume scale $S^{\prime}> S$. This indicates that a void is embedded in an overdense region and later on, due to the contracting overdense region, it will collapse at volume scale $S$ (Fig. \ref{fig:halovoid}). The trajectories that do not cross the collapse barrier indicate voids that merge or become mature without collapsing. These gradually merging or growing voids have sizes above a critical size in which case they are not effected by the overdense regions \citep{sw}. In Fig. \ref{fig:halovoid} two different random walks are given. These trajectories represent examples of void evolution processes. While the blue trajectory relates to void formation through the merging of voids, the red trajectory represents a collaping void evolution. As is seen from the figure, the trajectory is related with merging void events and associated random walk is in blue. This random walk shows that the present day void $V6$ corresponds to a larger volume than the smaller void $V2$ which merged into $V6$. The red random walk at largest $S$ concerns a location which at early times was found within a small void $V1$. This void, however, is embedded in an overdense halo $H1$ which later merges into a massive halo $H2$. Once this entity collapses into a massive virilized halo, the void with volume $V1$ will vanish.

\section{Lacey and Cole's Merging Tree Algorithm for Growing Voids}\label{sec:LC93V}
In this section, our aim is to formulate a merging tree algorithm of growing voids. This algorithm is constructed by applying the two-barrier volume scale distribution function (\ref{massfraction2fs}) in the halo merging tree algorithm \citep{lace} to the void merging tree algorithm.

\begin{figure}
\begin{center}
\includegraphics[width=0.55\textwidth]{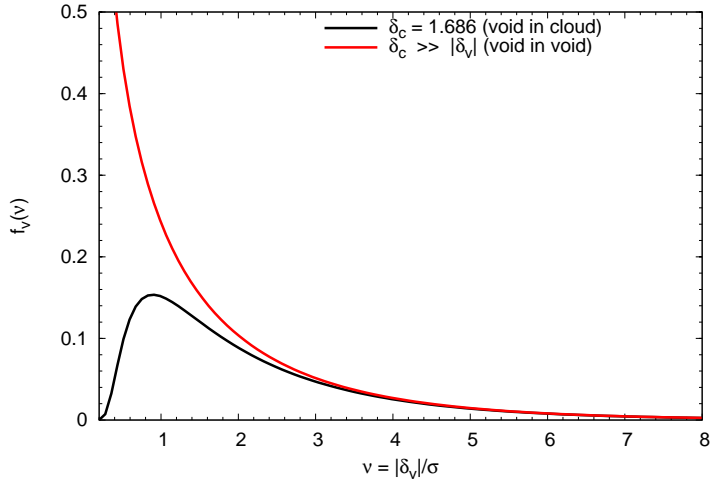}
\caption{This graphic illustrates the relation between volume fraction
function $f_{\mathbf{v}}(\nu)$ and the mass underdensity function $\nu$ in terms of the
critical overdensity value at $\delta_{c}=1.686$ and $\delta_{c}>>
|\delta_{\mathbf{v}}|=2.81$. When $\delta_{c}\rightarrow\infty$, the void in
cloud parameter becomes negligible, such that the mass fraction
function includes only the underdensity function $\delta_{\mathbf{v}}$ \citep[adaptation of][]{sw}.}
\label{fig:asymptotes}
\end{center}
\end{figure}

Here we limit ourselves to voids that grow in volume. The size criterion of voids that grow without vanishing is established by \cite{sw}. This criterion is based on the statement that there are no large scale voids embedded in large scale halos, on the scales where,

\begin{eqnarray}
\sigma \ll \left(\delta_{c}+|\delta_{\mathbf{v}}|\right),
\end{eqnarray}
\noindent
and here the collapse barrier $\delta_{c}$ does not have any effect on the void population. If we rearrange this statement in terms of the void size by using the relations (\ref{relation1}), we obtain a void size criterion,

\begin{eqnarray}\label{radiuscriteria}
\frac{R}{R_{*}} \gg \left(\delta_{c}+ |\delta_{\mathbf{v}}|\right)^{\frac{-1}{3}\frac{2}{\alpha}},
\end{eqnarray}
\noindent
where the characteristic void size is $R_{*}$ and in this study, it is chosen as $8 h^{-1}Mpc$. This size criterion leads to a classification between void sizes. Here we name voids with radius $< R/{R_{*}} $ as \emph{minor voids} that are most likely embedded in an overdense region. Due to the gravitational collapse of the overdense region, minor voids collapse and vanish. Voids with radii larger than the radius criterion (\ref{radiuscriteria}) grow in size and merge gradually. Here we name these voids as \emph{growing voids}. In the case of growing voids, the relation between the overdense and underdense linear densities is given by \cite{sw},

\begin{eqnarray}
\delta_{c}\gg \delta_{\mathbf{v}}.
\end{eqnarray}
\noindent
As a consequence of this, the void and cloud parameter in the volume fraction function (\ref{massfractionm}) vanishes ($\mathcal{D}=0$). When the void and cloud parameter tends to zero $\mathcal{D}\rightarrow 0$, the second exponential term, corresponding to the contribution of the embedded voids, in equation (\ref{massfraction2fs}) becomes unity. Due to the vanishing void and cloud parameter $\mathcal{D}$, the second exponential term in equation (\ref{massfraction2fs}) disappears. This means that the contribution of subvoids embedded in overdense regions becomes unimportant ( see Fig. \ref{fig:asymptotes}).

In this limit, the two-barrier mass fraction distribution (consisting of $\delta_{\mathbf{v}}$ and $\delta_{c}$) reduces to a single barrier at $\delta_{\mathbf{v}}$ \citep{sw}, as follows,

\begin{eqnarray}
f_{growing-void}(S,\delta_{\mathbf{v}}) dS =\frac{1}{{\sqrt{2\pi}}}\frac{|\delta_{\mathbf{v}}|}{S^{3/2}}\exp\left[-\frac{\delta^2_{v}}{2S}\right]dS.
\label{massfraction3}
\end{eqnarray}
\noindent
This mass fraction indicates that large voids are not affected by overdense regions, they are not squeezed under collapsing regions due to the lack of a collapse barrier. This is an important result because this fact provides a useful framework to construct a large void merging tree with one-barrier $|\delta_{\mathbf{v}}|$ called the \emph{void in void} process in the EPS formalism, which is analogous to the one-barrier \emph{cloud in cloud} process.

To construct the merging tree of large voids, we use the halo merging tree algorithm of LC93. This algorithm is originally an analytical description of merging virilized halos based on the EPS formalism and it can be applied to any hierarchical model in which structure grows via gravitational instability (LC93). This halo merging algorithm is also known as the binary method due to its choice of parent halo which splits into two (and only two) progenitors. Later on this algorithm has been modified because its assumption of binarity is an oversimplification \citep{kaufwh,sethlemson305,cole2000,somervillekolatt,Zhang08}. In this study, the reason to choose this algorithm is due to its simplicity in implementing a void merging tree, and its fast executable property. This algorithm provides a simple exercise to understand the complex void evolution in terms of the two-barrier EPS formalism. Therefore this method may lead us to apply the two-barrier EPS formalism of voids to more up to date merger algorithms to construct more realistic void merger trees.

\begin{figure}
\centering
\begin{tabular}{l}
\includegraphics[width=0.5\textwidth]{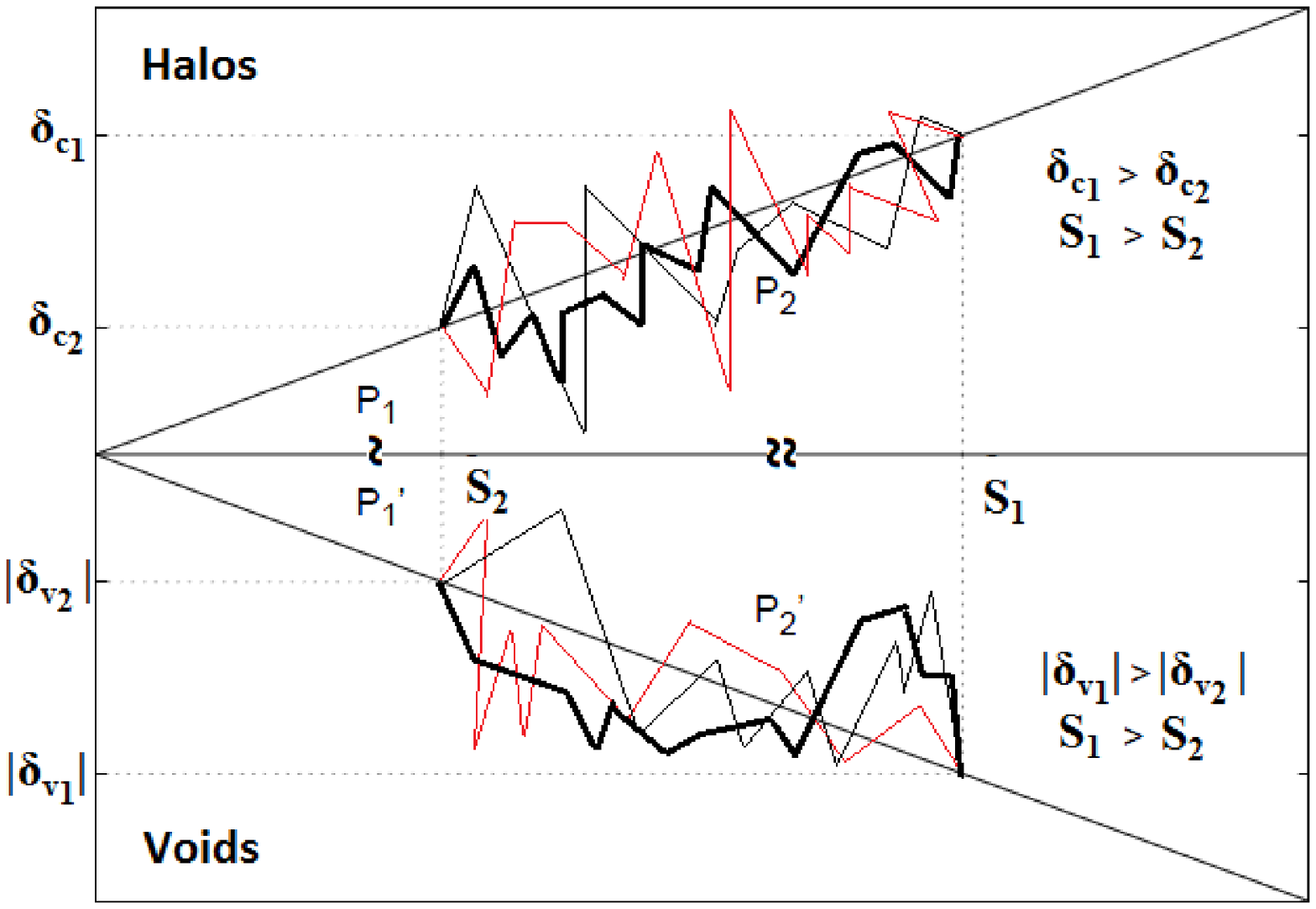}
\\
\includegraphics[width=0.5\textwidth]{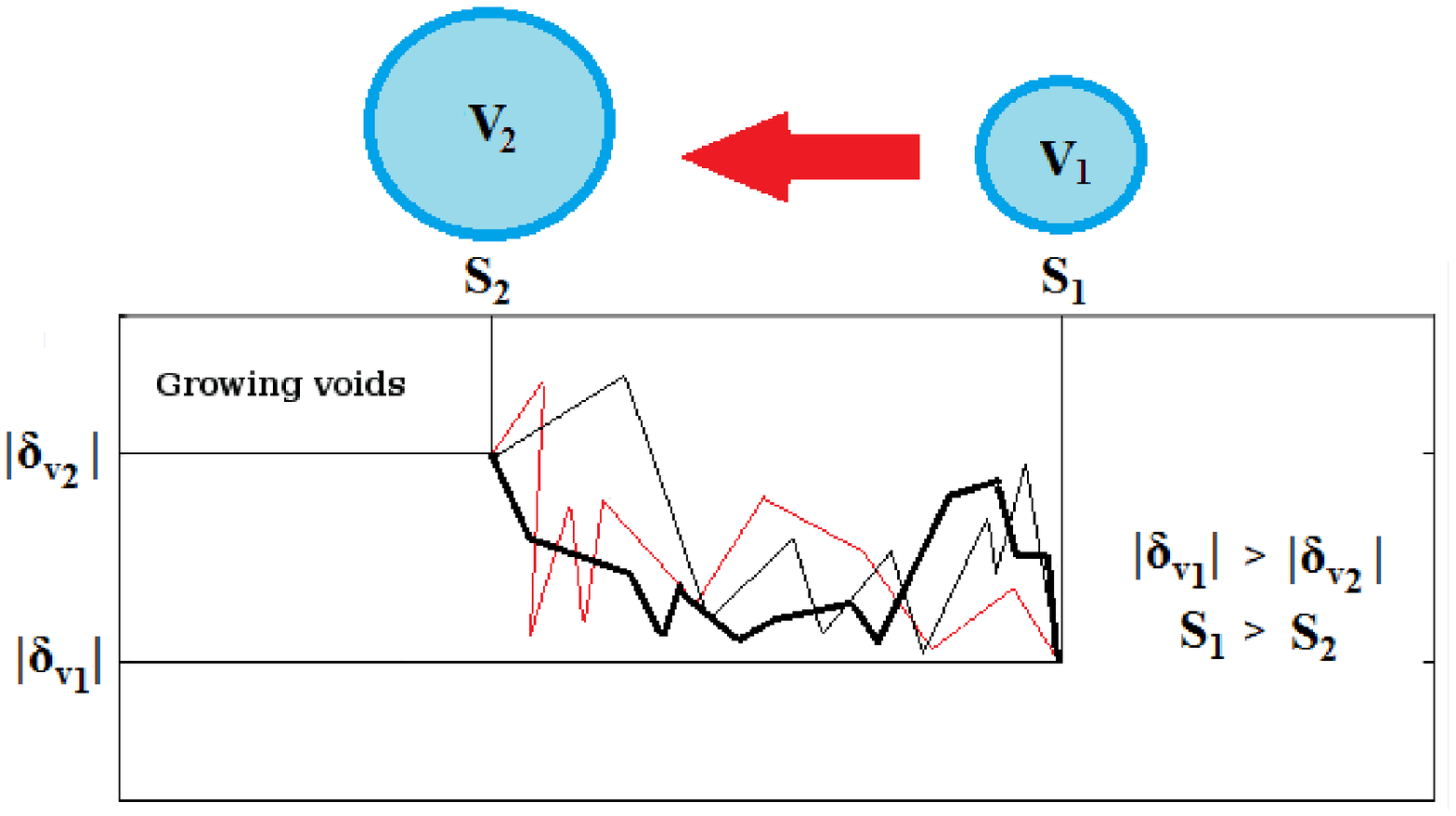}
\end{tabular}
\caption{Upper: The probability densities of the regions $P_{1}$, $P^{\prime}_{1}$ and $P_{2}$, $P^{\prime}_{2}$ which are equivalent to each other ($P_{1}=P^{\prime}_{1}$ and $P_{2}=P^{\prime}_{2}$). The upper part of the diagram shows the
simple illustration of the probability for one trajectory in LC93 for overdense regions, and the lower part is the mirror image of LC93
around the $x$-axis representing the underdense (void) regions. Note that there is an asymmetry in that $|\delta_{\mathbf{v}}|>\delta_{c}$, since the threshold values in linear theory are $|\delta_{\mathbf{v}}|=2.81$ and $\delta_{c}=1.686$. Lower: Void size/volume increases in terms of time while void volume scale decreases.}
\label{fig:model}
\end{figure}

We adapt the merging algorithm of LC93 to construct a void merging tree by taking into account the one-barrier EPS formalism. To do this, we take into account the symmetry properties of the probability densities of overdense and underdense regions in EPS formalisms. In Fig. ~\ref{fig:model}, halo and void random walks are represented in which we study merging bubbles instead of merging collapsed regions. The probabilities of these regions ($P_{1}$ and $P^{\prime}_{1}$, also $P_{2}$ and $P^{\prime}_{2}$) are analogous to each other. This analogy indicates that the trajectories at the lower part of the diagram represent large voids with mass scale $S_{1}(M_{1})$ which is equivalent to $S_{1}(V_{1})$ since $M\approx V$ at the time corresponding to the barrier $|\delta_{\mathbf{v}_{1}}|$. Later they will merge and construct the larger voids after reaching the second barrier $|\delta_{\mathbf{v}_{2}}|$ with scale $S_{2}(M_{2})\approx S_{2}(V_{2})$. Therefore we can see that the void merging tree and halo merging tree are analogous to each other. Note that the only difference between the halo and void merging comes from their barrier heights which are the linear extrapolated smoothed densities; their growth is only dependent on the growth factor $\mathcal{D}$ of the EdS Universe. As a result of different linearly extrapolated densities (\ref{twobarrierdensities})
void and halo merging events have slightly different timescales though they show the same merging characteristics.

The conditional probability $f_{S_{1}}(S_{1},|\delta_{\mathbf{v}_{1}}|\big|S_{2},|\delta_{\mathbf{v}_{2}}|)
d{S_{1}}$ that one of these trajectories makes its first upcrossing at $|\delta_{\mathbf{v}_{1}}|$ in the interval $S_{1}+dS_{1}$ can be obtained directly from equation (\ref{massfraction3}), but with a difference that the source of the trajectories has moved from the origin to the point $(S_{2},|\delta_{\mathbf{v}_{2}}|)$ (by following the algorithm derived by \cite{lace,bo}, also this formula was deduced by \cite{bower91}). The conditional probability density of a void whose trajectory is in the interval $S_{1}+ dS_{1}$ making its first upcrossing at $|\delta_{\mathbf{v}_{1}}|$ which later on crosses the point ($S_{2},\delta_{{\mathbf{v}_{2}}}$) is,

\begin{eqnarray}
f_{S_{1}}(S_{1},|\delta_{\mathbf{v}_{1}}|\big|S_{2},|\delta_{\mathbf{v}_{2}}|)
d{S_{1}}=
\frac{1}{\sqrt{{2\pi}}}\frac{|\delta_{\mathbf{v}_{1}}|-|\delta_{\mathbf{v}_{2}}|}{\left({S_{1}}-{S_{2}}\right)^{3/2}}
\exp\left[\frac{-\left(|\delta_{\mathbf{v}_{1}}|-|\delta_{\mathbf{v}_{2}}|\right)^{2}}{2\left(S_{1}-{S_{2}}\right)}\right]dS_{1},
\label{probabilitys1}
\end{eqnarray}
\noindent
where the void barriers are given by $|\delta_{\mathbf{v}_{1}}| > |\delta_{\mathbf{v}_{2}}|$ and the volume scales related with the void barriers should be $S_{1} > S_{2}$. This conditional probability function of voids is the same as equation ($2.15$) in LC93 which is derived for halos. The evolution of the two void barriers is defined by linear theory,

\begin{eqnarray}
\delta_{\mathbf{v}_{1}}(z_{1})=\delta_{lin,\mathbf{v}_{1}}(z_{1})= |{\delta_{\mathbf{v}}}|\left(1+z_{1}\right),\phantom{a}\delta_{\mathbf{v}_{2}}(z_{1})=\delta_{lin,\mathbf{v}_{2}}(z_{2}) = |{\delta_{\mathbf{v}}}|\left(1+z_{2}\right),
\label{twobarrierdensitiesvoidmerging}
\end{eqnarray}
\noindent
in which $|{\delta_{\mathbf{v}}}|=2.81$ is the threshold value of the spherical underdense perturbations in the EdS Universe. We can transform the conditional probability equation (\ref{probabilitys1}) of void volume scale distribution into the conditional probability of void size distribution for the $\Lambda$CDM and self similar models. For the CDM model, scale functions or variances $\sigma_{1}=S_{1}$ and $\sigma_{2}=S_{2}$ are obtained for $R_{1}$ and $R_{2}$ which is equal to $2R_{1}$ in the LC93 binary merging method, by using numerical integration. As a result, the void volume distribution function for the $\Lambda$CDM model in terms of size scale is,

\begin{eqnarray}
f_{S_{1}}(S_{1}(R_{1}),|\delta_{\mathbf{v}_{1}}|\big|S_{2}(R_{2}),|\delta_{\mathbf{v}_{2}}|)
\left|\frac{d S_{1}}{d R_{1}}\right| d{R_{1}}=
\frac{1}{\sqrt{{2\pi}}}\frac{|\delta_{\mathbf{v}_{1}}|-|\delta_{\mathbf{v}_{2}}|}{\left({S_{1}}(R_{1})-S_{2}(R_{2})\right)^{3/2}}\left|\frac{d S_{1}}{d R_{1}}\right|
\exp\left[\frac{-\left(|\delta_{\mathbf{v}_{1}}|-|\delta_{\mathbf{v}_{2}}|\right)^{2}}{2\left(S_{1}(R_{1})-S_{2}(R_{2})\right)}\right].
\label{sizedistributionCDM}
\end{eqnarray}
\noindent
The void size distribution function of the self similar models by using mass, the relation between volume and size scale by using equation (\ref{relation1}), is,

\begin{eqnarray}
\left|\frac{d\ln S}{d\ln M}\right|=\left|\frac{d\ln S}{d\ln V}\right|=\alpha,\phantom{a}\left|\frac{d\ln S}{d\ln R}\right|=3\alpha.
\label{differentiations}
\end{eqnarray}
\noindent
This leads to,
\begin{eqnarray}
f_{R_{1}}(R_{1},|\delta_{\mathbf{v}_{1}}|\big|R_{2},|\delta_{\mathbf{v}_{2}}|) \left|\frac{d S_{1}}{d R_{1}}\right| d{R_{1}}&=& \frac{3\alpha}{\sqrt{{2\pi}}}\frac{\delta^2_{\mathbf{v}_{1}}}{{R_{*}}} \left(\frac{R_{1}}{R_{*}}\right)^{-3\alpha-1}
\frac{|\delta_{\mathbf{v}_{1}}|-|\delta_{\mathbf{v}_{2}}|}{\left({\delta^{2}_{\mathbf{v}_{1}}}\left(\frac{R_{1}}{R_{*}}\right)^{-3\alpha}-
{{\delta^{2}_{\mathbf{v}_{2}}}\left(\frac{R_{2}}{R_{*}}\right)^{-3\alpha}}\right)^{3/2}}
\nonumber
\\ && \exp\left[\frac{-\left(|\delta_{\mathbf{v}_{1}}|-|\delta_{\mathbf{v}_{2}}|\right)^{2}}{2\left({{\delta^{2}_{\mathbf{v}_{1}}}\left(\frac{R_{1}}{R_{*}}\right)^{-3\alpha}}-
{{\delta^{2}_{\mathbf{v}_{2}}}\left(\frac{R_{2}}{R_{*}}\right)^{-3\alpha}}\right)}\right]d R_{1}.
\label{probabilitysize1}
\end{eqnarray}
\noindent
Equation (\ref{probabilitysize1}) indicates the void size probability distribution of a void corresponding to size $R_{1}$ at time $\delta_{\mathbf{v}_{1}}$ later on incorporates into another void corresponding to size $R_{2}$. Recall that here we follow the LC93 algorithm and this algorithm is a binary method. In the case of a binary method, we choose the initial void size $R_{1}$ which should be equal to its double size at $z_{2}$. Another important point that we should mention, is that LC93 assume $S\approx V$ by neglecting the time parameter, linear void density $\delta_{\mathbf{v}}$. In the above, we improve on this by taking into account time parameters. However after this point, we will follow the assumption of LC93 in order to obtain a merging tree formalism in their approximate analytical formalism.

\begin{figure}
\centering
\begin{tabular}{ll}
\includegraphics[width=0.45\textwidth]{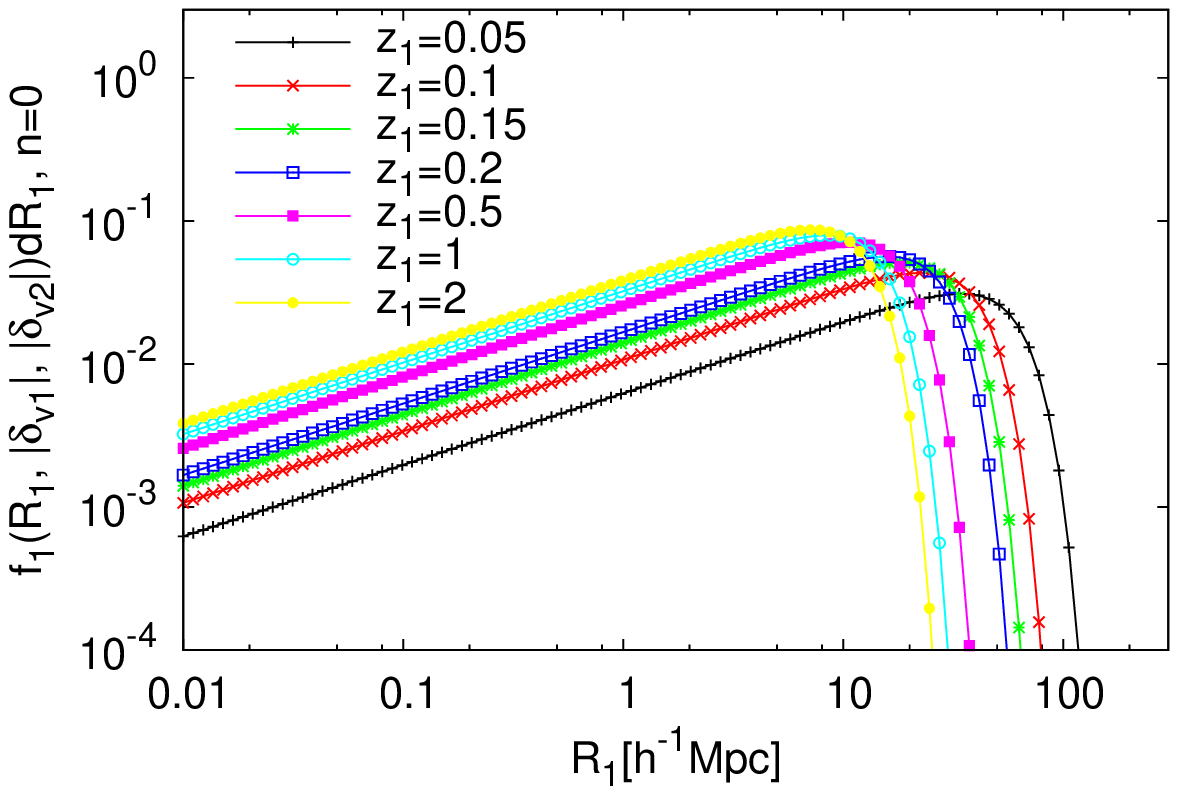}\\
\includegraphics[width=0.45\textwidth]{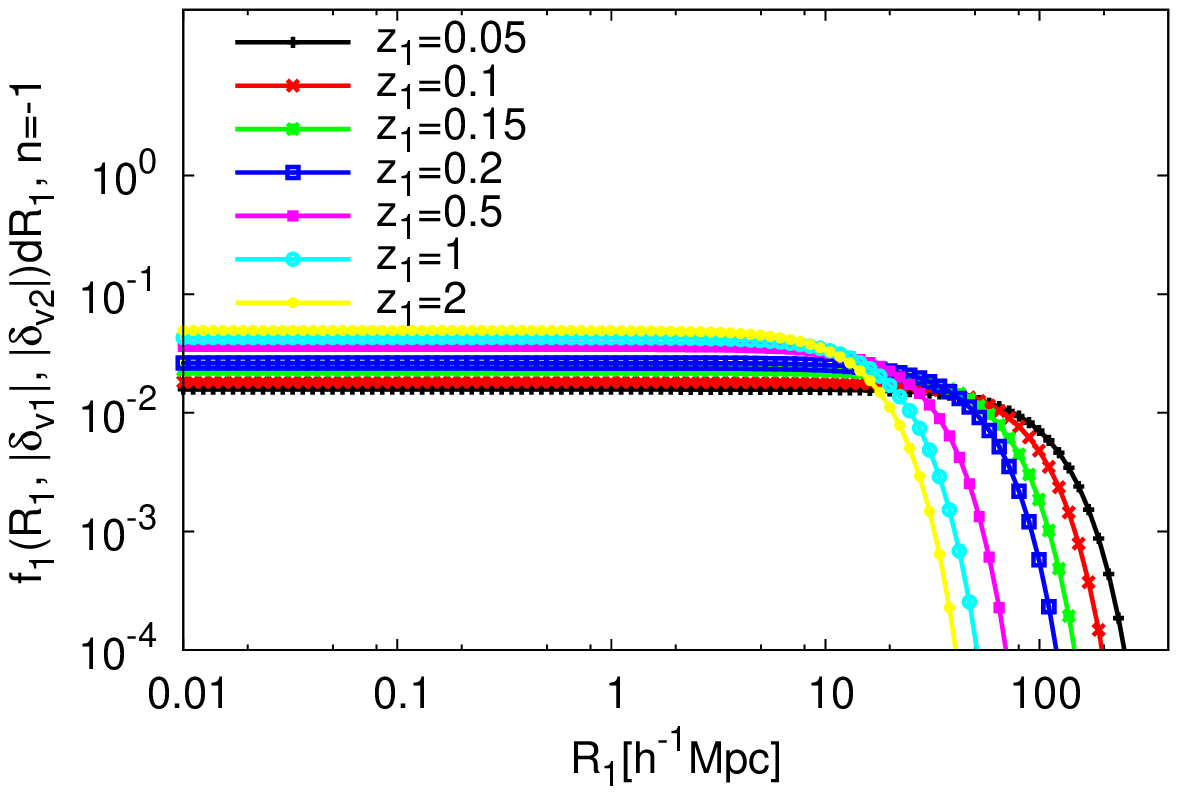}\\
\includegraphics[width=0.45\textwidth]{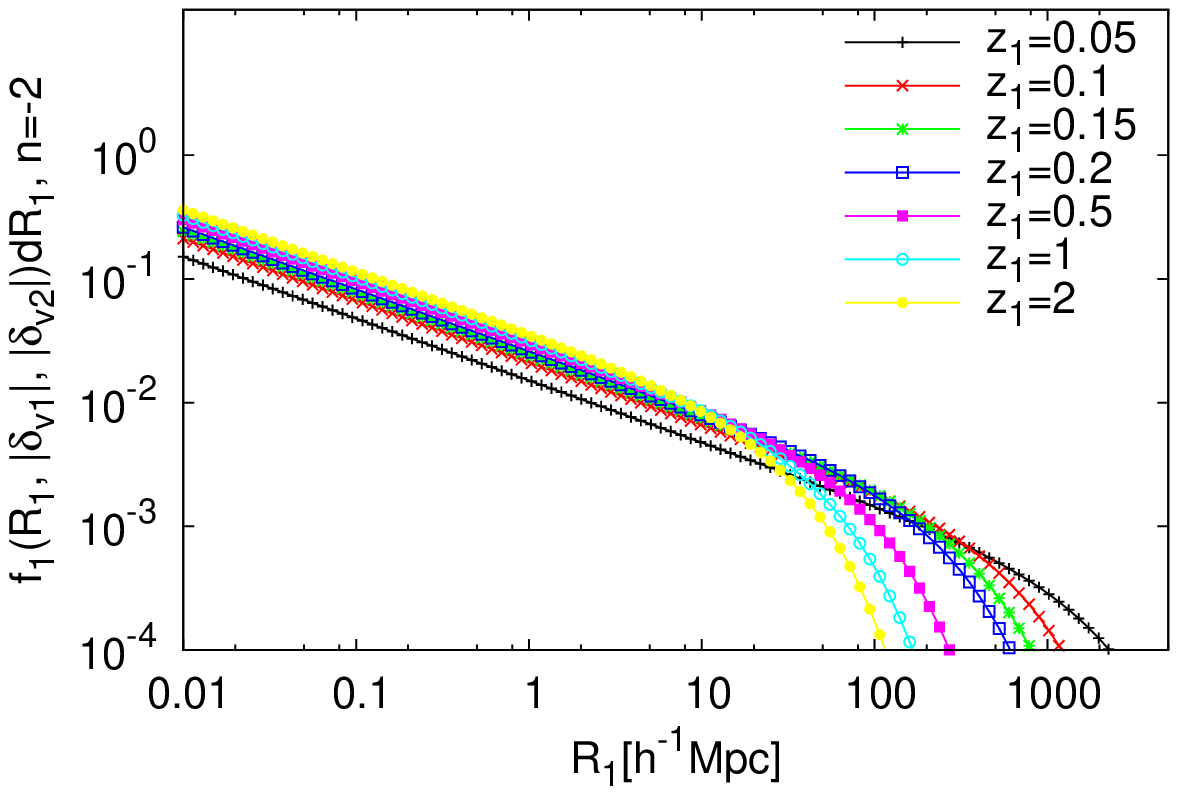}
\end{tabular}
\caption{The conditional void size distribution equation (\ref{probabilitysize1}) at given redshift $z_{1}$ in terms of $R_{1} [h^{-1} Mpc]$ with respect to the self similar models with index $n =$ $0$, $-1$, $-2$. In each case, the relation between a void with size $R_{1}$ at given redshift $z_{1}$ and the size of the system in which void $R_{1}$ is incorporated into, is given as $R_{2}= 2 R_{1}$. In each model, the characteristic void size $R_{*}=8 h^{-1}Mpc$ is obtained by equation (\ref{voidcharacteristicsize}). From left to right and top to bottom, at relatively small void sizes (left hand side of each plot), the void size distribution increases with decreasing spectral index. However, the void size distribution decreases in relatively larger voids as the spectral index decreases.}
\label{fig:f1distribution}
\end{figure}

\begin{figure}
\centering
\includegraphics[width=0.5\textwidth]{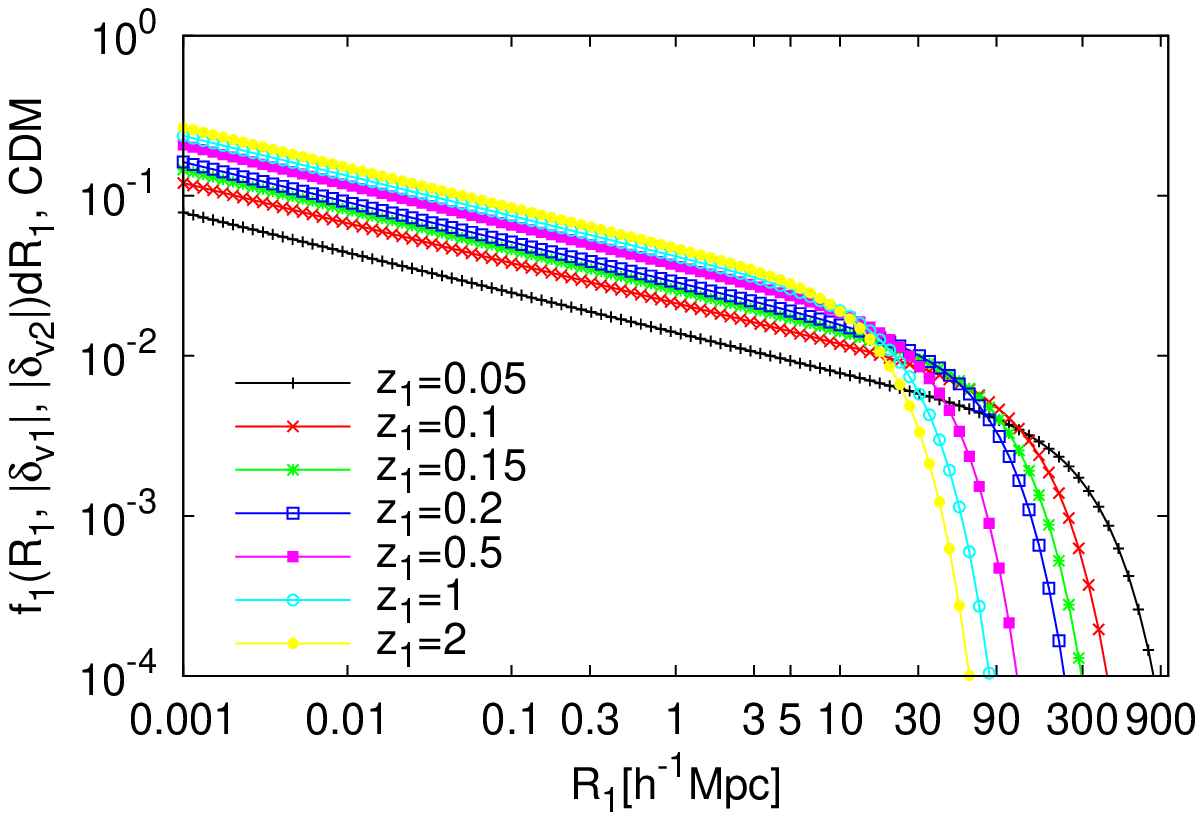}
%\hspace{-15mm}\includegraphics[scale=0.69]{Chapter3/CH3largevoiddenemen15.eps}
\caption{The conditional probability equation (\ref{probabilitysize1}) in terms of initial void size $R_{1} [h^{-1} Mpc]$ for the CDM model at given redshift values $z_{1} =$ $0.05$, $0.1$, $0.15$, $0.2$, $0.5$, $1$, $2$. In this model, CDM spectra on megaparsec scales
is obtained by choosing the spectral index $n=-1.5$ and by using the characteristic scale as $R_{*}=8 h^{-1}Mpc$ in equality (\ref{voidcharacteristicsize}). From left to right the void size distribution decreases towards relatively larger voids. The CDM model indicates that as redshift $z_{1}$ approaches the present day ($z_{1}\rightarrow z_{2}=0$) it is possible to see very large voids although they are less numerous.}
\label{fig:f1distributionCDM}
\end{figure}

Fig. \ref{fig:f1distribution} and Fig. \ref{fig:f1distributionCDM} show the conditional void size distribution equation (\ref{probabilitysize1}) for self similar models ($n=0, -1, -2$) and the CDM at given redshift ($z_{1}$). As a result of this, Figs. \ref{fig:f1distribution} and \ref{fig:f1distributionCDM} provide some interesting properties that give an insight into understanding void merging in terms of the size distribution based on toy models as well as the physical spectrum. In Fig. \ref{fig:f1distribution}, the void size distribution function or void size probability distribution for relatively small size voids increases with decreasing spectral index for self similar models. In all panels, the exponential cutoff in the size of the void distribution moves to very large sizes with decreasing index and decreasing redshift values. It is also obvious that in all models the small size void distribution is higher at high redshifts, $z_{1} \geq 0.5$ than at low redshift values. This may be an indication of a void hierarchy that is actually in agreement with the theoretical work of \cite{sw} in which they infer that the small size voids present at high redshifts must merge with each other to make larger voids that are present at lower redshifts. However large voids at low redshifts are less numerous. Similar to the self similar models, the void size distribution of the physical spectrum of small size voids at high redshift values $z_{1} \geq 0.5$ is higher than the one at lower redshift values (see In Fig. \ref{fig:f1distributionCDM}). Also large size voids are less dominant than small voids.

Note that since the conditional probability of the void size has a more complex expression than the distribution for volume scale equation (\ref{probabilitys1}), as of now, we will obtain the merging algorithm of voids in terms of scale function by following LC93. However in our plots we adopt the related distributions as a function of void size. That is why that for accuracy, we use the differentiation relations when the transformation is necessary.

Another probability density function that can be derived from the random walks is the probability of a trajectory first upcrossing $|\delta_{\mathbf{v}_{2}}|$ then $|\delta_{\mathbf{v}_{1}}|$ at $S_{1}$,

\begin{eqnarray}
f_{S_{2}}(S_{2},|\delta_{\mathbf{v}_{2}}|\big|S_{1},|\delta_{\mathbf{v}_{1}}|)
d{S_{2}}&=&\frac{|\delta_{\mathbf{v}_{2}}|\left(|\delta_{\mathbf{v}_{1}}|-|\delta_{\mathbf{v}_{2}}|\right)}{\sqrt{{2\pi}}|\delta_{\mathbf{v}_{1}}|}
{\left[\frac{S_{1}}
{S_{2}\left({S_{1}}-{S_{2}}\right)}\right]^{3/2}}
\exp\left[-\frac{\left(S_{1}|\delta_{\mathbf{v}_{2}}|-S_{2}|\delta_{\mathbf{v}_{1}}|\right)^{2}}{2S_{1}S_{2}\left(S_{1}-{S_{2}}\right)}\right]dS_{2},
\label{probabilitys2}
\end{eqnarray}
\noindent
which helps to obtain merging rates.

\section{Void Merging and Absorption Rates}\label{sec:VoidMerandAbs}
\begin{figure}
\centering
\begin{tabular}{l}
\includegraphics[width=0.5\textwidth]{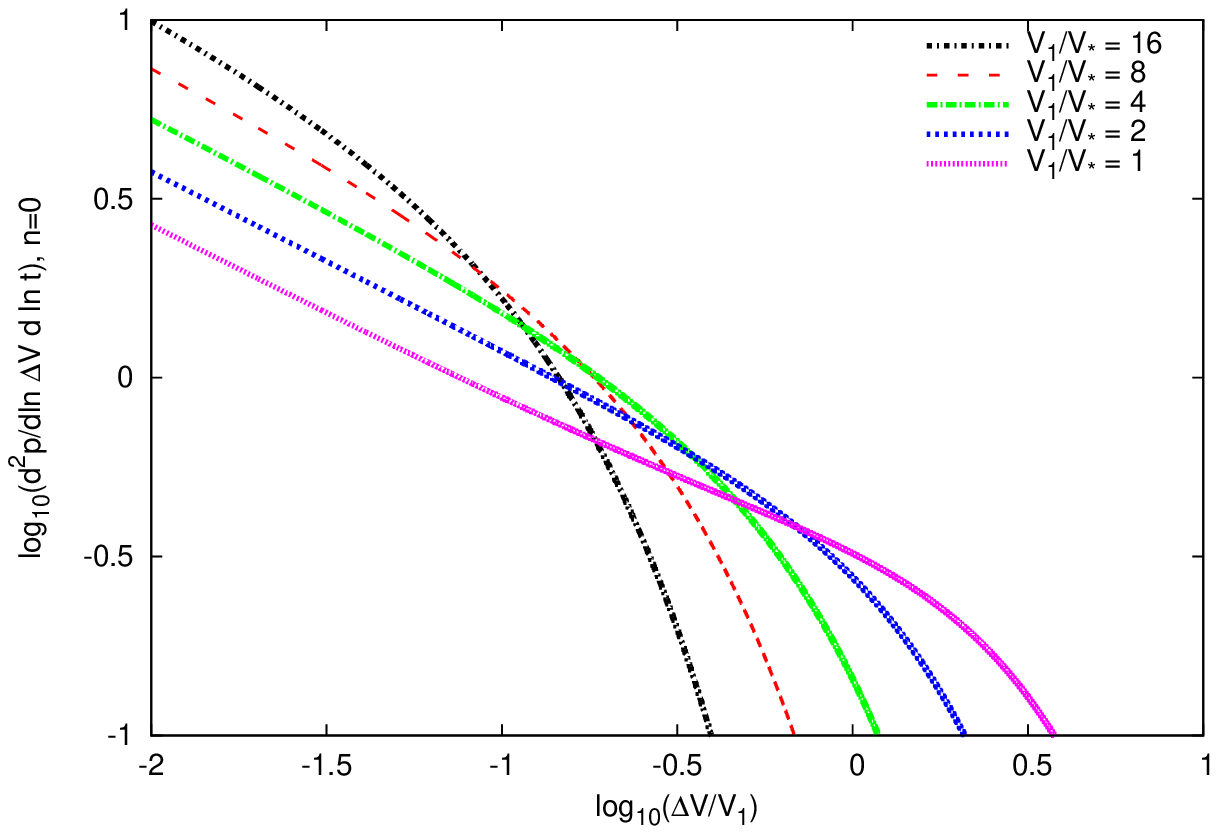}\\
\includegraphics[width=0.5\textwidth]{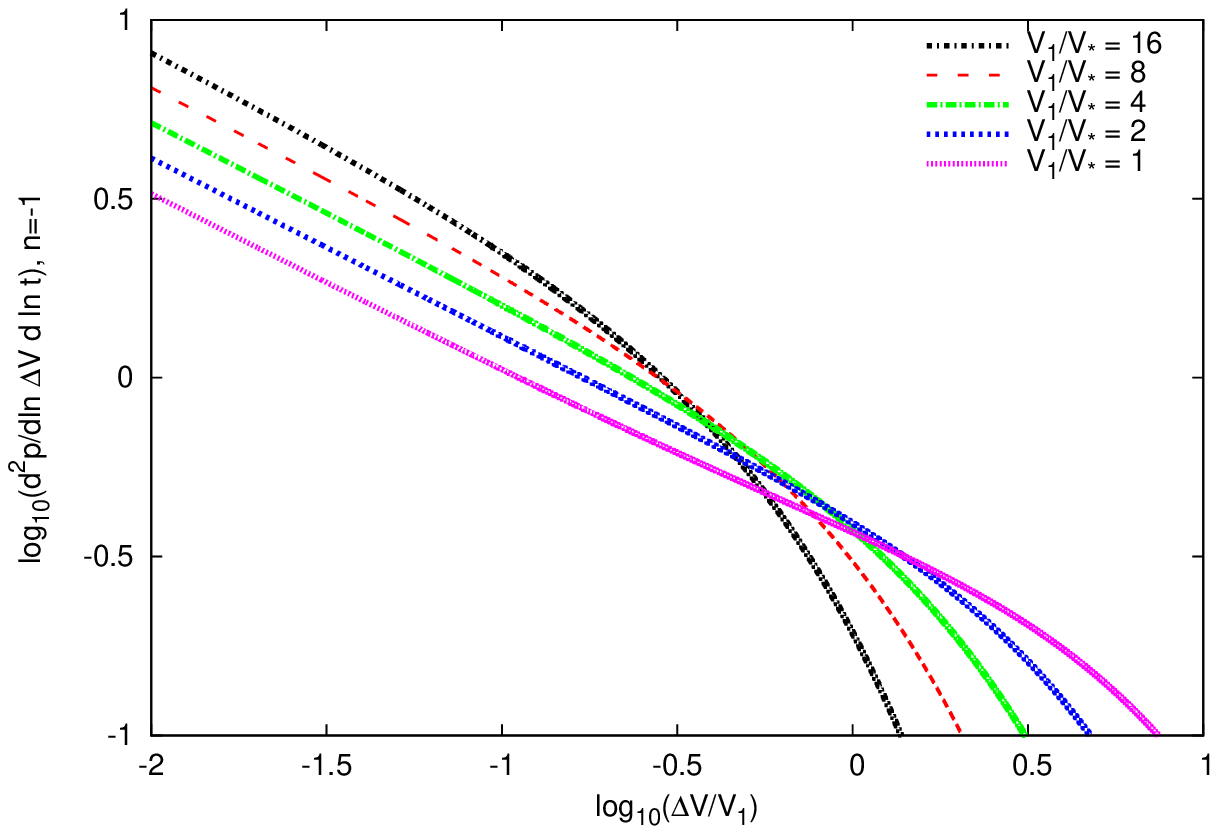}\\
\includegraphics[width=0.5\textwidth]{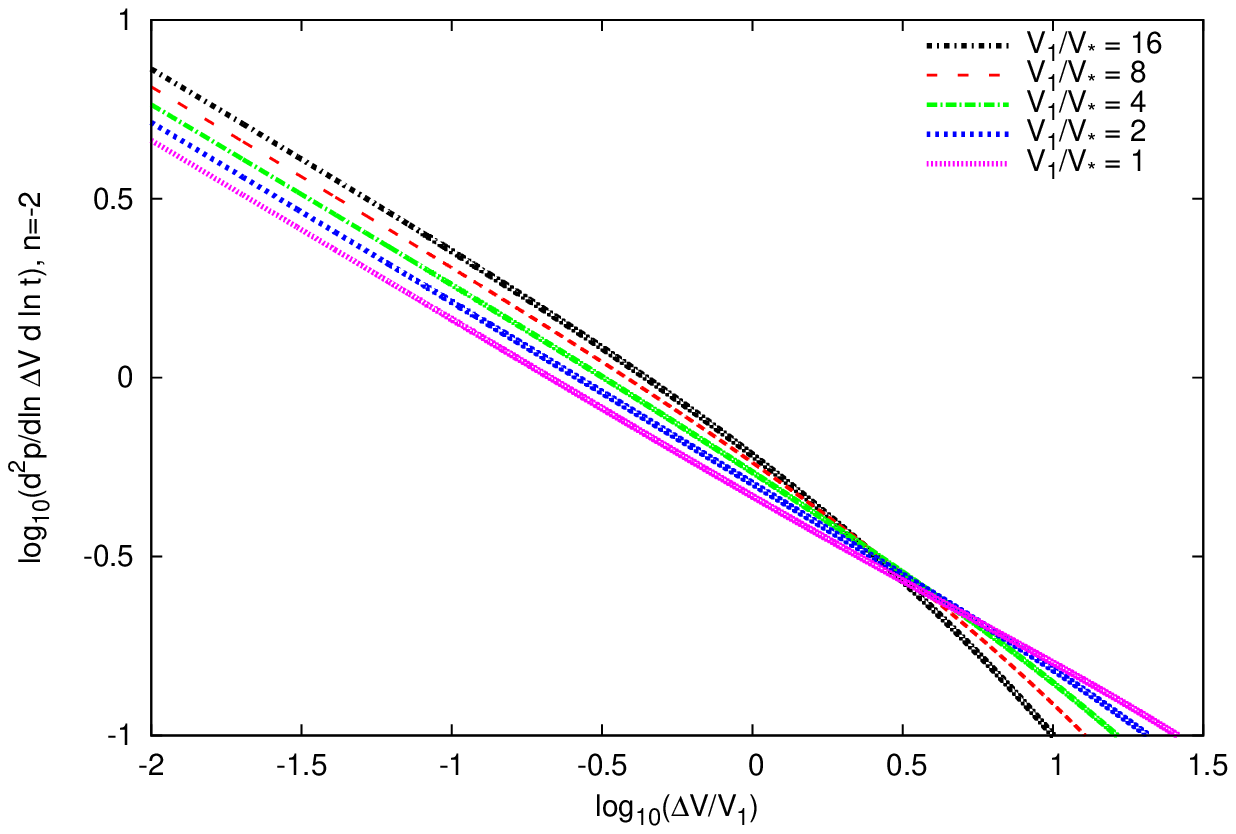}
%\hspace{0mm}\includegraphics[height=47mm,width=71mm]{Chapter3/LVz05merger.eps}
%\hspace{0mm}\includegraphics[scale=0.17]{Chapter3/LVz05merger.eps}
%\hspace{-10mm}\includegraphics[scale=0.165]{Chapter3/LVz1merger.eps}
\end{tabular}
\caption{The merger rates of the growing voids given by equation (\ref{mergingrate}) for the self similar models with index $n$ = $0$, $-1$,$-2$ In each panel, the curve which is the highest on the left of each plot, $V_{1}/V_{*} = 16$ and successive curves are $V_{1}/V_{*} =$ $8$, $4$, $2$, $1$.}
\label{fig:mergerrates}
\end{figure}

\begin{figure}
\centering
\begin{tabular}{l}
\includegraphics[width=0.62\textwidth]{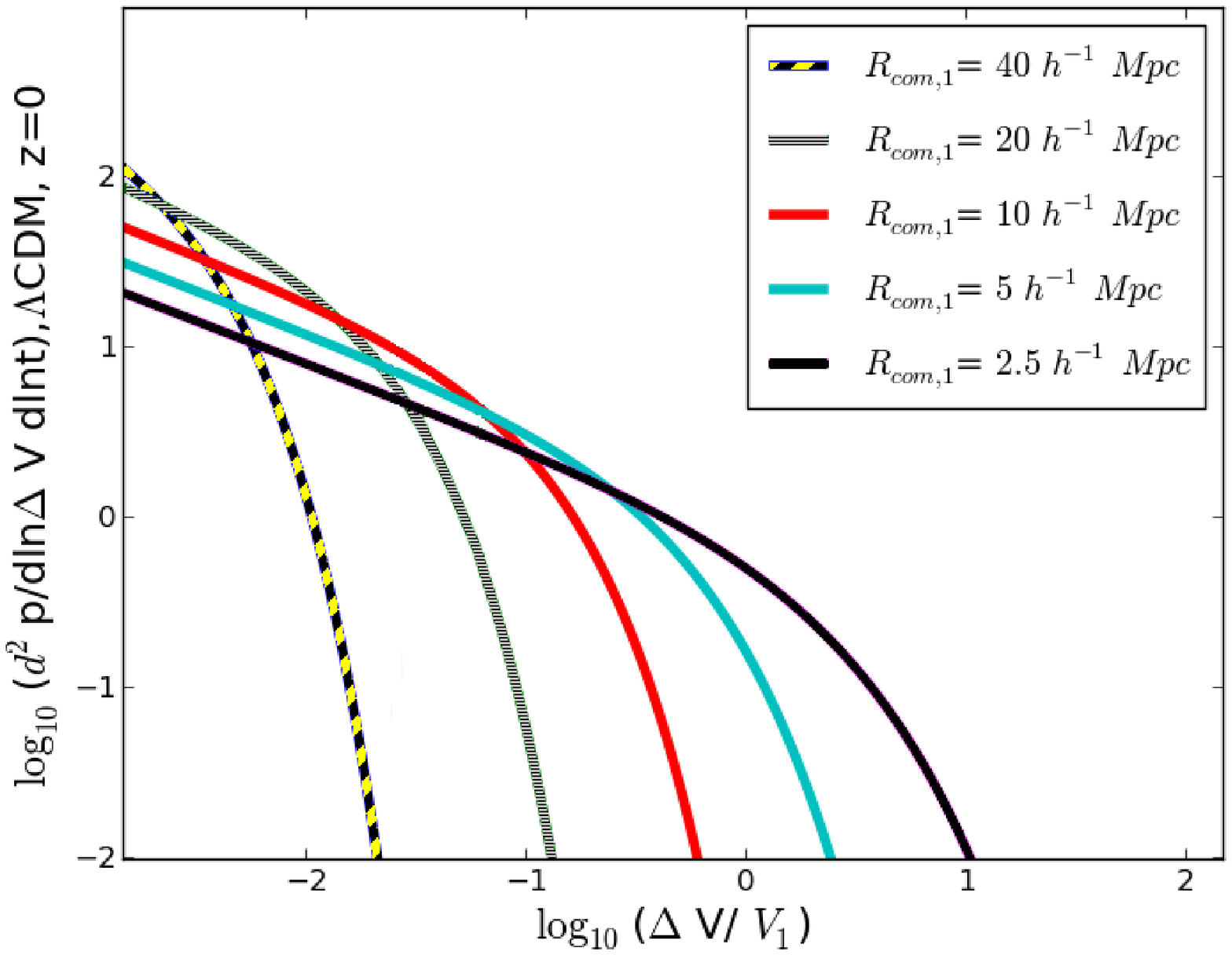}\\
\end{tabular}
\caption{The merger rate of the growing voids given by equation (\ref{mergingrate}) for the CDM model at redshift $z=0$. The successive curves represent the comoving volume with radii $R_{com,1} =$ $40$, $20$, $10$, $5$ and $2.5 h^{-1} Mpc$.}
\label{fig:mergerratesCDM}
\end{figure}

To obtain merging and absorption rates, first we derive the mean transition rate by taking the limit $|\delta_{\mathbf{v}_{2}}|\rightarrow |\delta_{\mathbf{v}_{1}}|=|\delta_{\mathbf{v}}|$ from equation (\ref{probabilitys2}),

\begin{eqnarray}
\frac{{d^2} p}{d{S_{2}}d|{{\delta_{\mathbf{v}}}}|}\left({S_{1}}\rightarrow
{S_{2}},|{\delta_{\mathbf{v}}}|\right) = \frac{1}{\sqrt{{2\pi}}}
{\left[\frac{S_{1}}{{S_{2}}\left({S_{1}}-{S_{2}}\right)}\right]^{3/2}}
\exp\left[-\frac{|{\delta_{\mathbf{v}}}|^2\left({S_{1}}-{S_{2}}\right)}{2{S_{1}}{S_{2}}}\right]
d{S_{2}}d|{\delta_{\mathbf{v}}}|.\label{meanrateb}
\end{eqnarray}
\noindent
This equation can be interpreted as one or more merging void events depending on the barrier $|\delta_{\mathbf{v}}|$, by following LC93. While any finite interval
of $\Delta \delta_{\mathbf{v}}$ at $\Delta{S}$ shows the cumulative effect of more than one merger, an infinitesimal interval $d\delta_{\mathbf{v}}$
at $d{S}$ indicates a single void merger event. Hence equation (\ref{meanrateb}) represents the probability of a void with volume scale $S_{1}$ at later times merging with another void of volume $\Delta V =V_{2}-V_{1}$. That is why we can define this probability transition function (\ref{meanrateb}) as the void probability transition. Thus, by following the halo merging algorithm by \cite{lace}, \emph{the so called merging rate of voids is defined as the rate of change in the transition probability of a void $V_{1}$ in terms of the total volume that increases due to merging events per unit time $t$},

\begin{eqnarray}
\frac{d^2 p}{d \Delta V dt}({V_{1}}\rightarrow V_{2},|t)= \rm{Void~Merging~Rate}.
\label{mergingrate1}
\end{eqnarray}
\noindent
Hence the explicit form of the void merging rate is given by,

\begin{eqnarray}
\frac{d^2 p}{d\ln \Delta V d\ln t}({V_{1}}\rightarrow V_{2},|t)
=\sqrt{\frac{2}{\pi}}\frac{\Delta
V}{V_{2}}\frac{|\delta_{\mathbf{v}}(t)|}{\sqrt{S_{2}}}
\left|\frac{d\ln {\sqrt{S_{2}}}}{d\ln
V_{2}}\right|\left|\frac{d\ln|\delta_{\mathbf{v}}(t)|}{d\ln
t}\right|
\frac{1}{\left(1-{S_{2}}/{{S_{1}}}\right)^{3/2}}
\exp\left[-\frac{|\delta_{\mathbf{v}}(t)|^2}{2}\left(\frac{1}{{S_{2}}}
-\frac{1}{{S_{1}}}\right)\right],
\label{mergingrate}
\end{eqnarray}
\noindent
in which $S_{1}= S(V_{1})$ and $S_{2}= S(V_{2})$ are the volume scale functions of voids smoothed on the scales of interest. The merging of the self similar models with index $n=$ $-2$, $-1$, $0$ and the CDM with redshift values $z=0$ are shown in Figs. \ref{fig:mergerrates} and \ref{fig:mergerratesCDM}. In Figs. \ref{fig:mergerrates} and \ref{fig:mergerratesCDM} each plot shows that voids with higher initial volume have higher merger rates. The merger rates of initially larger voids decrease when the ratio of progenitor and main volumes $V_{2}/V_{1}$ increases. In self similar models (see Fig. \ref{fig:mergerrates}), when the index decreases the merger rates of each curve representing different progenitor volumes get closer to each other.

Here we define another important parameter of void hierarchical build up, \emph{void absorption rate}. The void absorption rate is analogous to the accretion rate of a halo (LC93). \emph{The definition of absorption rate is given as the rate at which a void absorbs volume from surrounding small void(s)}.

\begin{eqnarray}
\frac{\Delta V}{V_{1}}\underbrace{\frac{d^2 p}{d\ln \Delta V d\ln t}({V_{1}}\rightarrow V_{2},|t)}_{\rm {Void~Merging~Rate}}= \rm{Void~Absorption~Rate}.
\label{mergingabsorbtion}
\end{eqnarray}
\noindent
In terms of the random walk concept, void absorption behaves as analogous to the accretion of halos. Void absorptions can be interpreted as small random walk steps in $S$ corresponding to upward steps in the trajectory of $\delta$ versus $S$. This behavior corresponds to incremental absorption events that add only a small amount of volume to the total merging rate. That is why the distinction between absorption and merger rates is strongly correlated with the resolution $\Delta{S}=S_{2}-S_{2}$.

\begin{figure}
\centering
\begin{tabular}{l}
\includegraphics[width=0.5\textwidth]{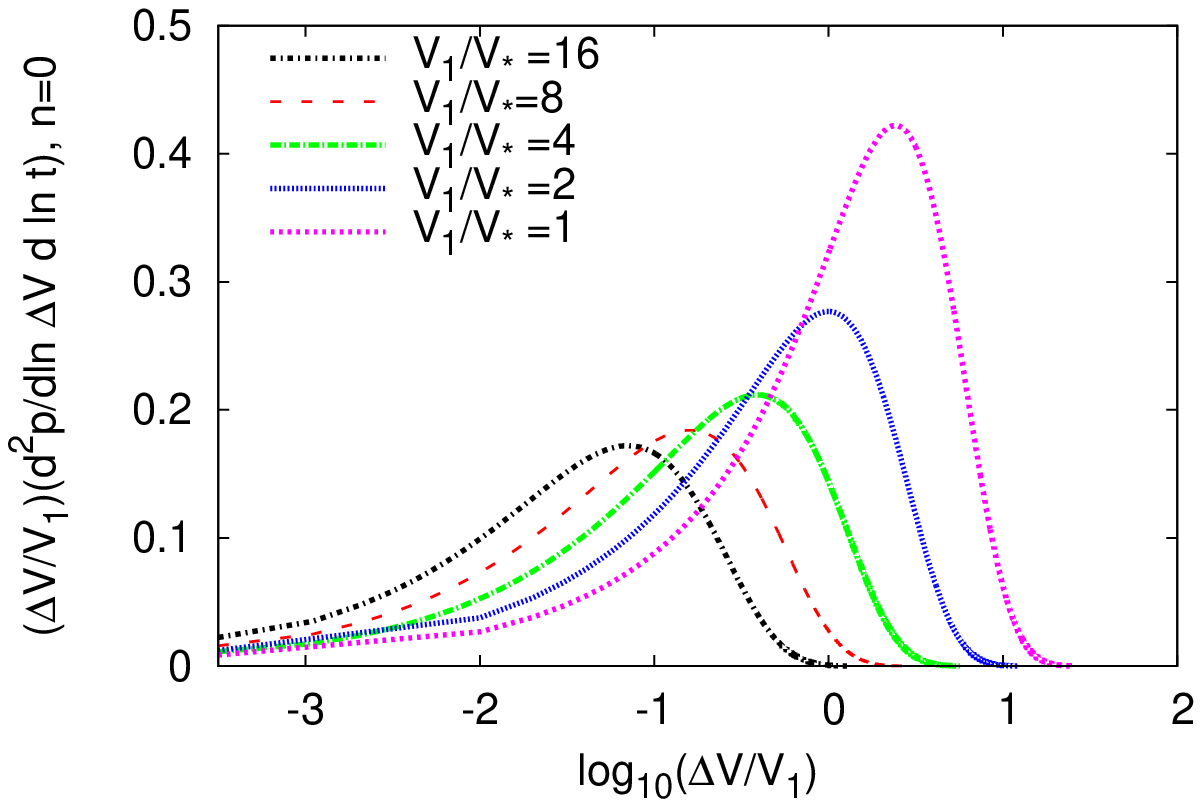}\\
\includegraphics[width=0.5\textwidth]{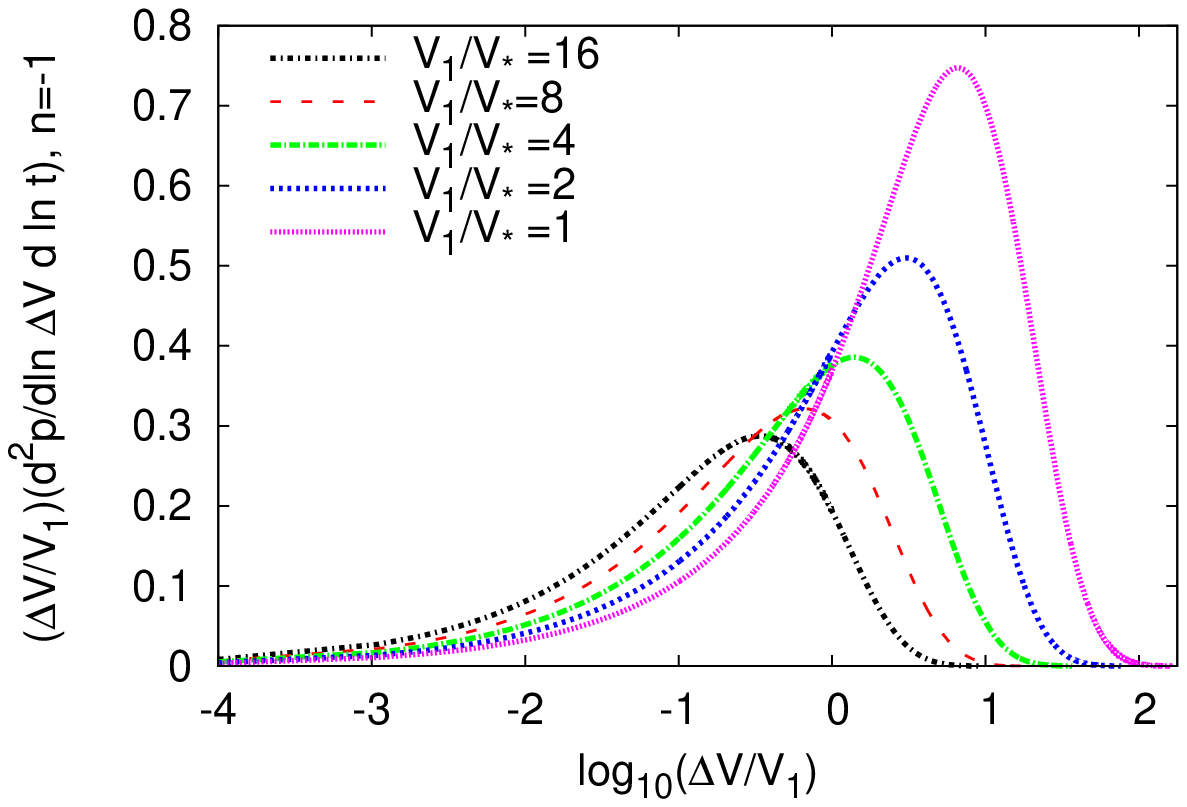}\\
\includegraphics[width=0.5\textwidth]{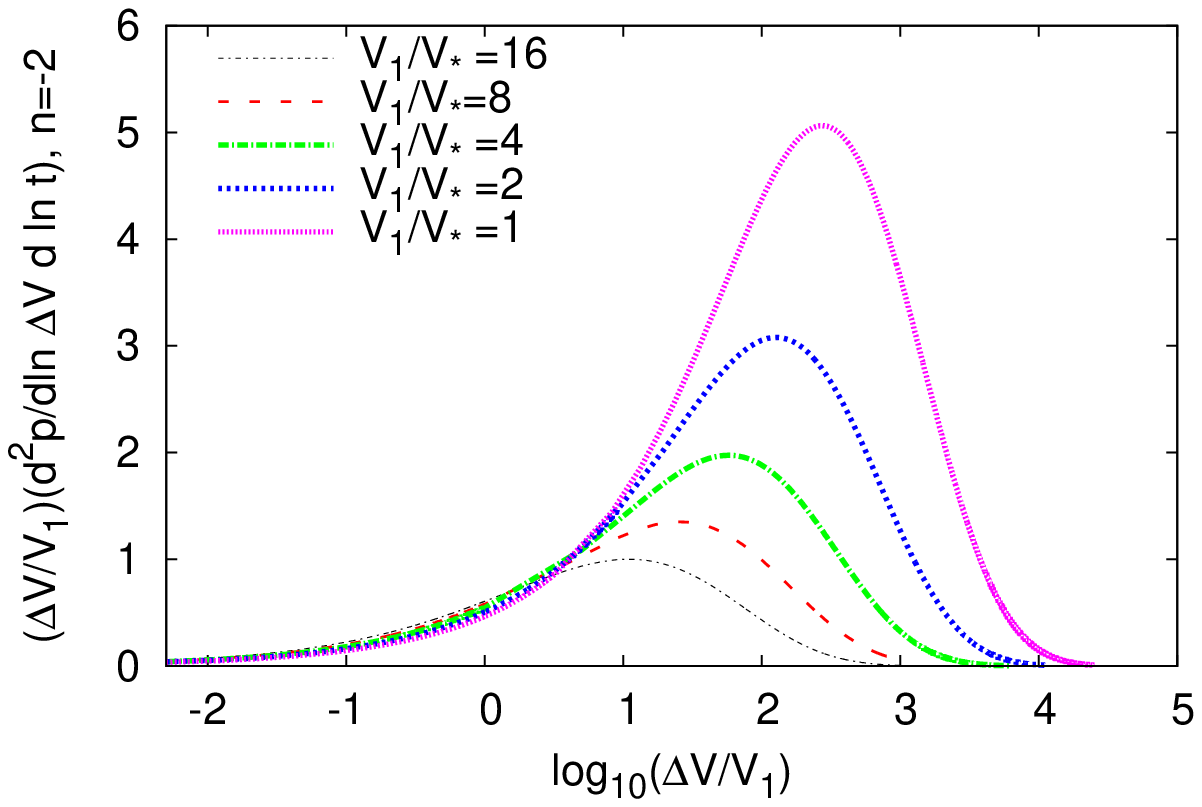}
%\hspace{-1mm}\includegraphics[scale=0.24]{Chapter3/LVz0lost.eps}
%\hspace{-4mm}\includegraphics[scale=0.3]{Chapter3/LVz05lost.eps}
%\hspace{-2mm}\includegraphics[scale=0.3]{Chapter3/LVz1lost.eps}
\end{tabular}
\caption{The absorption rates given by equation (\ref{mergingabsorbtion}) for the self similar models with the index $n=-1, -2, 0$. The quantity plotted is the fractional volume absorption per Hubble time, ($\Delta V/V_{1}$) ($d^2p/d\ln \Delta V d\ln t$). The curve which is the highest on the left of each plot has $V_{1}/V_{*}= 16$ and successive curves have $V_{1}/V_{*} =$ $8$, $4$, $2$, $1$.}
\label{fig:lossrates}
\end{figure}

\begin{figure}
\centering
\begin{tabular}{l}
\includegraphics[width=0.75\textwidth]{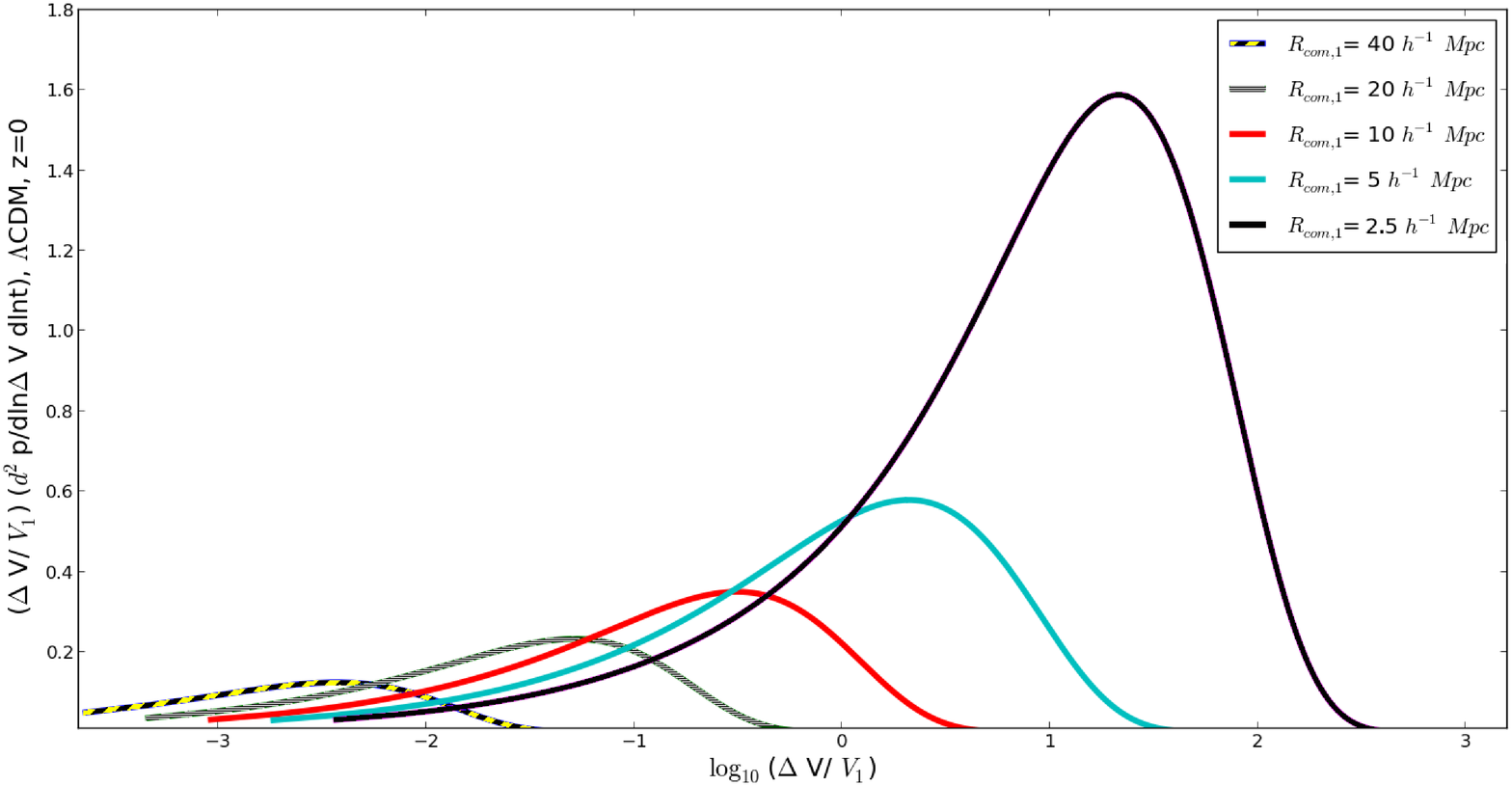}
\end{tabular}
\caption{The absorption rates given by equation (\ref{mergingabsorbtion}) for the CDM model at redshift value $z=0$. In the CDM model, the curves represent the comoving volume with radii $R_{com,1} =$ $40$, $20$, $10$, $5$ and $2.5 h^{-1} Mpc$.}
\label{fig:lossratesCDM}
\end{figure}

Figs. \ref{fig:lossrates} and \ref{fig:lossratesCDM} show the absorption rates of self similar and the CDM models. The figures indicate that mergers with small volume or small comoving radius (in the CDM model) dominate the absorption rate numerically while the merger rates are dominated by large voids as we see in Figs. \ref{fig:mergerrates} and \ref{fig:mergerratesCDM}. This indicates that the voids with relatively small volume are absorbed by their larger counterparts.

\section{Survival and Failure Times in The Growing Void Merging Tree Algorithm}\label{sec:SurvFail}
LC93's halo merging tree algorithm, halo survival and formation times are well defined within the context of probability theory. By following their method we obtain these times for growing spherical voids. Recall that LC93 define the barriers of the EPS formalism as time parameters of the hierarchical evolution such as $|\delta_{\mathbf{v}_{1}}|\approx z_{1}$ and $|\delta_{\mathbf{v}_{2}}|\approx z_{2}$.

\subsection{Void Survival and Failure Times}
\begin{figure}
\centering
\begin{tabular}{l}
\includegraphics[width=0.5\textwidth]{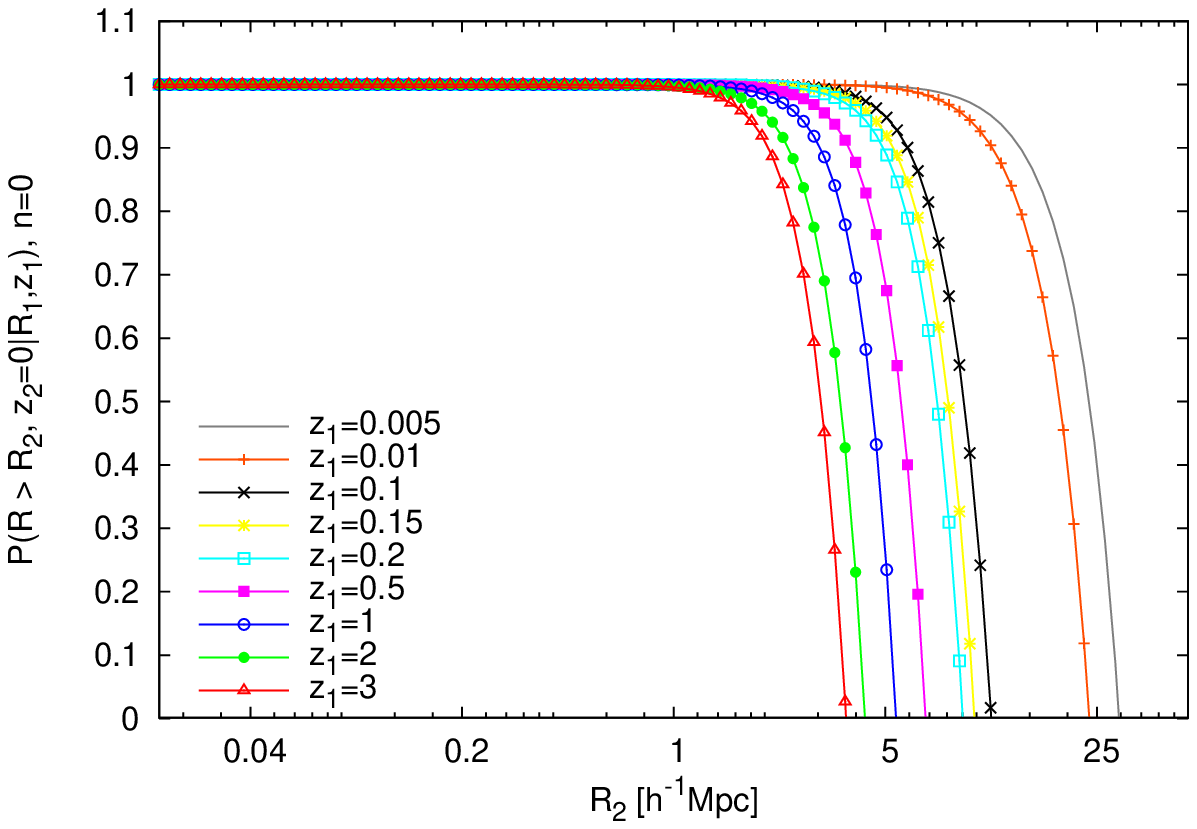}\\
\includegraphics[width=0.5\textwidth]{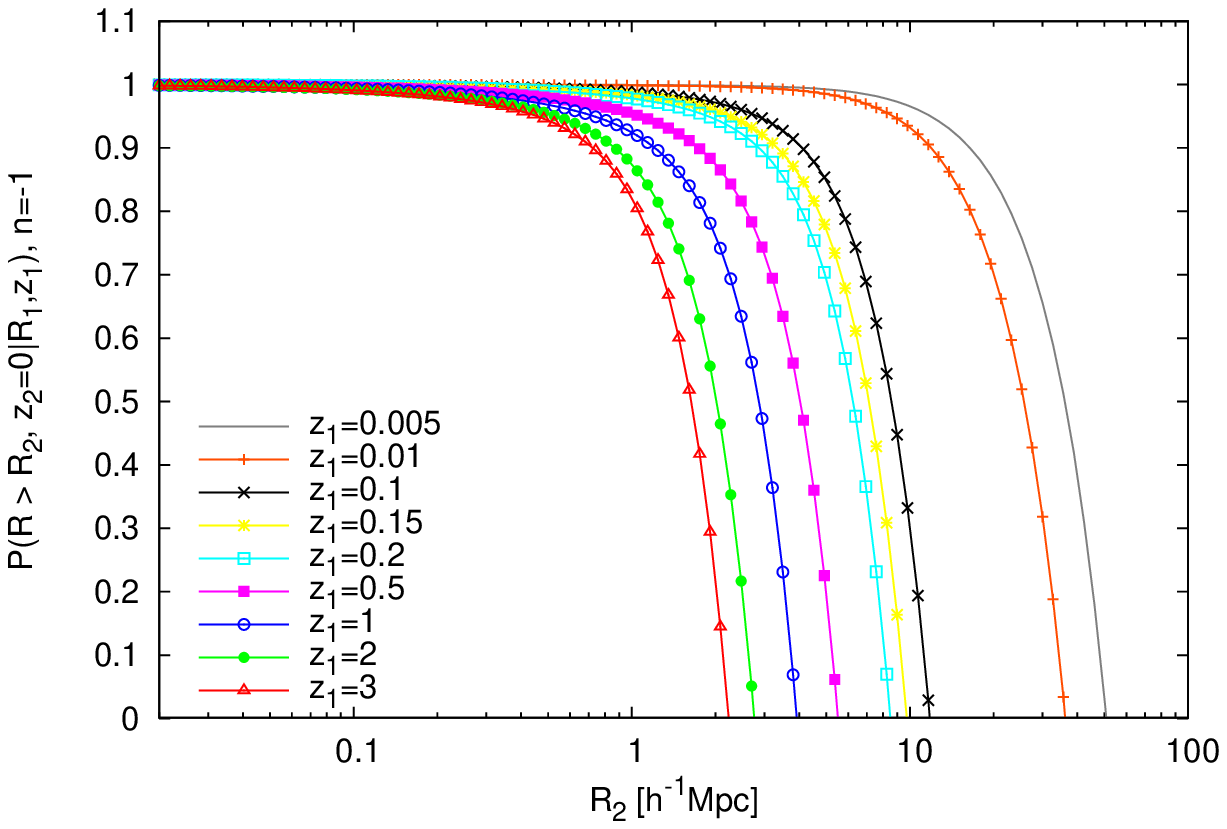}\\
\includegraphics[width=0.5\textwidth]{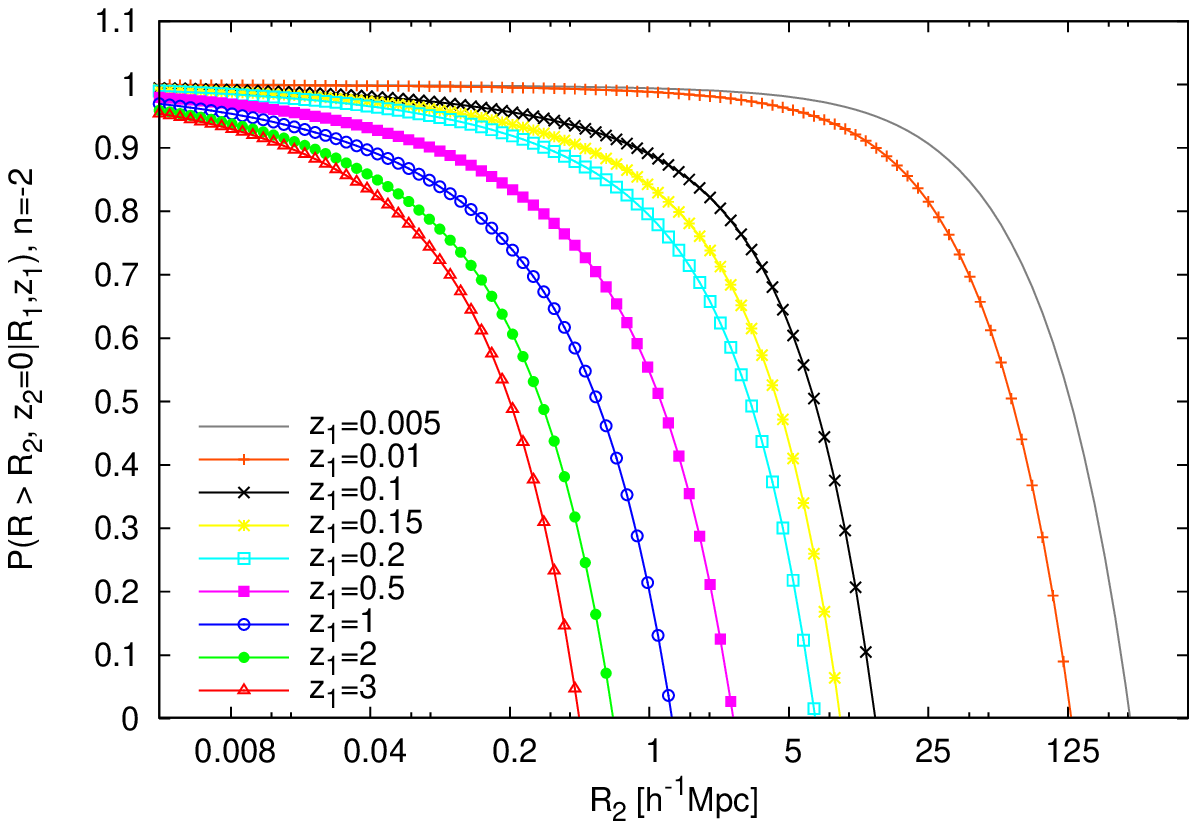}
\end{tabular}
\caption{The void size survival probabilities with respect to their present size $R_{2}$ in terms of self similar models with index $n$ = $0$, $-1$,$-2$.
According to this, a void with size $R_{1}$ at redshift $z_{1}$, later on, is incorporated into a void such that the size of the system becomes $R_{2}= 2R_{1}$ at present day $z_{2}=0$.}
\label{fig:analyticsurvivalvolumes}
\end{figure}

\begin{figure}
\centering
\begin{tabular}{l}
\includegraphics[width=0.5\textwidth]{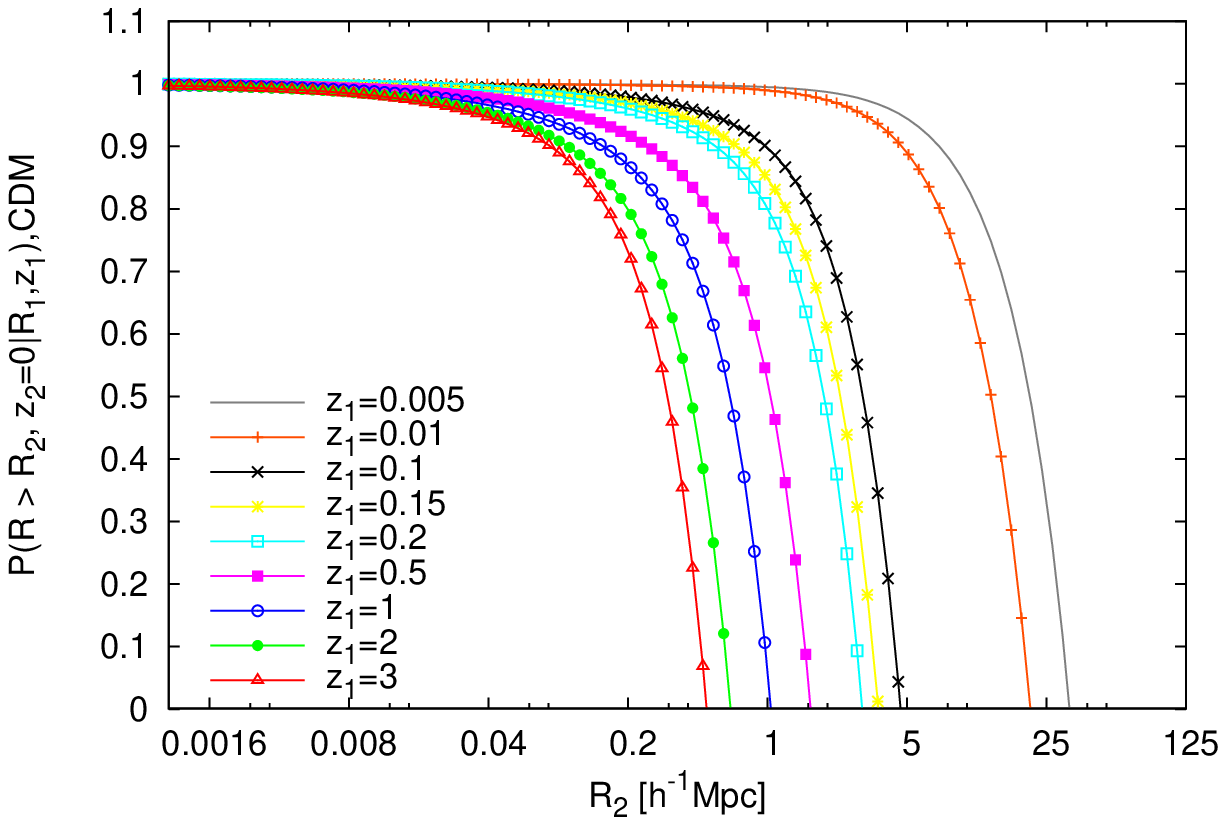}
\end{tabular}
\caption{The void size survival probabilities in terms of the approximated CDM model at given redshifts by setting $n=-1.5$.}
\label{fig:analyticsurvivalvolumesCDM}
\end{figure}

The survival time of a void is the time when a void with volume scale $S(V)$ lives before being incorporated into, or absorbed by, a larger void. The volume of this larger void is chosen to be double the size of the absorbed void at the volume scale $S(2 V)$ since LC93 is based on the binary method, as we mentioned before. This leads to the survival time of a void with volume $V$ being defined as the time when the volume gets doubled, $2V$ due to merging. The survival probability function of a void succeeding to merge into its double size is given by,

\begin{eqnarray}\label{defsurv}
P\left[S> S_{2}\right] &=& 1-P\left[S < S_{2}\right]\nonumber\\
\underbrace{P(S>S_{2},{\delta_{\mathbf{v}_{2}}}|S_{1},\delta_{\mathbf{v}_{1}})}_{\text{probability of survival beyond}\phantom{a}S_{2}} &=& 1-\underbrace{P(S<S_{2},{\delta_{\mathbf{v}_{2}}}|S_{1},\delta_{\mathbf{v}_{1}})}_{\text{probability of a void failing before reaching}\phantom{a} S_{2}},
\end{eqnarray}
\noindent
where the survival probability distribution $P(S>S_{2},\delta_{c}|S_{1},\tilde{\delta}_{v})$ varies between one and zero indicating survival and death
processes, while $P(S<S_{2},{\delta_{\mathbf{v}_{2}}}|S_{1},\delta_{\mathbf{v}_{1}})$ indicates the probability that a void with volume scale $S(V_{1})$ at time $\delta_{\mathbf{v}_{1}}$ cannot merge into its double volume $V_{2}$ at time $\delta_{\mathbf{v}_{2}}$. Therefore, this probability is called failure or failing probability since the void at $S(V_{1})$ can not merge into a void with volume scale $S(V_{2})$ at barrier $\delta > {\delta_{\mathbf{v}_{2}}}$, and its explicit form is given by,

\small
\begin{eqnarray}
P(S< S_{2},{\delta_{\mathbf{v}_{2}}}|S_{1},\delta_{\mathbf{v}_{1}})&=&\int^{S_{2}}_{0}f_{S_{2}}\left(S^{\prime}_{2},|\delta_{\mathbf{v}_{2}}||
S_{1},|\delta_{\mathbf{v}_{1}}|\right)dS^{\prime}_{2}.
\label{failuresolprobs2a}
\end{eqnarray}
\normalsize
\noindent
We can then obtain the survival probability of merging voids by using equation (\ref{failuresolprobs2a}),

\small
\begin{eqnarray}
P(S> S_{2},{\delta_{\mathbf{v}_{2}}}|S_{1},\delta_{\mathbf{v}_{1}})&=&1-\int^{S_{2}}_{0}f_{S_{2}}\left(S^{\prime}_{2},|\delta_{\mathbf{v}_{2}}||
S_{1},|\delta_{\mathbf{v}_{1}}|\right)dS^{\prime}_{2}\nonumber\\
&=&1-\sqrt{\frac{2}{\pi}}|\delta_{\mathbf{v}_{1}}-\delta_{\mathbf{v}_{2}}|\left|\frac{\delta_{\mathbf{v}_{2}}}{\delta_{\mathbf{v}_{1}}}\right|
\left[\sqrt{\frac{S_{2}}{S_{1}\left(S_{1}-S_{2}\right)}}e^{-\frac{\delta^{2}_{\mathbf{v}_{1}}}{2}\left(\frac{1}{S_{1}}-\frac{1}{S_{2}}\right)}\right.\nonumber\\
&+& \left.\sqrt{\frac{\pi}{2}}\frac{1}{|\delta_{\mathbf{v}_{1}}|}\left(\frac{\delta^2_{\mathbf{v}_{1}}}{S_{1}}-1\right)
\text{erf}\sqrt{\frac{|\delta^2_{\mathbf{v}_{1}}|}{2}\left(\frac{1}{S_{1}}-\frac{1}{S_{2}}\right)}\right]\label{solprobs2a},
\end{eqnarray}
\normalsize
\noindent
in which the scales should be set as $S_{1}\approx S(V)$, $S_{2}\approx S(2V)$, $\delta_{\mathbf{v}_{1}}= \delta_{\mathbf{v}}(t)$ and ${\delta_{\mathbf{v}_{2}}}= \delta_{\mathbf{v}}(t_{surv})$.
The survival probability function $P(S>S_{2},{\delta_{\mathbf{v}_{2}}}|S_{1},\delta_{\mathbf{v}_{1}})$ is usually assumed to approach zero as volume increases while volume scale decreases as $V_{2}\rightarrow \infty$ (see Fig.~\ref{fig:analyticsurvivalvolumes} and Fig.~\ref{fig:analyticsurvivalvolumesCDM}). Figs.~\ref{fig:analyticsurvivalvolumes} and \ref{fig:analyticsurvivalvolumesCDM} show the survival probabilities of void size distributions based on equation (\ref{solprobs2a}) in terms of self similar and $\Lambda$CDM models. As is seen, as redshift decreases the range of surviving void sizes increases while the small size voids have more chance to survive than their larger counterparts. This means that at high redshifts, a large size void ($\gg 10 h^{-1}Mpc$), is unlikely to survive until present day. Instead, there are very small size surviving voids with a narrow range of radii depending on the model. Please note that this result clearly depends on the definition of a void, and the
sparsity of the model that we use in our calculations, which is based on LC93 for dark matter halos in which a binary system is used. As we know,
large density depressions may exist even in a Gaussian field with small fluctuations, depending on the definition of density depression.

In Fig.~\ref{fig:analyticsurvivalvolumes}, when the spectral index is decreasing as an indication of hierarchical clustering, the size of surviving voids at high redshifts becomes lower and their survival probabilities decrease. Similar behavior is seen in the CDM model as well (Fig.~\ref{fig:analyticsurvivalvolumes}).

\begin{figure}
\centering
\begin{tabular}{l}
\includegraphics[width=0.5\textwidth]{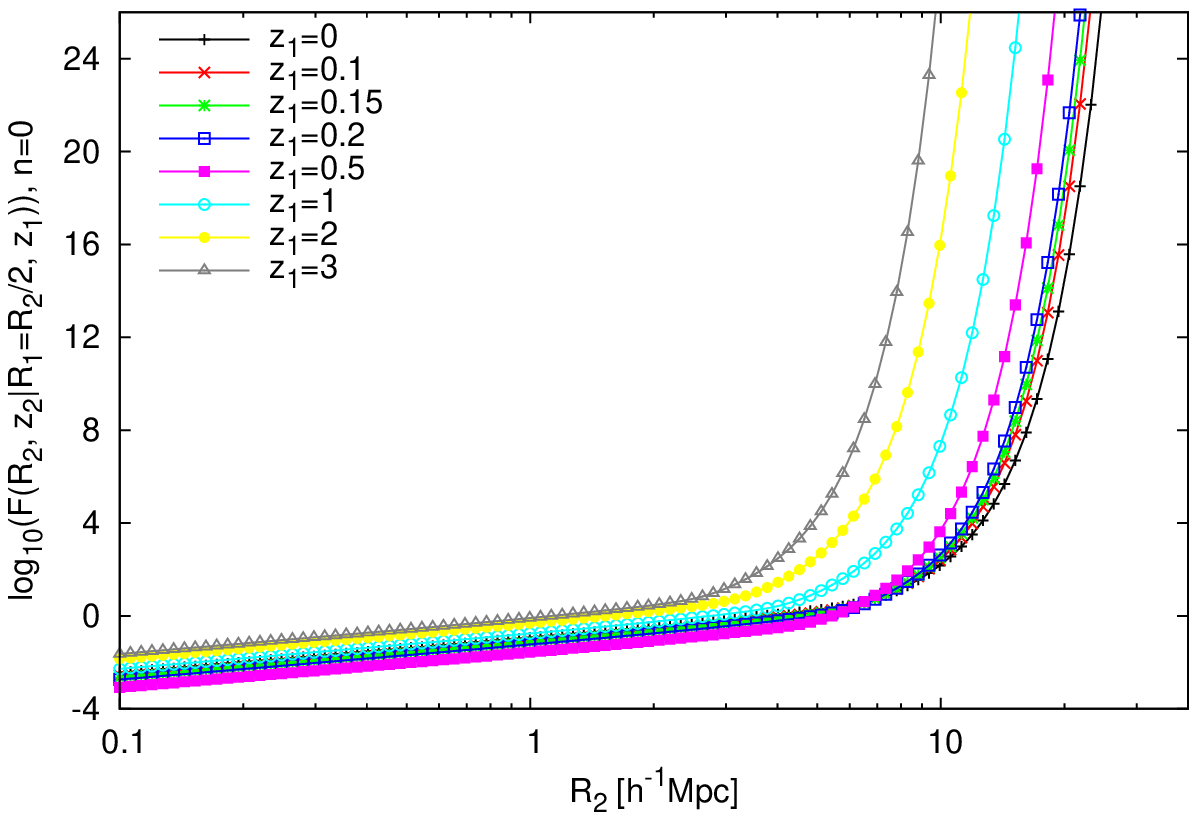}\\
\includegraphics[width=0.5\textwidth]{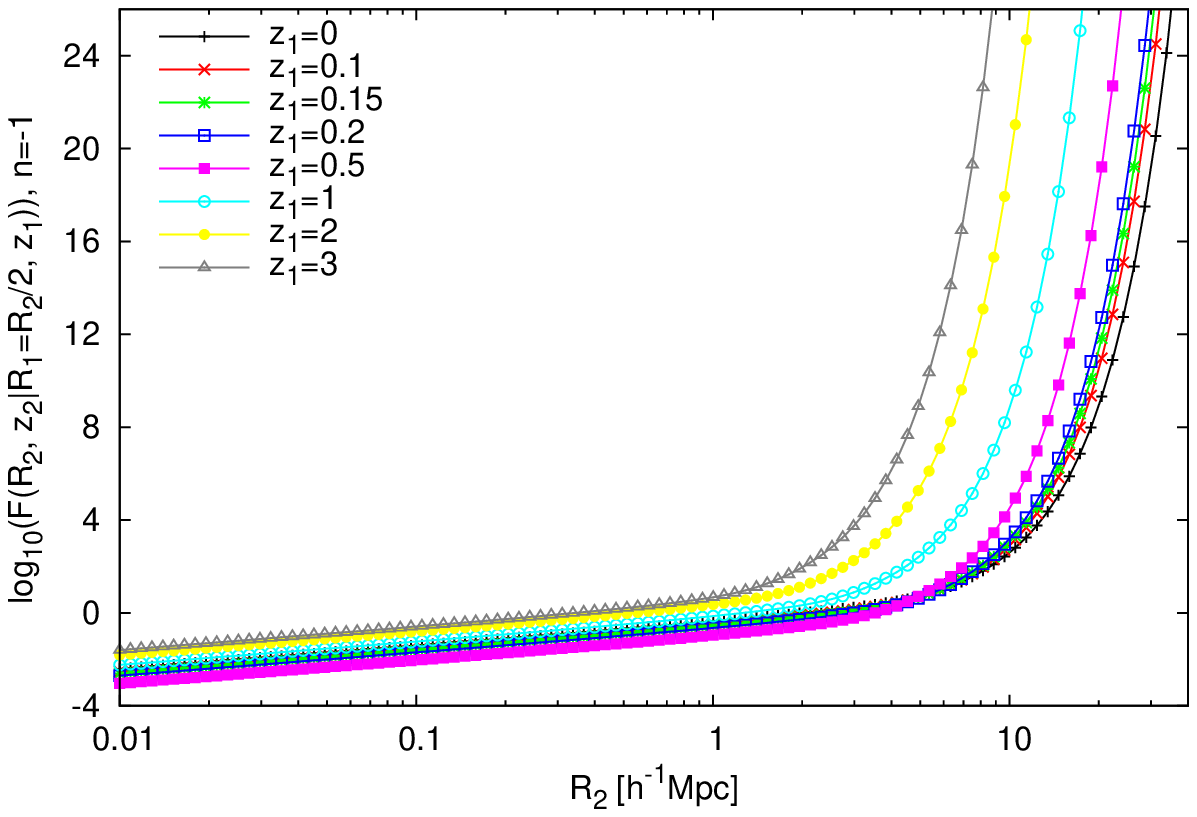}\\
\includegraphics[width=0.5\textwidth]{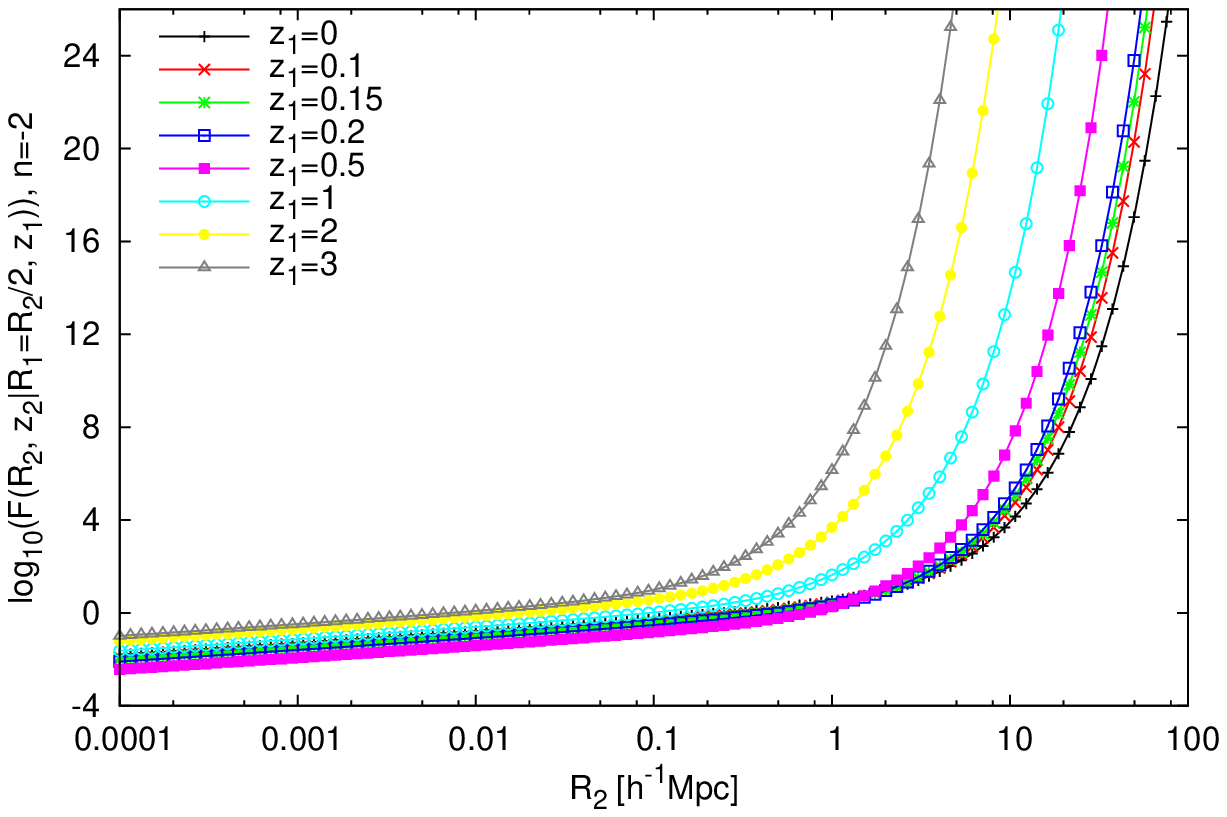}
%\hspace{-1mm}\includegraphics[scale=0.65]{Chapter3/timedistributionCDM.eps}
%\hspace{-15mm}\includegraphics[scale=0.65]{Chapter3/timedistributionn15yeni.eps}
\end{tabular}
\caption{The failure rate of voids with respect to the incorporated size $R_{2}$ (present size) at redshift interval $[z_{1}, z_{2}=0]$ for a void with size $R_{1}$ at a given redshift $z_{1}$ in terms of self similar models with index $n$ = $0$, $-1$,$-2$.}
\label{fig:yenitimedistribution}
\end{figure}

\begin{figure}
\centering
\begin{tabular}{l}
\hspace{-1mm}\includegraphics[width=0.5\textwidth]{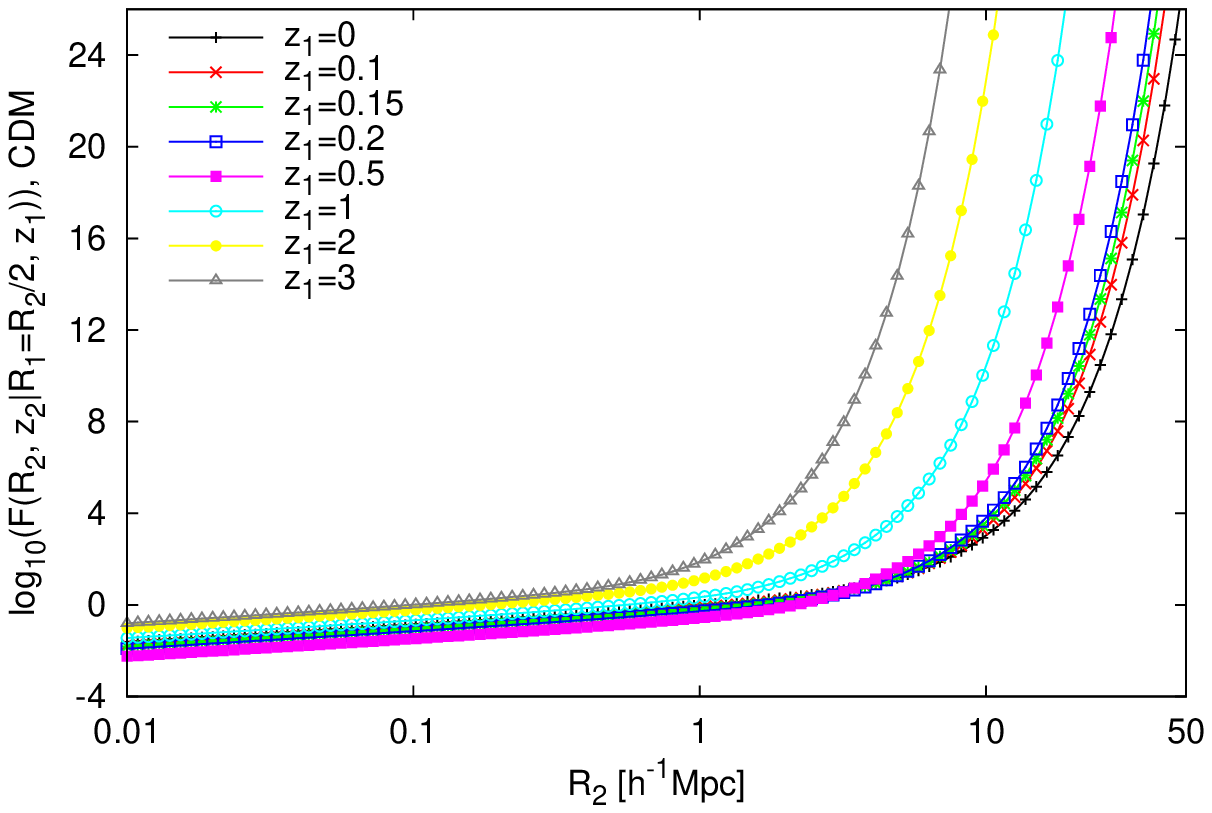}
%\hspace{-15mm}\includegraphics[scale=0.65]{timedistributionn15yeni.eps}
\end{tabular}
\caption{The failure rate of voids with respect to the incorporated size $R_{2}$ (present size) at redshift interval $[z_{1}, z_{2}=0]$ for a void with size $R_{1}$ at a given redshift $z_{1}$ in terms of the CDM model.}
\label{fig:yenitimedistributionCDM}
\end{figure}

Apart from the survival probability, LC93 obtain the survival probability time distribution. Based on their definition, we explain the void survival probability time distribution as the probability of a void with volume scale $S_{1}=S(V_{1})$ being incorporated into a system of volume larger than the corresponding scale $S_{2}=S(V_{2})$ in the time interval $d\tilde{\delta}_{\mathbf{v}_{2}}$. This distribution gives an insight into how long a growing void may survive or continue to merge. By adapting this to void populations, the void survival time distribution is given by the following expression,

\begin{eqnarray}
{F}_{|\delta_{\mathbf{v}_{2}}|}&=& -d|\delta_{\mathbf{v}_{2}}|\left(\frac{\partial P(S<S_{2},|\delta_{\mathbf{v}_{2}}||S_{1},|\delta_{\mathbf{v}_{1}}|)}{\partial|\delta_{\mathbf{v}_{2}}|}\right)\nonumber\\
&=&\sqrt{\frac{2}{\pi}}\left|1-2\frac{\delta_{\mathbf{v}_{2}}}{\delta_{\mathbf{v}_{1}}}\right|
\left[\sqrt{\frac{S_{2}}{S_{1}\left(S_{1}-S_{2}\right)}}e^{-\frac{\delta^{2}_{\mathbf{v}_{1}}}{2}\left(\frac{1}{S_{1}}-\frac{1}{S_{2}}\right)}
+ \sqrt{\frac{\pi}{2}}\frac{1}{|\delta_{\mathbf{v}_{1}}|}\left(\frac{\delta^2_{\mathbf{v}_{1}}}{S_{1}}-1\right)
\text{erf}\sqrt{\frac{|\delta^2_{\mathbf{v}_{1}}|}{2}\left(\frac{1}{S_{1}}-\frac{1}{S_{2}}\right)}\right].
\label{solprobssurvt}
\end{eqnarray}
\noindent
If we multiply the distribution (\ref{solprobssurvt}) with minus sign, it becomes the conditional failure rate or hazard function
of Statistical Mathematics. This function or rate measures the failure rate of void radii that could not merge at a given redshift, or measures
the failure rate of voids not merging/growing for a given size with respect to time interval,

\begin{eqnarray}
\rm{Failure~Rate}&=& d|\delta_{\mathbf{v}_{2}}|\left(\frac{\partial P(S<S_{2},|\delta_{\mathbf{v}_{2}}||S_{1},|\delta_{\mathbf{v}_{1}}|)}{\partial|\delta_{\mathbf{v}_{2}}|}\right).
\label{failurerate}
\end{eqnarray}
\noindent
Equation (\ref{failurerate}) is called the instantaneous rate of failure. In contrast to the survival probability which varies between zero and one, the failure rate can vary between zero and infinity. Over time, the failure rate can increase, decrease, remain constant, or even take more serpentine shapes \citep{Cleves}. The failure rate measures the rate at which risk of a void will not double its size is accumulated. Hence, the failure rate of merging voids, provides important information by showing at what time/redshift interval what size of voids fail to double their size due to merging or growing.
Figs. \ref{fig:yenitimedistribution} and \ref{fig:yenitimedistributionCDM} illustrate the failing rate, depending on the incorporated void size $R_{2}$, defined by equation (\ref{solprobssurvt}) in the self similar ($n=0, -1, -2$) and the CDM models. These figures tell us at what size voids fail to merge/grow at a given redshift interval $[z_{1}, z_{2}]$ where we choose $z_{2}=0$ as an example. As is seen, in each model, the failure rate shows the same behavior. According to this, in all models, the failure rate of not merging/growing for a void with radius $R_{1}=R_{2}/2$ dramatically increases until reaching an asymptotic value at a certain radius at a given redshift. The radius in which failure rate approaches to its asymptote is named as asymptote Radius $R_{asym}$. $R_{asym}$ decreases with increasing redshift and decreasing spectral index. In addition, the failure rate is very low with small size voids $<< R_{asym}$ while the risk increases with large size voids. Another feature we can see from Figs. \ref{fig:yenitimedistribution} and \ref{fig:yenitimedistributionCDM} is that, the failure rate for a void that growing or merging until the asymptote radius $R_{asym}$ will stop growing or merging after this radius. This feature indicates that if we have a void with radius $> R_{asym}$ at a certain redshift, this void will never fail to survive, merge or grow with $100\%$ confidence. As is seen from Figs. \ref{fig:yenitimedistribution} and \ref{fig:yenitimedistributionCDM}, $R_{asym}$ values at $z_{2}=0$, for the self similar models with index $n=0, -1, -2$, are $\sim 35, \sim 45 \sim 90 h^{-1}Mpc$ while the asymptote radius for the CDM model is $\sim 50 h^{-1}Mpc$.

\begin{figure}
\centering
\begin{tabular}{l}
\includegraphics[width=0.5\textwidth]{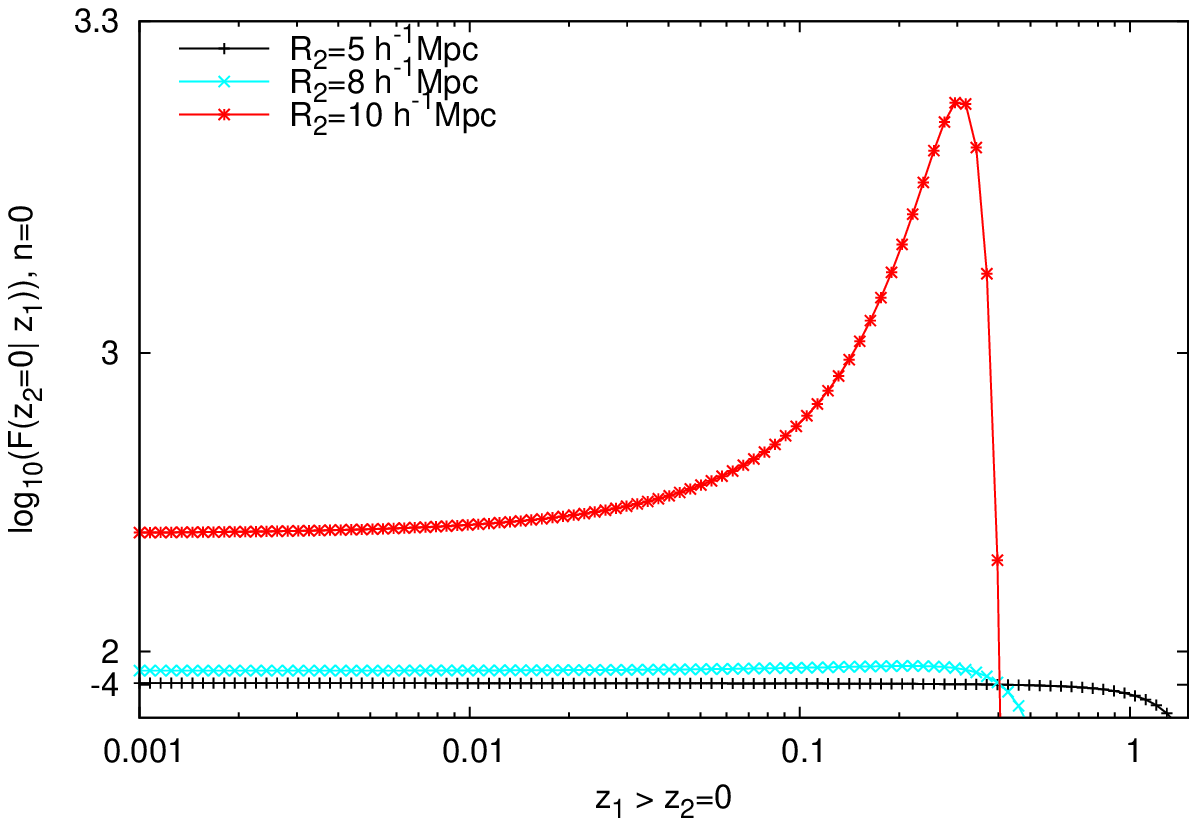}\\
\includegraphics[width=0.5\textwidth]{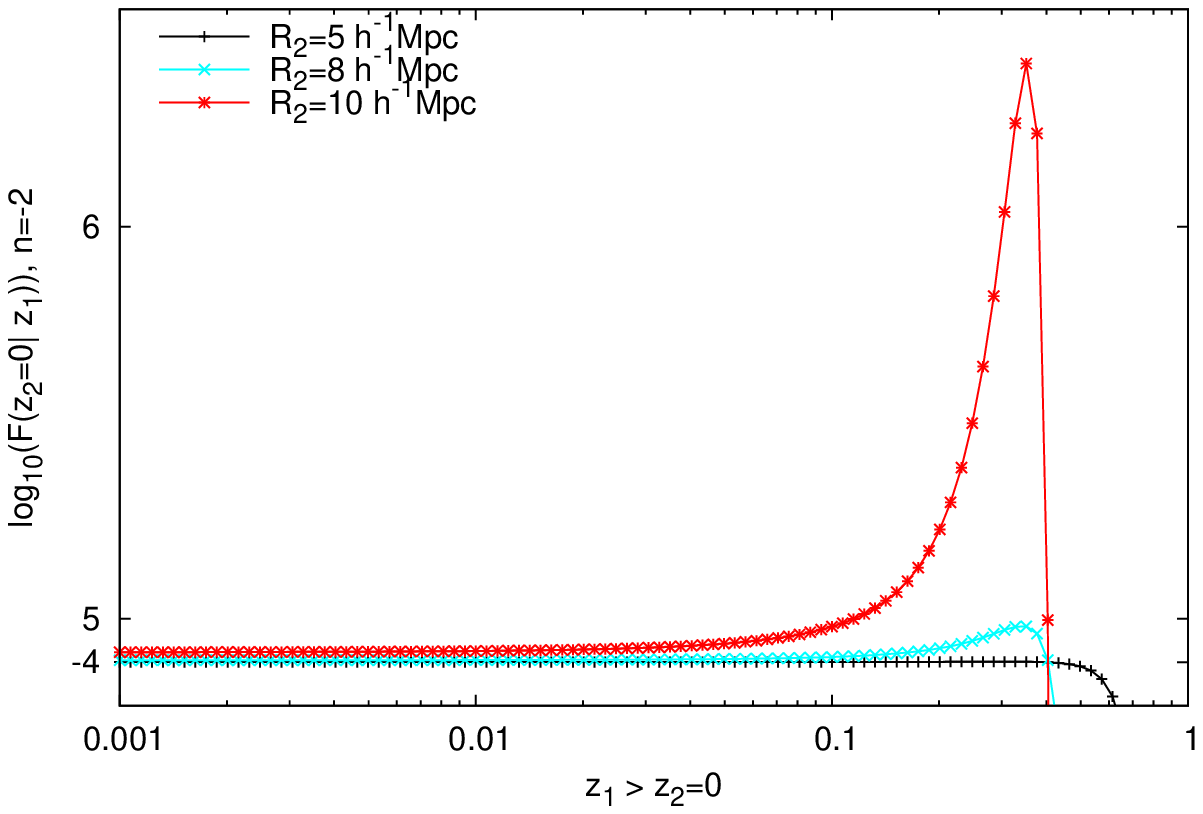}
\end{tabular}
\caption{The failure rate of voids for a given size $R_{1} = $ $10$, $8$, $5 h^{-1}Mpc$ in terms of redshift for self similar models $n=0, -2$.}
\label{fig:yenitimedistributionyeni}
\end{figure}

\begin{figure}
\centering
\includegraphics[scale=0.7]{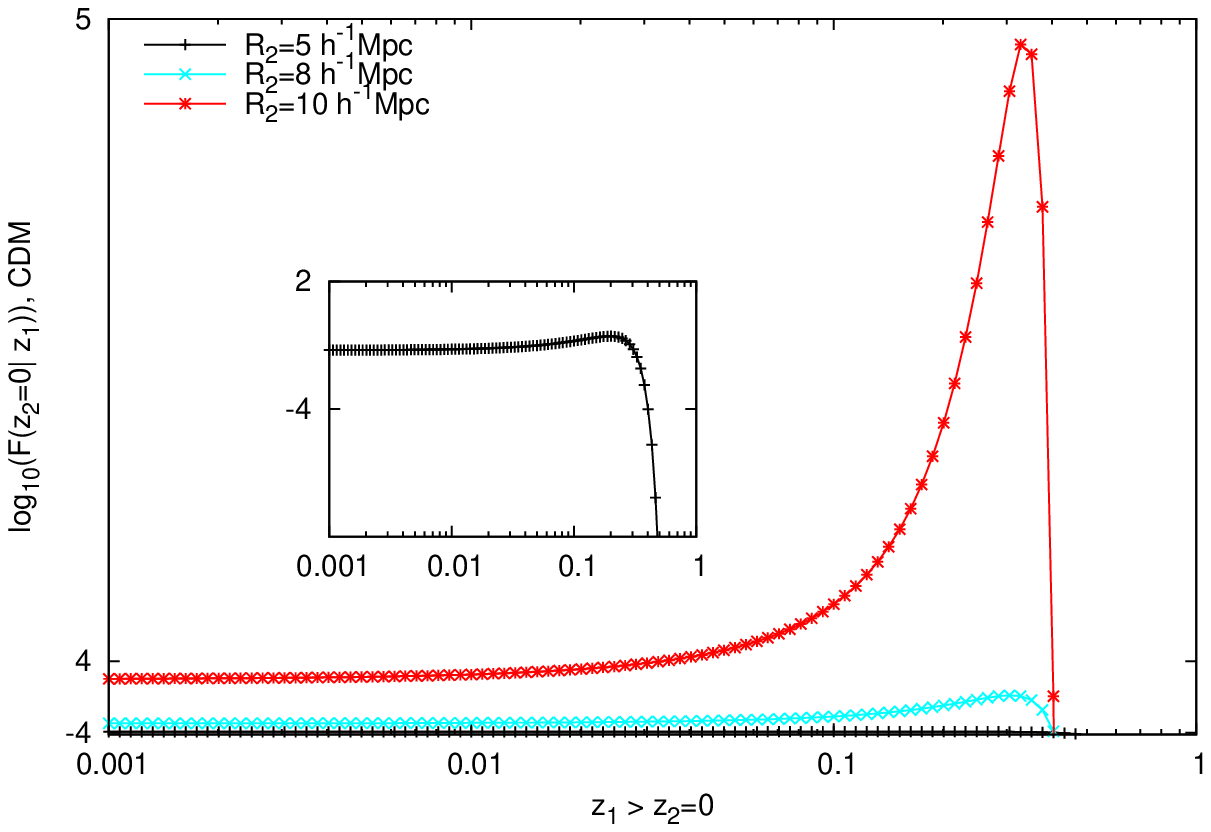}
\caption{The failure rate of voids for a given size $R_{1} =$ $10$, $8$, $5 h^{-1}Mpc$ for the CDM models in terms of redshift interval [$z_{1}$:$z_{2}=0$].}
\label{fig:yenitimedistributionyeniCDM}
\end{figure}

Apart from obtaining the size distribution of failing voids at a given redshift value, we can obtain the redshift distribution of failure rate for a void with a given radius. Fig. \ref{fig:yenitimedistributionyeni} and Fig. \ref{fig:yenitimedistributionyeniCDM} represent the failure rates of three different size voids $5 h^{-1}Mpc$, $10 h^{-1}Mpc$, $8 h^{-1}Mpc$ in terms of redshift for two self similar $n=0,-2$ and the $\Lambda$CDM models. In all models, the failure rate of voids with $10 h^{-1}Mpc$ size have a distinctive peak at a redshift $\sim z_{1}=0.3$. Moreover, in all models large size voids have higher failure rates than small ones, however have constant failure rates up until a critical redshift. At redshifts greater than this critical value, the failure decreases sharply. This indicates that at higher redshift values than the critical redshift value for a given size, voids have very high merging/growing rates. Voids within the same size range at redshifts lower than this critical redshift value, do show same level of failure rate.

\section{Volume Formation Time and Growing Void Merging Tree}\label{sec:volumeformation}
Before giving the details of the void counting method, it is useful to define a very important property of the void hierarchical build-up process; their formation history. By following and adapting previous studies \citep{lace,LemsonKauffmann,Bosch02,Gao2005,Wechsler2006} on halos to voids, the void formation history is characterized by a single parameter which is the formation time $\delta_{f}\sim z_{f}$. \emph{The formation time is the time when a void has acquired half of its final volume ${V_{2}}/2 = {V_{1}}$} (based on LC93). The formation time indicates when the main body/progenitor is accumulated.
Based on the LC93 algorithm, after the formation time, the choice of the largest volume progenitor as the main progenitor defines a continuous
track through the merging tree. It is obvious that formation times have key importance to construct a merging tree of voids as well as halos. Obtaining formation times from random walks is more problematic than obtaining the survival times. This is because the halo volume assigns more to a particle by tracking its density $\delta$, and is not its actual volume but is an approximate value (see LC93). However this fact does not lead to any
self inconsistency in merger rates and survival times. In addition to this, it has been shown that the Monte Carlo method and analytical counting
argument of generating merging histories provide similar results (LC93). We discuss these methods from the void perspective in the following.

\subsection{Void Counting, Analytical Method to Void Merging Tree:}\label{subsubsection:Void counting}
The void counting is based on defining the number density of voids in a given volume range which evolve into a larger range at later times. This number density allows us to obtain the probability distribution of a void with volume ${V_{2}}$ which had a parent in the volume range ${V_{2}}/2 < {V_{1}} < {V_{2}}$ at ${{\delta_{\mathbf{v}_{1}}}}$. This probability equals the probability that its formation time is earlier than ${{\delta_{\mathbf{v}_{1}}}}> {{\delta_{f}}}$. The counting method provides analytical solutions in terms of self similar models which can be extended into the CDM model numerically. After giving the general idea of this method, the details can be given as follows.

The number density of voids $\left(V_{1}, V_{1} + dV_{1}\right)$ at time ${{\delta_{\mathbf{v}_{1}}}}$, which is incorporated into voids of volume $\left(V_{2},
V_{2}+dV_{2}\right)$ at time ${{\delta_{\mathbf{v}_{2}}}}>{{\delta_{\mathbf{v}_{1}}}}$ is,

\begin{eqnarray}
{d^2}n=\frac{d n}{d V_{1}}(V_{1},|{{\delta_{\mathbf{v}_{1}}}}|)d V_{1}
f_{S_{2}}\left(S_{2},|{\delta_{\mathbf{v}_{2}}}||S_{1},|\delta_{\mathbf{v}_{1}}|\right){d S}_{2}.
\label{probhalof}
\end{eqnarray}
\noindent
So long as $V_{2}\geq V_{1} > V_{2}/2$ each trajectory must connect unique voids because there cannot be two paths each of which contain more than half of
the final volume. However, it is possible that a volume of a void, $V_{2}$ at ${{\delta_{\mathbf{v}_{2}}}}$ has no progenitor of mass $< V_{2}/2$ at time ${{\delta_{\mathbf{v}_{1}}}}$. The probability that a void with volume $V_{2}$ at ${{\delta_{\mathbf{v}_{2}}}}$ has a progenitor in the volume range $V_{2}/2 < V_{1} < V_{2}$ at time ${{\delta_{\mathbf{v}_{1}}}}$ is then given by the ratio of voids that evolve into another volume $V_{2}$ relative to the total number of voids in volume $V_{1}$,

\begin{eqnarray}
\frac{d P \left({V_{1}},{\delta_{\mathbf{v}_{1}}}|{V_{2}},|{\delta_{\mathbf{v}_{2}}}|\right)}{d{V_{1}}}=\left(\frac{d n({V_{1}})/ d {V_{1}}}
{d n({V_{2}})/ d {V_{2}}}\right)
{f_{S_{1}}}\left({S_{1}},|{\delta_{\mathbf{v}_{1}}}||{S_{2}},|{{\delta_{\mathbf{v}_{2}}}}|\right)\left|\frac{d{S_{1}}}{d V_{2}}\right|,
\label{probhalovoida}
\end{eqnarray}
\noindent
which leads to,

\begin{eqnarray}
\frac{d P\left({V_{1}},{\delta_{\mathbf{v}_{1}}}|{V_{2}},|{\delta_{\mathbf{v}_{2}}}|\right)}{d{V_{1}}}d{V_{1}}=\left(\frac{{V_{2}}}
{{V_{1}}}\right) {f_{S_{1}}}\left({S_{1}},|{\delta_{\mathbf{v}_{1}}}||{S_{2}},|{{\delta_{\mathbf{v}_{2}}}}|\right)d{S_{1}}.
\label{probhalovoidb}
\end{eqnarray}
\noindent
Integration of equation (\ref{probhalovoidb}) over the volume range ${V_{2}}/2 < {V_{1}} < {V_{2}}$ gives the probability distribution of void ${V_{2}}$ having a parent in this volume range at time ${{\delta_{\mathbf{v}_{1}}}}$. This equals the probability that its formation time is earlier than this,

\begin{eqnarray}
P\left(\delta_{f} < |{{\delta_{\mathbf{v}_{1}}}}| |V_{2},|{{\delta_{\mathbf{v}_{2}}}}|\right)&=&P\left(V_{1}< V_{2}/2 |{{\delta_{\mathbf{v}_{1}}}}||V_{2}, |{{\delta_{\mathbf{v}_{2}}}}|\right)\nonumber\\
&=& \int^{S_{h}=S_{2}({V_{2}}/2)}_{S_{2}}\left(\frac{V_{2}}{V_{1}}\right)f_{S_{1}}\left(S_{1},|\delta_{\mathbf{v}_{1}}||S_{2},|\delta_{\mathbf{v}_{2}}|\right)d S_{1},
\label{probhalovoidc}
\end{eqnarray}
\noindent
where ${V_{2}}/{V_{1}}$ is the weighting factor and $S_{h}=S({V_{2}}/2)$. Interestingly, the integral (\ref{probhalovoidb}) is the expectation. In terms of self similar models we derive the exact solutions of the probability function for $n= 1, 0, -1.5, -2$ (see~Appendix \ref{appendix:exactsolutiononmergertree}). However the probability in terms of self similar models with index $n=-1$ does only have a numerical solution.

The probability distribution of formation times can be given by the following expression (based on LC93),

\begin{eqnarray}
P\left( > \delta_{f}\right)&=&\int^{1}_{0}\frac{1}{2\pi}\left[\tilde{S}\left(2^{\alpha}-1\right)+1\right]^{1/\alpha}\frac{\delta_{f}}{\tilde{S}^{3/2}}
\exp\left(-\frac{1}{2}\frac{\delta^2_{f}}{\tilde{S}}\right) d {\tilde{S}},\nonumber\\
\text{where}\phantom{a}\alpha &=&\frac{n+3}{3},
\label{reducedprobability}
\end{eqnarray}
\noindent
and where the parameters are defined as,
\begin{eqnarray}
{\tilde{S}}\equiv\frac{S-S_{2}}{S_{h}-S},\phantom{a} {\delta_{f}}\equiv\frac{\delta-{\delta_{\mathbf{v}_{2}}}}{\sqrt{S_{h}-S_{2}}}.
\label{newparameters}
\end{eqnarray}
\noindent
It is crucial to mention that \cite{lace} suggest that since the scaled volume $\tilde S$ varies slowly between $1$ and $0$ ($0\leq\tilde{S}\leq 1$) in
equation (\ref{reducedprobability}), this probability equation can be extended to CDM models due to the slowly varying CDM power spectra.

\begin{figure}
\centering
\begin{tabular}{ll}
\includegraphics[width=0.4\textwidth]{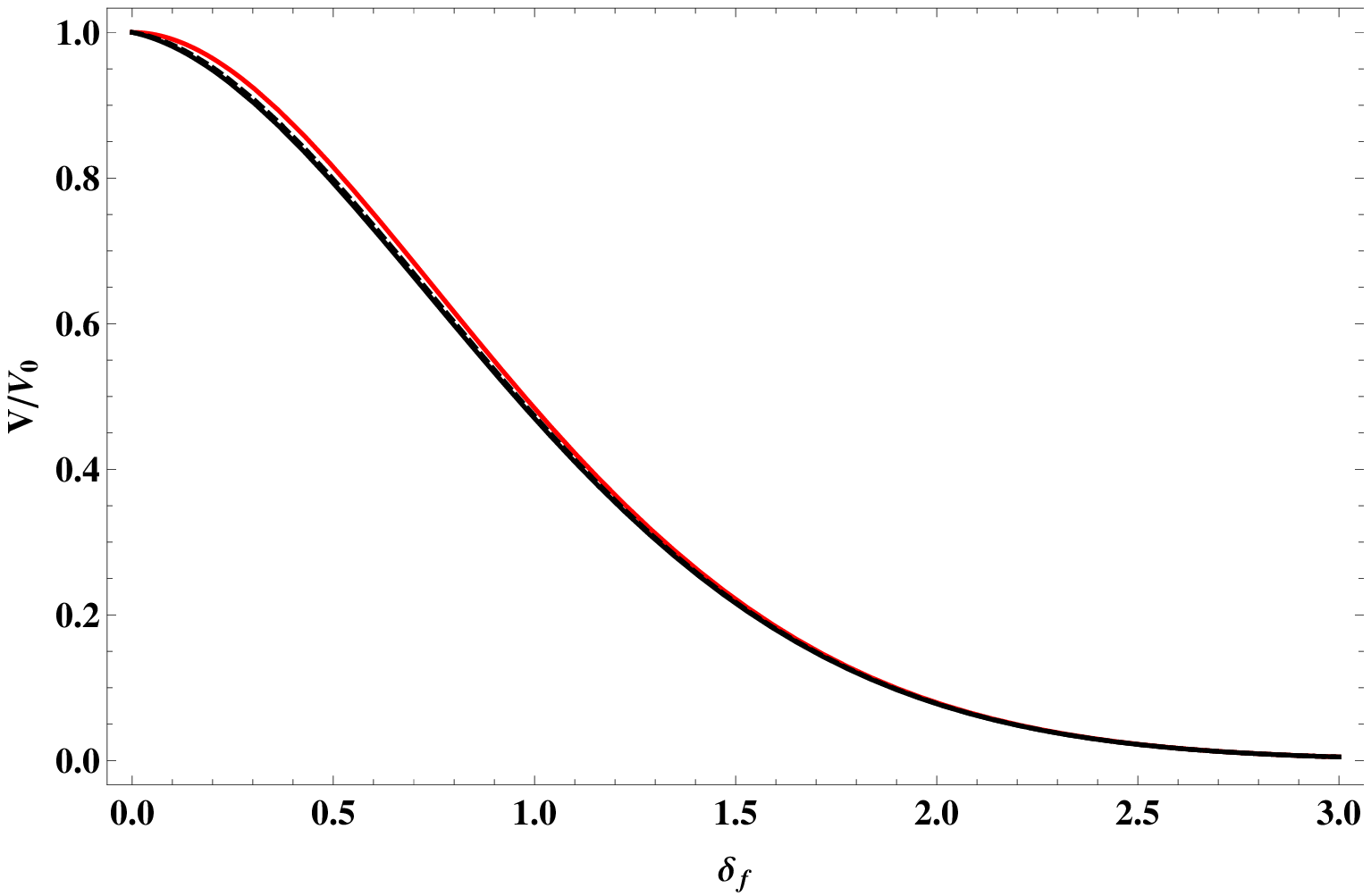}
\includegraphics[width=0.4\textwidth]{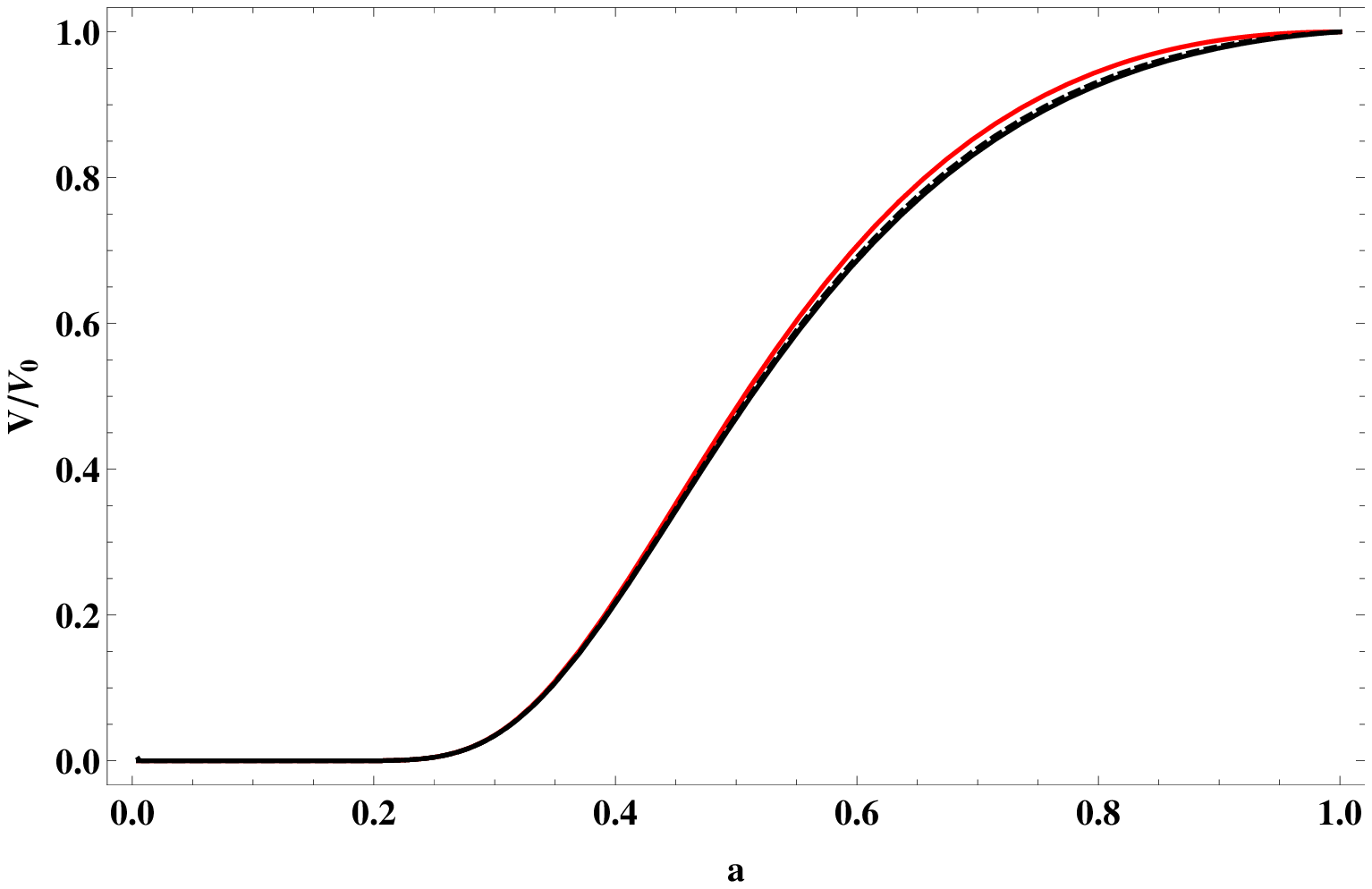}
\end{tabular}
\caption{Analytical merging history of voids based on analytical solutions with index $n =$ $0$, $-1.5$, $-2$, represented by red, dotted and black lines, respectively. The left panel shows the distribution as a function of threshold height $\delta_{f}\sim z_{f}$ and the right panel shows the same probability but in terms of scale factor. The difference between models is very small that it can be negligible since the scale parameter $\tilde{S}$ varies slowly between $1$ and $0$.}
\label{fig:analyticalProbmerger}
\end{figure}

Here, the analytical solutions are derived from the distribution equation (\ref{reducedprobability}) (see~Appendix \ref{appendix:exactsolutiononmergertree}) for the self similar
models, although LC93 point out that there is only one analytical solution to the probability distribution of the formation times for $n=0$.
Differing from the solutions derived from equation (\ref{probhalovoidc}), the analytical results of equation (\ref{reducedprobability}) are only dependent on the barrier height/time $\delta_{f}\approx z_{f}$. In this sense we can say that the probability distribution of voids can be represented in terms of redshift or both redshift and scale parameters. Analytical solutions of void volume distributions (see Appendix \ref{appendix:exactsolutiononmergertree}) provide
the merger histories of large voids (Fig. ~\ref{fig:analyticalProbmerger}). Note that they are the exact solutions based on the rough approximations of
LC93. According to this, when formation barrier height/redshift increases the probability distribution of void ${V_{2}}$ which had a parent in the volume
range ${V_{2}}/2 < {V_{ 1}} < {V_{2}}$ at ${{\delta_{\mathbf{v}_{1}}}}$ decreases. From the probability of the most obvious feature of formation times, this shows
that small voids form early and large voids form relatively late in accordance with our general expectation in hierarchical models of structure formation.
However, at least they can provide an idea of the behavior of merging history.

\begin{figure}
\centering
\begin{tabular}{ll}
\includegraphics[width=0.4\textwidth]{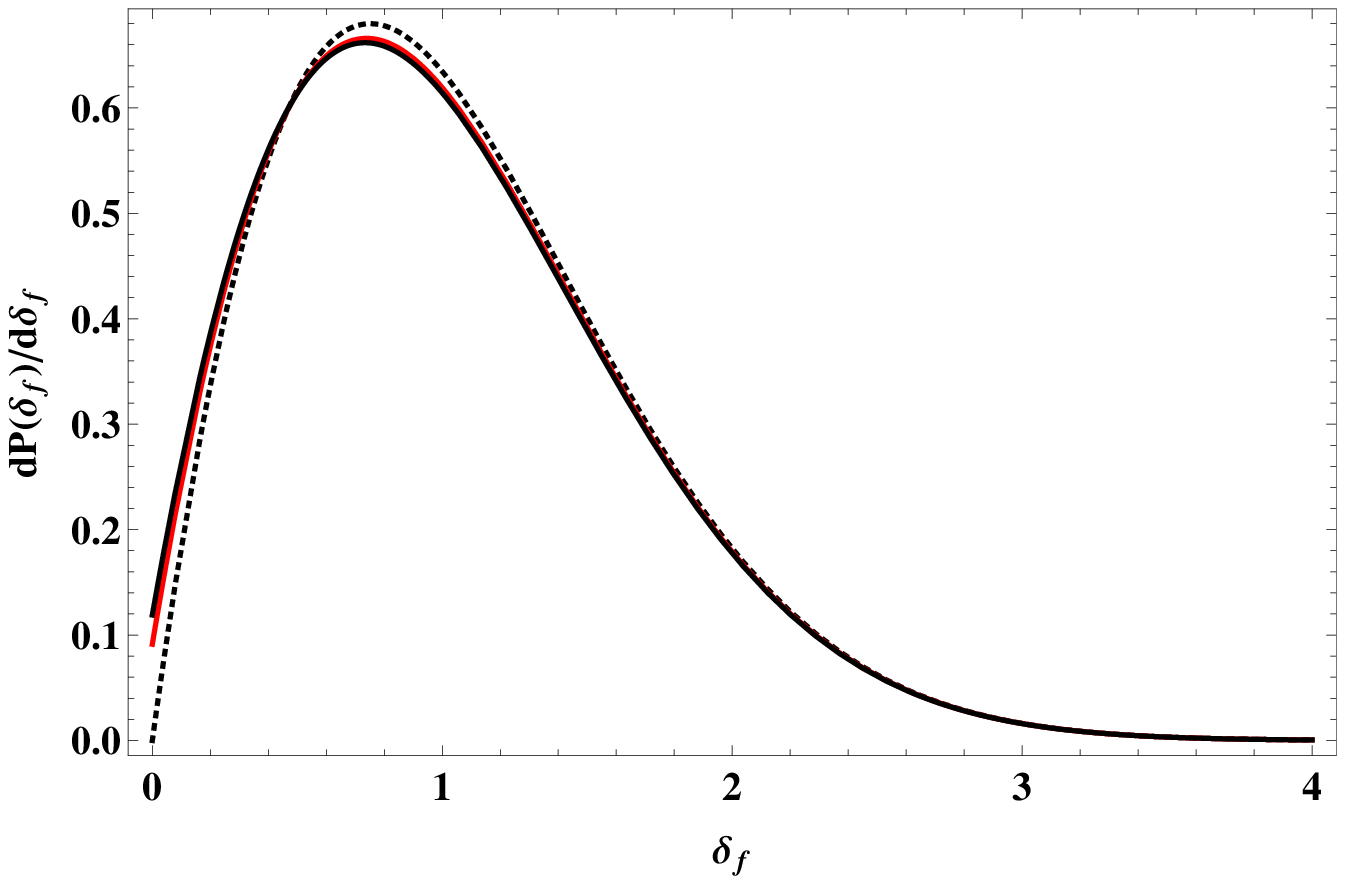}
\includegraphics[width=0.4\textwidth]{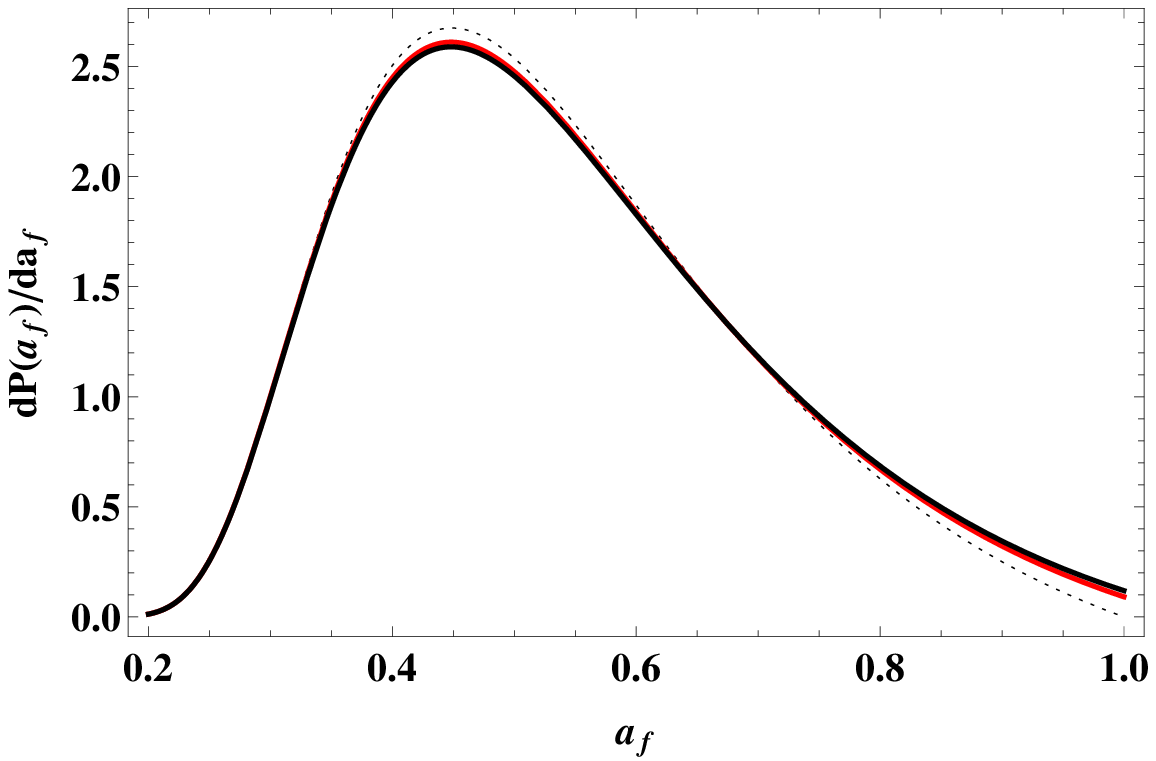}
\end{tabular}
\caption{The probability density distributions of void merging histories in the self similar models $n =$ $0$, $-1.5$, $-2$ (dotted, red and black lines, respectively) based on analytical solutions. The left panel shows the distribution as a function of threshold height $\delta_{f}$, while the right panel shows the probability distribution of formation scale factors $a_{f}$.}
\label{fig:analyticalP}
\end{figure}

Formation time probability densities computed from analytical solutions (\ref{probanalyticsol}) are shown in Fig. \ref{fig:analyticalP} for different self similar models. Fig. \ref{fig:analyticalP} indicates the probability densities in terms of barrier height and scale factor. These plots depict the fact that models have similar distributions and their differences are so small that they can be negligible. This is because of a slowly varying $\tilde{S}$. Here, care should be taken with respect to probability theory terminology. These are the probability densities, not the actual probability functions of the merger history. The cumulative probability functions can be defined as the integral of its density function in a given interval; that is why the probability density functions can have values greater than unity.

\subsection{Monte Carlo Method to Void Merging Tree:}
In the preceding section, we applied the probability of first upcrossing at a second barrier given a particular starting point to derive relations about various aspects of void survival and formation in terms of time. So far, these quantities have been averages or probability distributions. Based on the LC93 halo algorithm, the key element of the merging tree of voids is the conditional probability function which is a transition between barriers,

\begin{eqnarray}
f_{S_{1}}(S_{1},|\delta_{\mathbf{v}_{1}}|\big|S_{2},|\delta_{\mathbf{v}_{2}}|)
d{S_{1}}=
\frac{1}{\sqrt{{2\pi}}}\frac{\Delta\delta}{\left(\Delta S\right)^{3/2}}
\exp\left[\frac{-\left(\Delta\delta\right)^{2}}{2\left(\Delta S\right)}\right]dS_{1},\nonumber
\end{eqnarray}
\noindent
where the void barriers are given as $|\delta_{\mathbf{v}_{1}}| > |\delta_{\mathbf{v}_{2}}|$. By using this transition function one can draw specific volume scale increments $\Delta S$ probabilities repeatedly for a number of time intervals. As a result of this, for each trajectory we can obtain trajectories of volume versus cosmic expansion factor, $z_{f}$ or the formation time $\delta_{f}$. This enables one to follow the fragmentation of voids on an object by object basis. This is the logic behind the algorithm for the generation of Monte Carlo merger trees described by \cite{Cka88,Cole1991,lace} based on halos.

The trajectories have structure on arbitrarily small scales corresponding to mergers with small size voids. This is a restatement of the divergence of the mean number of transitions with small volume changes. Consequently, it is necessary to examine trajectories with a particular resolution in order tosmooth over the mergers with many small size voids. In practice, each application to the prediction of a specific physical quantity has a minimum volume scale of interest $V_{min}$, so it is natural to set the resolution with which trajectories are computed according to this minimum scale.

The trajectories $\Delta S$ have structure on arbitrarily small scales corresponding to mergers with very small volume. This is a restatement of the divergence of the mean number of transitions with small volume changes shown in Fig. \ref{fig:mergerrates} and \ref{fig:lossrates}.
Consequently, it is necessary to examine trajectories with a particular resolution in order to smooth the mergers with numerous small sizes. In practice, each application to the prediction of a specific physical quantity has a minimum volume scale of interest $V_{min}\approx M_{min}$, so it is natural to set the resolution with which trajectories are computed according to this minimum scale. They consider equation (\ref{probabilitys1}) in the limit $\delta_{f} /\sqrt{\Delta S}\ll 1$ which leads to a transition probability,

\begin{eqnarray}
f=\frac{\delta_{f}}{\sqrt{2\pi} {\Delta S}^{3/2}},
\end{eqnarray}
\noindent
where $\delta_{f}$ is equal to $\Delta \delta$. This transition probability is directly proportional to the time interval $\Delta \delta$. LC93 interpret this as indicative of a probability of a single merger event. This is the reason $\delta_{f} /\sqrt{\Delta S}\ll 1$ is used to obtain binary regime and this limit leads to define a particular choice of step size $\delta_{f}$ which is given by,

\begin{eqnarray}
{\delta_{f}}\leq \sqrt{\left|\frac{dS(V)}{d V}\right| V_{min}}.
\label{timesteps}
\end{eqnarray}
\noindent
The LC93 algorithm for generating merger histories is as follows. First, determine the appropriate time step using equation (\ref{timesteps}). Second, select a transition $\Delta{S}$ from the probability distribution of (\ref{probabilitys1}) and invert the $S(V)$ relation to obtain both the change in volume $\Delta V$ and the new main progenitor volume $V_{new} = V -\Delta V$. One then repeats this procedure at the new values of ${\delta_{f}}$ and $S(V_{new})$ to obtain the next fragmentation of the main progenitor. This process continues until the remaining volume is less than $V_{min}$.

The trajectory for the main progenitor, defined as the largest progenitor at each timestep according to this algorithm, may form the trunk of a merger
tree. For each volume $\Delta V$ above the threshold $V_{min}$, one can generate an independent history for the infalling branch on the tree using the same algorithm. The process of constructing a volume accretion history or merger tree in this way must be repeated numerous times in order to sample the variety of ways in which a void of a fixed volume at a fixed time might build up its volume. This is a Monte Carlo method for exploring the various volume accretion histories. It is conventional to refer to each individual tree generated in this way as a particular realization in an ensemble of merger histories.

\begin{figure}
\centering
\includegraphics[width=0.4\textwidth]{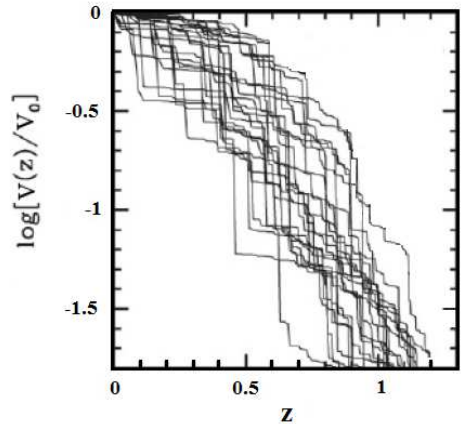}
\caption{Seventeen examples of volume merging histories as a function of redshift $\sim \delta$ based on the EPS formalism by adopting LC93 halo merging algorithm in the EdS Universe. The solid lines correspond to voids with radii $10 h^{-1}Mpc$ at $z=0$ \citep[adaptation of][]{Bosch02}.}
\label{fig:mergerhistory}
\end{figure}

The algorithm for generating Monte Carlo merger trees described above is convenient because of its simplicity. Unfortunately, while this algorithm
conserves volume by construction, it overpredicts the number of progenitor voids with volume $V > V_{min}$ at previous timesteps relative to the analytic distribution of equation (\ref{probhalof}). This point has been emphasized by \cite{somervillekolatt} for halos. In other words, the Monte Carlo procedure does not lead to a mean population of voids at high redshift that is consistent with the excursion set relations of the previous sections. In Fig. (\ref{fig:mergerhistory}) different merger trees are represented for self similar model $n=-1.5$ for $10 h^{-1}Mpc$ size voids. Here, we can see the problematic part of the EPS formalism that at higher redshifts/earlier times ($t/t_{0}$) merging activity is drastically low which we do not expect from the hierarchical evolution scenario. However the analytic halo counting method provides an insight into the hierarchical build up of a growing void population. Fig. \ref{fig:analyticalProbmerger} shows the approximate analytical merging history of growing voids in terms of formation time/redshift for self similar models. As is seen, growing void merging events start at high redshifts/$\delta_{f}$ and merging events increase in time as an expected procedure from a merging tree algorithm. However, the Monte Carlo method of the LC93 merging algorithm does not show any sign of merging events at high redshifts. To treat this problem for obtaining a physically reasonable void merging algorithm, the different Monte Carlo methods should be considered and checked for consistency in the context of the excursion set void hierarchy. Alternatively, taking into account the full two-barrier void distribution function for merging and collapsing void populations as well as a more convenient Monte Carlo method for merging voids can be the solution.

\section{Summary and Discussion}
In this study, we formulate a void merging tree model of spherical growing voids based on one-barrier excursion set theory by following up on \cite{sw}. Here, we limit this study to the void merging process called the `void in void problem' in excursion set theory. The interpretation of this process is in analogy with
the merging of overdense regions/halos. To construct a merging tree model of growing voids, we have used the two-barrier void distribution function in the context of the two-barrier excursion set. This distribution function is then reduced to the one-barrier excursion set by showing the minor void contribution to the distribution function is negligible in the case of growing voids, by following \cite{sw}. By taking into account this reduced void distribution, a merging tree algorithm is constructed based on a halo merging tree algorithm derived by \cite{lace}. On the basis of the algorithm, we define and obtain merging and absorption rates of void hierarchical built-up. In addition to this, the cumulative void size distribution as well as the void survival probability and failure rate times are formulated analytically. Analytical approximations obtained in these merging tree formalisms have key importance to understand the evolution of voids in different Universe models. The void merging tree algorithm in this study is based on an approximate analytical formalism, in which time elements are given by the linear density in the EdS Universe. Here, we give our main results which are compared to previous results:

\begin{enumerate}
%The redshift evolution of the void population is shown
%in Figure 6. As expected from theoretical arguments(Sheth & van de Weygaert 2004), we find that voids expand
%as time proceeds. As shown in the top-left panel of Figure
%6, from high- to low-redshift the abundance of large
%voids increases, while at the same time, smaller voids become
%less numerous. Also the shape parameters display a
%significant evolution with redshift, as shown in the top-right

\item We obtain the conditional void size distribution in terms of self similar models ($n=0, -1, -2$) and the CDM model at a given redshift. In the self similar models, the exponential cutoff in the size of the void distribution moves to very large sizes with decreasing index and decreasing redshift values. In contrast to this tendency, in all models the small size void distribution increases from lower to higher redshifts. We see these tendencies of decreasing small and increasing large void distributions towards lower redshifts in the CDM model as well. This can be an indication of a void hierarchy that is actually in agreement with the theoretical work of \cite{sw} and in the numerical study of \cite{2013MNRAS.434.1192R}. In particular, \cite{sw} infer that the small size voids present at high redshifts must merge with each other to make larger voids that are present at lower redshifts. Related with this, the statistical study of \cite{conroy2005} point out that voids are smaller at $z\approx 1$ compared to the present day voids. Agreeing with \cite{conroy2005}, as we mentioned above the void size distribution at high redshifts is small which indicates small size voids. In the same study, \cite{conroy2005} also point out that small size voids are rare at high redshifts. Opposite to this statement of \cite{conroy2005}, we show that smaller voids dominate higher redshifts compared to lower redshifts in the self similar models, as well as the CDM model.

\item We define and provide analytical descriptions of survival growing void probability in terms of self similar and approximated CDM models. The survival probability provides information on what size voids manage to double their size due to merging/growing at a given redshift. Following this, we show that the probability of surviving void size range, increases towards low redshifts in the self similar and the CDM models. On the other hand, the survival probability of small size voids is higher than their larger counterparts at a given redshift, again in all models. This result is in agreement with the results from studies by \cite{sw,Shang2007,Viel2008,2013MNRAS.428.3409A}.
    We also find that when the spectral index is decreasing, the size of surviving voids at high redshifts becomes smaller and also void survival probabilities decrease. This indicates that detecting very large voids at high redshifts is unlikely. However we should not forget that this result clearly depends on the definition of a void, and the sparsity of the model that we use in our calculations, which is based on LC93 for dark matter halos, in which the binary system is used. As we know, large density depressions may exist even in a Gaussian field with small fluctuations, depending on the definition of density depression.

\item Another parameter that we derive and define in this study is the failure rate of the growing void population in terms of the self similar and the CDM models. This rate is defined as accumulated risk of a void that will not incorporate into another void during a given redshift interval. Using the analytical failure rate that we derive, we show that in all models, the failure rate of a void not doubling size increases until an asymptotic radius at a given redshift. This asymptotic radius can be interpreted as a special void size void will definitely fail to merge/grow. Therefore, we name this asymptotic size as the asymptotic radius $R_{asym}$. In self similar models, $R_{asym}$ decreases with increasing redshift and decreasing spectral index. For example, $R_{asym}$ at $z=3$ has a smaller size compared to the one at redshift $z=0$. Generally speaking, the failure rate of small size voids before reaching their $R_{asym}$ tends to be lower than the failure rate of relatively large size voids in self similar and CDM models. Another common feature in all models that the failure rates are infinite at $R_{asym}$. This feature shows that the voids that reach a size greater than $R_{asym}$ in a given model may merge continuously. This result allows us to obtain $R_{asym}$ values at different redshifts. In this study we provide asymptote radii at present day $(z=0)$ for the self similar models with index $n=0, -1, -2$, which are $\sim 35, \sim 45 \sim 90 h^{-1}Mpc$ respectively while the asymptote radius for the CDM model is $\sim 50 h^{-1}Mpc$.

    As well as the size distribution of voids that cannot merge/grow at a given redshift, we obtain some results on the redshift distribution of the failure rate of voids with a given radius for self similar $n=0,-2$ and the $\Lambda$CDM models. Interestingly, the failure rate of voids with $10 h^{-1}Mpc$ radius show a distinctive peak at a redshift $\sim z=0.3$. This indicates that at redshift $z=0.3$, voids with radii $10 h^{-1}Mpc$ will show very low merging activities. In addition, the void failure time distribution indicates that especially at $z=0.3$, voids do not show high merging/growing behavior. This result is in particular agreement with \cite{hoylevogley2004} and \cite{conroy2005}. They point out that there is no strong void evolution between $z=0.1$ and $z=0.3$ for voids of size approximately $R=10 {h^{-1}Mpc}$.

\item Finally, the approximate analytical void formation probabilities (see Appendix \ref{probanalyticsol}) and the merging history of growing voids (see Appendix \ref{probanalyticsol2}) for self similar models are obtained. The formation probabilities of the self similar models provide useful predictions for what to expect in the Monte Carlo method of voids. As we mentioned before, void progenitors are overpredicted due to the simplicity of the LC93 Monte Carlo merging tree algorithm. Apart from this, it is important to mention that LC93 suggest that analytical solutions for the expected void distribution for self similar models can be extended to CDM models due to the slowly varying dark matter power spectrum. However, this assumption is not based on a physical understanding of the merging tree problem. The reason is the mathematical tractability rather than a physical understanding. On the other hand, we cannot deny the importance of the LC93 algorithm due to its simplicity and efficiency as the first halo merging algorithm.
\end{enumerate}
The main challenge is to compare our model with observations and numerical simulations to assess the agreement between these and the void merging algorithm. Our model parameters could then be refined to fit the observations more accurately. Which quantity that we obtain here is observationally detectable/measurable? The void merging algorithm provides us some approximate analytical formulae of growing spherical voids as a function of redshift and volume/size. These parameters, as we mentioned in the results, are the conditional probabilities of merging voids based on the one-barrier excursion set, merging and absorption rates, survival probability and failure rates. We know that LC93 show that the mass fraction function of the one-barrier excursion set and the conditional distribution are in agreement with simulations for the limited range halo. Following this idea, by taking into account that growing void algorithm formulae are analogues to halo merging of LC93, one may compare the volume fraction function and its conditional distribution with N-body simulations. As a result we can check whether these quantities are in agreement with the numerical studies or not. In addition, we can obtain the void number density for a given redshift and we can compare this quantity with N-body simulations.

A possible comparison of quantities of the growing void algorithm and N-body simulations/observations comes from the analytical form of survival probability and failure rate depending on cosmological parameters and time via underdensities/barriers ($\delta_{\mathbf{v}_{1}}$ and $\delta_{\mathbf{v}_{2}}$), providing a framework to calculate the percentage of voids with certain radii at given redshifts for a given model and density profile. It is possible to obtain what size voids will show high and low merging events/activities at a given redshift. On the other hand, even advanced large scale surveys cannot reach high redshift ranges. Another difficulty from the observational point of view is that detecting voids smaller that $10$ $h^{-1}$ $Mpc$ in size is not easy due the signal to noise problem \citep{hoylevogley2004}. Of course, this fact changes depending on the definition of a void and the void finding algorithm. A possible solution to test the failure rate and survival probability for an observationally given redshift is to obtain the void distribution up to void sizes which are limited observationally. Following this, one can obtain a limited part of the observational survival distribution due to the limited size range by counting voids that are observed. Therefore it is possible to compare this observational distribution with the theoretical survival distribution for the same redshift value for the limited void sizes. As is seen it can be checked that there is an agreement or not between the theoretical distribution and the observational one for a limited size of voids at a given redshift.

We must note that the growing void merging model takes into account voids that just reaches shellcrossing. \cite{blumenthal} discuss that the voids that are observed in the galaxy distribution are identified with primordial underdensities that have only just reached shellcrossing. Shellcrossing happens when a density depression in which a void is born reaches a linearly extrapolated underdensity $\delta_{\mathbf{v}}=-2.81$ in the EdS Universe ($\Omega_{0}=1$). However, there is no void finder that explicitly tests for shellcrossing. Almost all voidfinders simply find density depressions, and when the sampling is dialed down, these can be nearly arbitrarily large. For example, \cite{2008ApJ...683L..99G} use a  parameterfree algorithm called ZOBOV \citep{marc2008} in order to find supervoids (density depressions) in the galaxy sample. In our calculations and figures, we take into account the initial comoving size $R$ (see equation \ref{voidcharacteristicsize}) of a region that is identified as a depression/void in the growing void algorithm by following \cite{sw}. This means that a void just reaches shellcrossing and final size rv of the void after shellcrosing has not been used in our estimations and calculations. Therefore the void merging algorithm provides an useful framework which can be used in numerical studies as well as observations. However the main challenge is to construct an algorithm that can deal with nonlinear evolution by taking care of full void evolution processes.

Another possible application of the growing void algorithm can be constructing the dynamics of reionization bubbles. There are few attempts to combine excursion set theory and dynamics of reionization bubbles. An interesting work comes from the work of \cite{Shang2007}, in which a simple model is made for the spatial distribution of preheated regions. The model assumes spherical ionized bubbles around collapsed dark matter halos and allows these spheres to merge into larger superbubbles. Also \cite{D'Aloisio2007} present analytic estimates of galaxy void sizes at redshifts $z\sim 5-10$ using the excursion set formalism. As a result, it is possible to obtain a toy model based on the growing void merging tree, which would describe the evolution of the network of reionization bubbles.

In short, the growing void merging algorithm allows us to construct a relatively simple approximate void merging formalism based on the EPS formalism. However, the key issues are how we would translate this into real time such as the Hubble time or the expansion factor, and how to do this for non EdS Universes in which this assumption is not valid for later evolutionary times. These issues lead us to look for more advanced techniques to construct more realistic void merging algorithm(s). However, as a first step to obtain an insight into understanding a void merging algorithm, this is a particularly important model due to the often ill-defined nature of voids. We should  not forget that a proper full understanding of the formation and dynamics of the Cosmic Web is not possible without understanding the structure and evolution of voids \citep{sw} since voids are a good probe of cosmological parameters. For example, it is possible to constrain the dark energy equation of state due to the fact that matter flow in voids consists of information about dark energy $\Omega_{\Lambda}$ and matter density $\Omega_{m}$ \citep{1997MNRAS.290..566B,2004MNRAS.353..162S,2006MNRAS.367.1629S,2007PhRvL..98h1301P,2012MNRAS.426..440B,2013arXiv1304.5239S}. Currently, our void merging tree formalism may provide a guideline to define and trace the evolution of voids as seen in $N$-body simulations.

\section*{Acknowledgments}
Russell would like to thank Johan Hidding for providing the simulation based on the Adhesion model in subsection \ref{subsec:evolution} and Prof. Ravi Sheth for useful discussions on merging voids. Also Russell is grateful for the insightful comments and suggestions of the anonymous referee.
\bibliographystyle{mn2e}
\bibliography{bib}
\appendix
\section{Analytical Derivations In the Void Merging Tree} \label{appendix:exactsolutiononmergertree}
Analytical solutions derived from formation time probabilities, equations (\ref{probhalovoidc}) and (\ref{reducedprobability}), and in the section \ref{subsubsection:Void counting}, are as follows,

\begin{eqnarray}
P\left(\delta_{f} < {{\delta_{\mathbf{v}_{1}}}} |V_{2},{{\delta_{\mathbf{v}_{2}}}}\right)&=&P\left(V_{1}< V_{2}/2 {{\delta_{\mathbf{v}_{1}}}}|V_{2},{{\delta_{\mathbf{v}_{2}}}}\right)\nonumber\\
&=&\int^{S_{h}=S_{2}({V_{2}}/2)}_{S_{2}}\left(\frac{V_{2}}
{V_{1}}\right)f_{S_{1}}\left(S_{1},|\delta_{\mathbf{v}_{1}}||S_{2},|\delta_{\mathbf{v}_{2}}|\right)d S_{1},
\nonumber
\end{eqnarray}
\noindent
where ${V_{2}}/{V_{1}}$ is the weighting factor and $S_{h}=S({V_{2}}/2)$. The exact solutions of this probability function in terms of self similar models $n= 1,0, -1.5, -2$ ($n=-1$ has a numerical solution) are as follows,

\small
\begin{eqnarray}
P_{+1}&=&\sqrt{\frac{2}{\pi}}\frac{k^2}{S^{3/4}_{2}}\left(S_h-S_2\right) e^{-\frac{k^{2}}{2\left(S_h-S_2\right)}}\left[k+e^{\frac{k^{2}}{2\left(S_h-S_2\right)}}\sqrt{\frac{\pi}{2\left(S_h-S_2\right)}}
\left(k^2-S_{2}\right)\text{erf}
\left(\frac{k}{\sqrt{{2\left(S_h-S_2\right)}}}\right)\right],\\
P_{0}&=&\frac{1}{S_{2}}\left[\text{erf}\left(\frac{k}{\sqrt{2} \sqrt{S_h-S_2}}\right) \left(k^2-S_2\right)-\sqrt{\frac{2}{\pi }}k \sqrt{S_{h}-S_2} e^{\frac{-k^2}{\left(S_h-S_2\right)}}\right],\\
P_{-1.5}&=&\frac{k}{S^{2}_{2}}\left(\text{erf}\left(\frac{k}{\sqrt{2} \sqrt{S_{h}-S_2}}\right)
\left(2 k S_{2}-\frac{k^3}{3} -\frac{S^{2}_{2}} {k}\right)+\frac{1}{3}\sqrt{\frac{2}{\pi}}\sqrt{S_{h}-S_{2}}\left(S_{h}+5 S_{2}-k^2\right)e^{- \frac{k^2}{2\left(S_{h}-S_{2}\right)}}\right),\\
P_{-2}&=&\frac{k}{S^{3}_{2}}\left(\text{erf}\left(\frac{k}{\sqrt{2} \sqrt{S_{h}-S_2}}\right)
\left(\frac{k^5}{15}-{k^3} S_{2}+3 k S^{2}_{2}-\frac{S^{3}_{2}} {k}\right)+\frac{1}{15}\sqrt{\frac{2}{\pi}}\sqrt{S_{h}-S_{2}}\left(k^4-k^2 S_{h}+3 S^{2}_{h}\right.\right.\nonumber\\&+& \left.\left.S_{2}(9S_{h}-14k^2)+33 S^2_{2}\right)e^{- \frac{k^2}{2\left(S_{h}-S_{2}\right)}}\right),
\label{probanalyticsol}
\end{eqnarray}
\normalsize
\noindent
in which parameter $k$ is defined as,

\begin{eqnarray}
k\equiv|\delta_{\mathbf{v}_{1}}|-|\delta_{\mathbf{v}_{2}}|.
\label{deltaparameter}
\end{eqnarray}
\noindent
The solutions of the probability distribution of formation times is given by the following expression,

\begin{eqnarray}
P\left( > \delta_{f}\right)=\int^{1}_{0}\frac{1}{2\pi}\left[\tilde{S}\left(2^{\alpha}-1\right)+1\right]^{1/\alpha}\frac{\delta_{f}}{\tilde{S}^{3/2}}
\exp\left(-\frac{1}{2}\frac{\delta^2_{f}}{\tilde{S}}\right) d {\tilde{S}},
\nonumber
\end{eqnarray}
\noindent
where the parameters $\tilde{S}$ and $\delta_{f}$ are given by,

\begin{eqnarray}
{\tilde{S}}\equiv\frac{S-S_{2}}{S_{h}-S},\phantom{a} {\delta_{f}}\equiv\frac{\delta-{\delta_{\mathbf{v}_{2}}}}{\sqrt{S_{h}-S_{2}}}.
\nonumber
\end{eqnarray}
\noindent
Hence, analytical solutions of this probability distribution in terms of self similar models $n= 1, 0, -1.5, -2$ are given by,
\small
\begin{eqnarray}
P_{+1}(\delta_{f})&=&\sqrt{\frac{2}{\pi}}\frac{1}{\delta_{f}} e^{-\frac{\delta^{2}_{f}}{2}}\left[1.52 \delta^{2}_{f}+\sqrt{\frac{\pi}{2}}\frac{\delta_{f}}e^{\frac{\delta^{2}_{f}}{2}}\left(1.52 \delta^2_{f}-1\right)\text{erf} \left(\frac{\delta_{f}}{\sqrt{2}}\right)\right],\\
P_{0}(\delta_{f})&=&\delta_{f} \left({\frac{1}{\delta_{f}}}\left(1- \delta^2_{f}\right)+e^{-\frac{\delta^2_{f}}{2}} \sqrt{\frac{2}{\pi }}\right)+ \text{erf}\left[\frac{\delta_{f}}{\sqrt{2}}\right] \left(\delta^2_{f}-1\right),\\
P_{-1.5} (\delta_{f})&=&\frac{1}{3}\left(\text{erf}\left[\frac{\delta_{{f}}}{\sqrt{2}}\right] \left(-3+6 \left(-1+\sqrt{2}\right) \delta^2_{{f}}+\left(-3+2 \sqrt{2}\right) \delta^4_{{f}}\right)+\frac{e^{-\frac{\delta^2_{{f}}}{2}}\delta_{{f}}}{\pi} \left(8-3 \sqrt{2}+3 e^{\frac{\delta^2_{{f}}}{2}} \sqrt{\pi }
-\delta_{{f}} \left(-4+3 \sqrt{2}\right)\right.\right.\nonumber\\
&+& \left. \left. \left. e^{\frac{\delta^2_{{f}}}{2}} \sqrt{\pi}
{\frac{1}{\delta_{{f}}}} \left(6 \left(-1+\sqrt{2}\right)+\left(-3+2 \sqrt{2}\right) \delta^2_{{f}}\right)\right)\right)\right),\\
P_{-2}(\delta_{f})&=&e^{-0.5 \delta^2_{{f}}} \left(37.6 e^{0.5\delta _{{f}}^2}+
\left(25.53 \sqrt{\frac{1}{\delta^2_{{f}}}}-29.32 e^{0.5 \delta^2_{{f}}}\right)\delta^2_{{f}}
+\delta^4_{{f}}\left(2.54 e^{0.5 \delta^2_{{f}}}-2.06 \sqrt{\frac{1}{\delta ^2_{{f}}}}\right)\right.\nonumber\\
&+&\left.\left(\left(-0.044 e^{0.5 \delta^2_{{f}}}+0.04{\frac{1}{\delta_{{f}}}}\right) \delta^2_{{f}}+e^{0.5 \delta^2_{{f}}} \text{erf}\left[0.71 \delta_{{f}}\right]\left(29.32 \delta^2_{{f}}-2.54 \delta^4_{{f}}+0.044 \delta^6 _{{f}}-37.6\right)\right)\right).
\label{probanalyticsol2}
\end{eqnarray}
\bsp
\label{lastpage}
\end{document}